%% file: noutflows-comp.mnras.tex
\documentclass[useAMS,usenatbib]{mnras}

\usepackage{natbib}
\usepackage{graphicx}	
\usepackage{amsmath}
\usepackage[T1]{fontenc}		
\usepackage{color}			
\usepackage[normalem]{ulem}
\usepackage[protrusion=true,expansion=false]{microtype} 
\usepackage[percent]{overpic}

\input{commands.tex}

\title[Molecular gas in four active galaxies]{Properties of cold molecular gas in four type-1 active galaxies hosting outflows}
\author[J. C. Runnoe et al.]
{\parbox{\textwidth}{Jessie C. Runnoe,$^{1}$\thanks{E-mail: \texttt{jessie.c.runnoe@vanderbilt.edu}}
Kayhan G{\"u}ltekin,$^{2}$
David Rupke,$^{3}$
and Ana L\'opez-Sepulcre$^{4,5}$}\vspace{0.4cm}\\
\parbox{\textwidth}{$^{1}$Department of Physics and Astronomy, Vanderbilt University, Nashville, TN 37235, USA\\
$^{2}$Department of Astronomy, University of Michigan, 1085 S. University Ave., Ann Arbor, MI 48109, USA \\
$^{3}$Department of Physics, Rhodes College, 2000 N. Parkway, Memphis, TN 38104, USA \\
$^{4}$Univ. Grenoble Alpes, IPAG, F-38000 Grenoble, France \\
$^{5}$Institut de RadioAstronomie Millim{\'e}trique (IRAM), 300 rue de la Piscine, F-38406 Saint-Martin d'H{\`e}res, Grenoble, France}}

\begin{document}		

\date{Preprint 2020 July 27}

\pagerange{\pageref{firstpage}--\pageref{lastpage}} \pubyear{2020}

\maketitle

\label{firstpage}
	

\begin{abstract}
Feedback from active galactic nuclei (AGN) has proven to be a critical ingredient in the current picture of galaxy assembly and growth.  However, observational constraints on AGN-driven outflows face technical challenges and as a result, the cold molecular gas outflow properties of type-1 AGN are not well known.  We present new IRAM NOrthern Extended Milimeter Array (NOEMA) observations of \COIO\ in F07599+6508, Z11598$-$0112,  F13342+3932, and PG1440+356, all nearby type-1 AGN and ultraluminous infrared galaxies (ULIRGs).  We achieve spatial resolution of 1--3\arcsec\ corresponding to physical scales of 2--8~kpc and spectral resolution of 15--60~\kms\ which enables updated \COIO\ redshifts and a detailed morphological view of the cold molecular gas in these sources.  The \COIO\ luminosities, $L_{CO}^{\prime}$, are in the range 2--12$\times 10^9$\,K~km~s$^{-1}$~pc$^{2}$ and inferred molecular gas masses, $M$(H$_2$), are in the range 2--9$\times 10^9$~\Msun.  The velocity fields and gas distributions do not unambiguously identify any of these sources as having outflows.  However, Z11598$-$0112 has signs of infalling material and after the subtraction of a rotating disk model PG\,1440+356 shows complex kinematics in the residuals that may indicate an outflow or warped disk.  
\end{abstract}


\section{Introduction}
Gas outflows driven by active galactic nuclei (AGN) and stellar processes can be an important mode for transporting energy in galaxies \citep[e.g.,][]{veilleux05,harrison18}. Large-scale AGN-driven outflows are a manifestation of quasar-mode feedback, which is invoked in semi-analytic models and simulations \citep{silk98,king03,dimatteo05} to solve problems with massive galaxy cooling and star formation quenching in a cosmological context \citep[e.g., see][]{somerville08,zubovas12}.  However, the degree to which AGN actually drive large-scale outflows \citep[e.g.,][]{karouzos16,villar-martin16,fischer17,fischer18}, their relative importance compared to stellar processes \citep{wylezalek18}, and especially their impact on star formation in the host galaxy \citep[e.g.,][]{scholtz20} remain unclear. 

The observational detection of AGN-driven outflows is challenging due primarily to difficulty of spatially resolving the outflow \citep[e.g.,][]{rupke17}, the difficulty in separating it from the central point source and host galaxy \citep{lutz20}, and the ambiguity of identifying its physical driver and impact on the host \citep{scholtz20}. Nearby type-2 AGN have proven to be excellent laboratories for investigating outflows because of the achievable spatial resolution and improved contrast to the central engine \citep{rupke13a,harrison14,mcelroy15,kang18,leung19}.  The outflows in type-1 AGN are less well studied, due to the challenge of observing them in the presence of a bright point source \citep[see][]{rupke17}.  Since some type-1 AGN may differ from type-2 by more than just orientation \citep{alexander12,dipompeo14b,dipompeo15a,dipompeo18}, both are important.  Relatively recent efforts show that outflows in type-1 AGN are common; they have been observed in ionized \citep{westmoquette12,arribas14,harrison14,rupke17}, neutral \citep{morganti05,rupke05,cazzoli16,morganti16,rupke17}, and molecular \citep{feruglio10,rupke13b,cicone14,carniani15,garcia-burillo15,fiore17} gas phases.  

The practice of observing outflows in all gas phases (which is also applied to type-2 sources) is especially critical to constrain the global outflow properties and thus their true impact \citep{cicone18}.  The molecular gas phase is a constituent of the cold interstellar medium (ISM) in galaxies, which directly fuels star formation and holds a large fraction of the mass outflow rate in AGN \citep{fluetsch20}.  Molecular outflows \citep[for a review, see][]{veilleux20} have been identified in a number of individual objects \citep[e.g.,][]{feruglio10,fischer10,alatalo11,dasyra12,nesvadba10,combes13,pereira-santaella16,salak16,slater19,alonso-herrero19}, typically via P-Cygni profiles, interferometric obserations, or the high-velocity wings of emission lines like \COIO.  Using the same techniques, samples of objects reveal that while stellar processes and AGN can both drive outflows, the outflow properties do correlate with properties of the central engine \citep{sturm11,spoon13,veilleux13b,cicone14,gonzalez-alfonso17,pereira-santaella18,fluetsch19,combes19,ramakrishnan19,lutz20}.  However, these samples are often small, and cold molecular outflows are particularly poorly statistically characterized.  

In this paper, we present new interferometric observations of \COIO\ in 4 nearby ($z<0.3$) Type-1 quasars ($L>10^{45}$~erg~s$^{-1}$) that have observations of warm molecular, neutral, or ionized outflows.  This is part of an ongoing effort to characterize the multi-phase outflow properties in type-1 AGN \citep{rupke17}, and this installment was undertaken with the aim of quantifying the cold molecular gas properties with increased spatial resolution and identifying outflows if they are present.

This paper is organized as follows:  In Section~\ref{sec:data} we introduce the targets, present the new observations, and describe the data reduction process.  Section~\ref{sec:analysis} details the analysis of the \COIO\ data cube, including measurements of the \COIO\ properties.  Section~\ref{sec:obj} we discuss the details of individual objects, including a detailed calculation of the kinematics in one object in Section~\ref{sec:outflow}.  We discuss this work in the context of previous results and summarize our findings in Section~\ref{sec:discussion}.  Throughout this work, we adopt a cosmology of $H_0 = 73\;{\rm km\; s^{-1}\;Mpc^{-1}}$, $\Omega_{\Lambda} = 0.73$, and $\Omega_{m} = 0.27$.

\section{Targets, observations, and data reduction}
\label{sec:data}
In this section we describe the acquisition and analysis of \COIO\ ($\nu_{\mathrm{rest}} = 115.271$~GHz) observations for F07599+6508, Z11598$-$0112,  F13342+3932, and PG1440+356. These sources were chosen from the larger Quasar and ULIRG Evolution Study sample \citep[QUEST][]{schweitzer06,veilleux06,veilleux09a,veilleux09b,netzer07}, which comprises local ultraluminous infrared galaxies (ULIRGs) from the 1~Jy sample \citep{kim98} and Palomar-Green (PG) quasars \citep{schmidt83}.  The specific targets were chosen based on a qualitative combination of observability, expected signal-to-noise ratio based on existing single-dish observations, achievable spatial resolution, and the availability of multi-wavelength data to determine physical and outflow properties in other gas phases.  The heterogenous nature of the sources reflects our ongoing efforts to observe the multi-phase outflow properties of the QUEST sample.

We observed the targets with the IRAM NOrthern Extended Millimeter Array (NOEMA) between January 2016 and January 2019. During the time of our observations, the telescope underwent substantial hardware upgrades, including three new antennas and a new correlator.  We document these changes and summarize other technical details of the observations for each object in Table~\ref{tab:obslog}.

\input{./tables/obslog_mnras.tex}

The data calibration was performed using the \textsc{gildas} package \textsc{clic} (versions oct16--aug19a).  Although each object was observed in multiple tracks, the tracks often shared flux calibrators.  Notably, there was very good agreement between the fluxes obtained for calibrators observed on multiple nights.  Following standard practices, the accuracy of the absolute flux calibration in the 3~mm band is better than 10\% \citep{castro-carrizo10}.  

We used the WideX correlator and then the new PolyFiX correlator after it was installed in 2017.  WideX has a bandwidth of 3.6~GHz and spectral resolution of 1.95~MHz.  PolyFiX has a configurable spectral window with the possibility of distributing high spectral resolution chunks in areas of interest.  For consistency with WideX, we used the wideband setup to obtain a 3.9~GHz baseband window with a spectral resolution of 2~MHz centered on the \COIO\ emission line to maximize signal.  At the tuning frequency, the spectral resolution corresponds to 5.5--7~\kms.  We binned the $uv$ tables by 2--11 channels to obtain spectral resolution of 60, 15, 21, and 25~\kms\ for F07599+6508, Z11598$-$0112,  F13342+3932, and PG1440+356, respectively. The range in spectral resolution is a practical consequence of the fact that the observations were designed and scheduled completely independently based on known multi-wavelength properties of the individual sources. The velocity scale was initially set using the redshift values listed in Table~\ref{tab:obslog}.  These were determined based on optical emission lines, with one exception.  The redshift for F07599+6508 was measured by \citet{strauss92} based on the H$\alpha$ and [\ion{N}{2}]~$\lambda$6583 emission lines.  The redshift for F13342+3932 is based on a fit to optical integral field spectroscopy (IFS) of the host galaxy \citep{rupke17}.  In PG\,1440+356 and Z11598$-$0112 the \COIO\ redshift was not well matched to the optical redshift.  The velocity scales for these objects were set based on visual inspection of the NOEMA spectrum during the data reduction to put the center of the line profile at the systemic velocity.  The redshift that achieves this for PG\,1440+356 is consistent with but more precise than the one obtained from single-dish \COIO\ observations and optical spectroscopy \citep{evans09,veilleux13b}. For Z11598$-$0112, the \citet{strauss92} optical redshift of 0.15069$\pm0.0002$ is not consistent with the NOEMA redshift. Based on visual inspection, we identified two line-free continuum regions per object.  These were used to generate a continuum $uv$ table, which was subtracted to isolate a $uv$ table for \COIO\ \citep[e.g.,][]{cicone14}.

The image cleaning and analysis was done with the \textsc{mapping} package in \textsc{gildas}. The absolute astrometric uncertainty, including both telescope and observational contributions, is of order $10^{-1}$ the beam size, or 0\farcs15--0\farcs32.  We generated a $1024\times1024$ pixel map with pixel size recommended by the \textsc{gildas} software based on the synthesized beam size: 0\farcs163, 0\farcs16, 0\farcs17, 0\farcs1524 for F07599+6508, Z11598$-$0112,  F13342+3932, and PG1440+356, respectively. Observations were made with a single pointing per target, so the largest angular scale is determined by the shortest baseline given in Table~\ref{tab:obslog}. The image was cleaned with natural weighting down to 3 times the noise determined from an initial cleaned version of the image, using a $25\arcsec\times25\arcsec$ square support mask around the center of the field.  The rms noise level in the cleaned, continuum-subtracted cubes is a function of frequency, with an average flux uncertainty of 0.259--0.367~mJy~beam$^{-1}$ across all channels.

We present several visualizations (channel maps, spectral maps, and integrated spectra) of the reduced \COIO\ data cubes in Appendix~\ref{app:maps}. All visualizations are centered on the coordinates in Table~\ref{tab:obslog} and the \COIO\ redshift in Table~\ref{tab:prop}. We discuss these in the context of further multi-wavelength data for each object in Section~\ref{sec:obj}.

\begin{figure*}
\vspace{-.5cm}
\centerline{
\includegraphics[width=5.1cm]{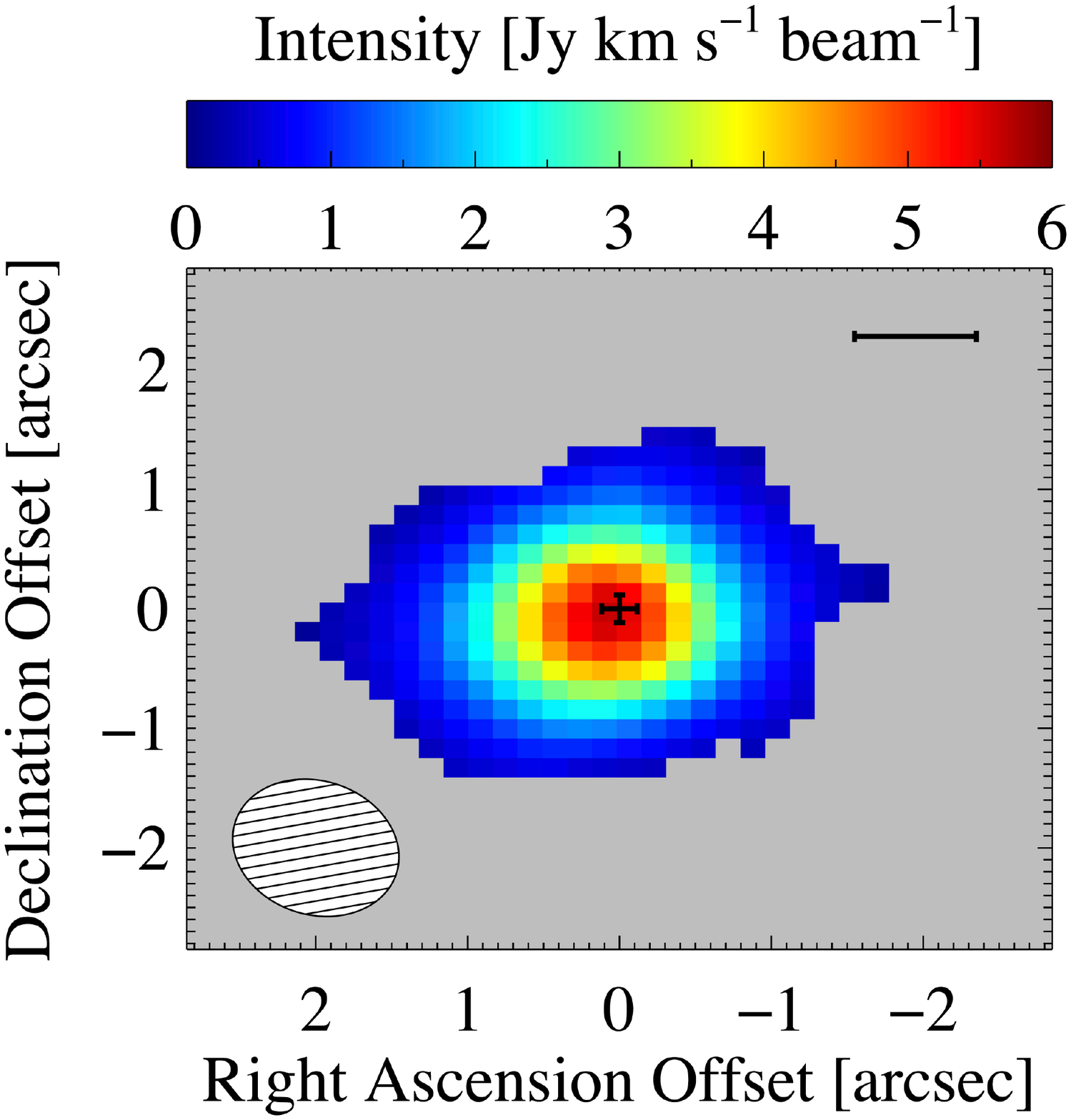}
\hspace{-0.55cm}
\includegraphics[width=5.1cm]{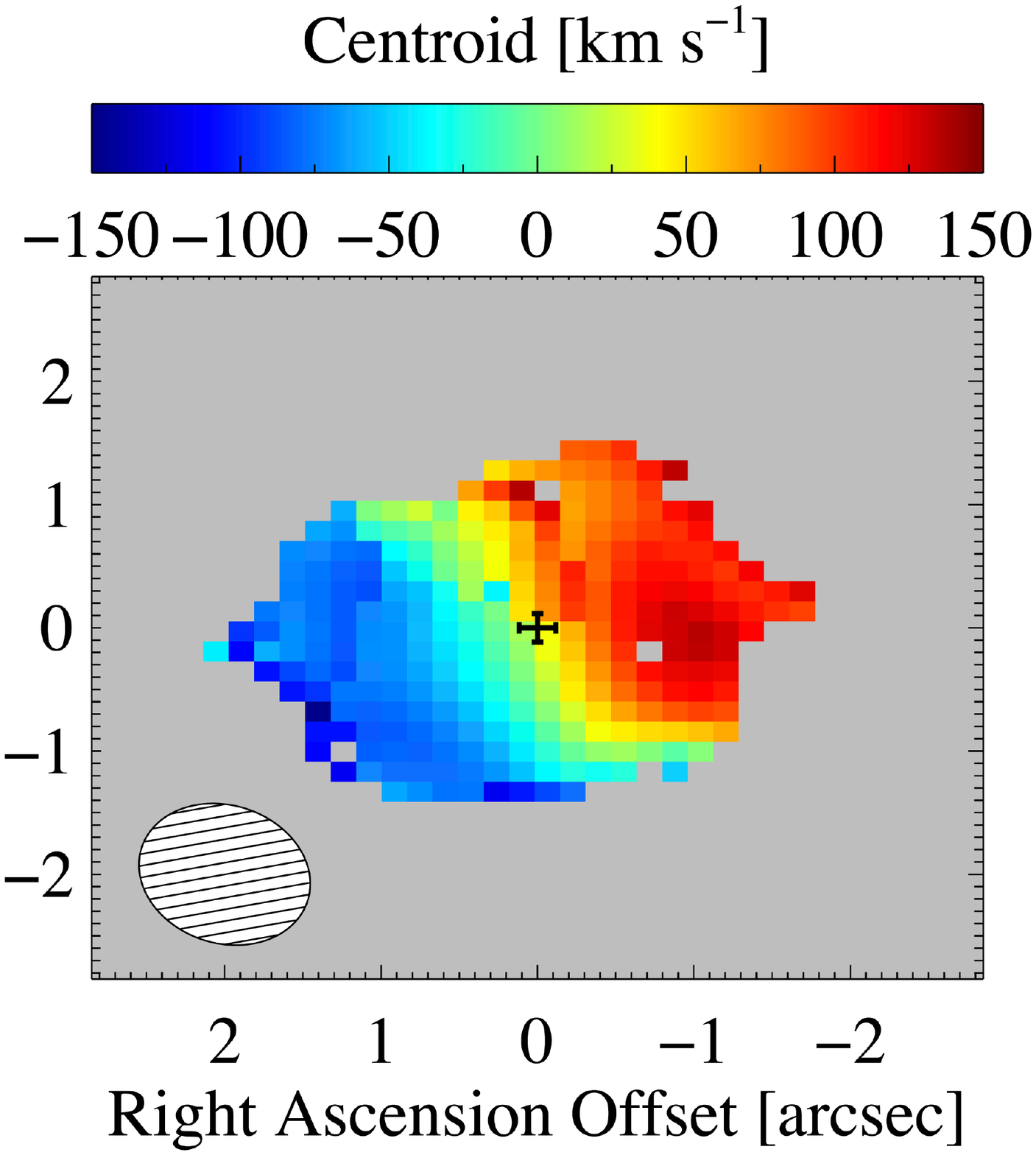}
\hspace{-0.85cm}
\begin{overpic}[width=5.1cm]{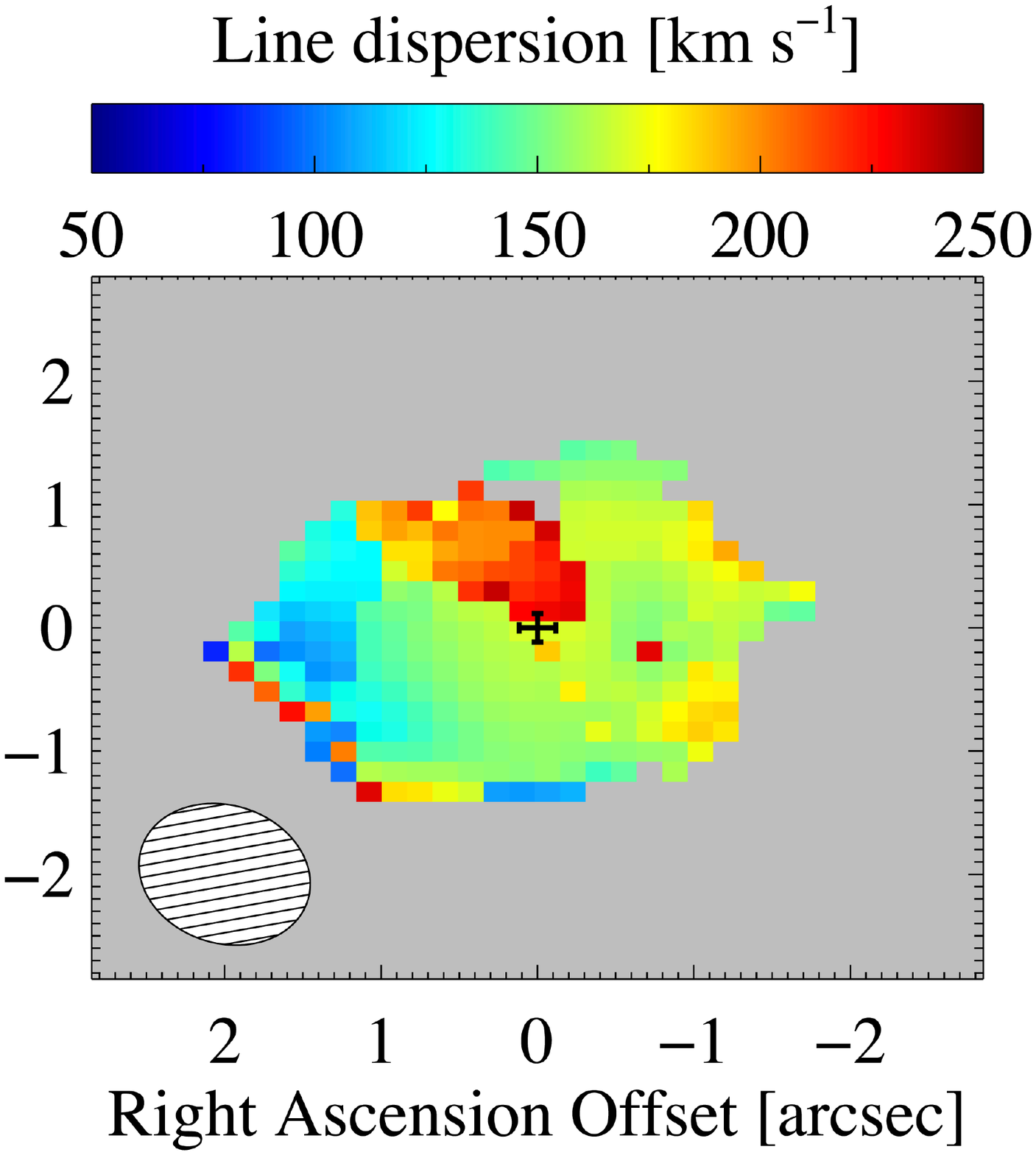}
  \put (67,69.5) {\rotatebox{-90}{\large{F07599+6508}}}
\end{overpic}
}
\vspace{-1.2cm}
\centerline{
\includegraphics[width=5.1cm]{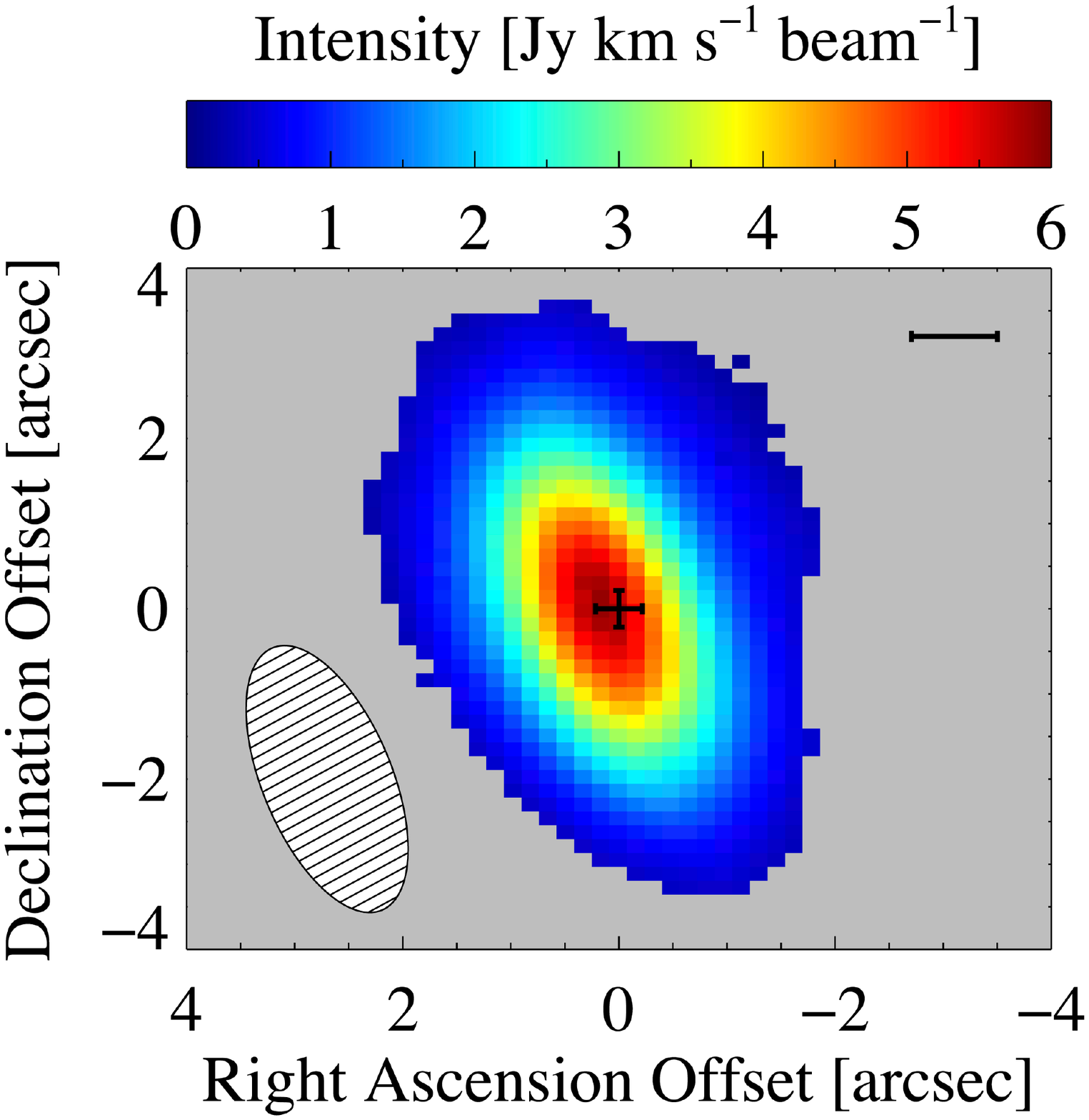}
\hspace{-0.55cm}
\includegraphics[width=5.1cm]{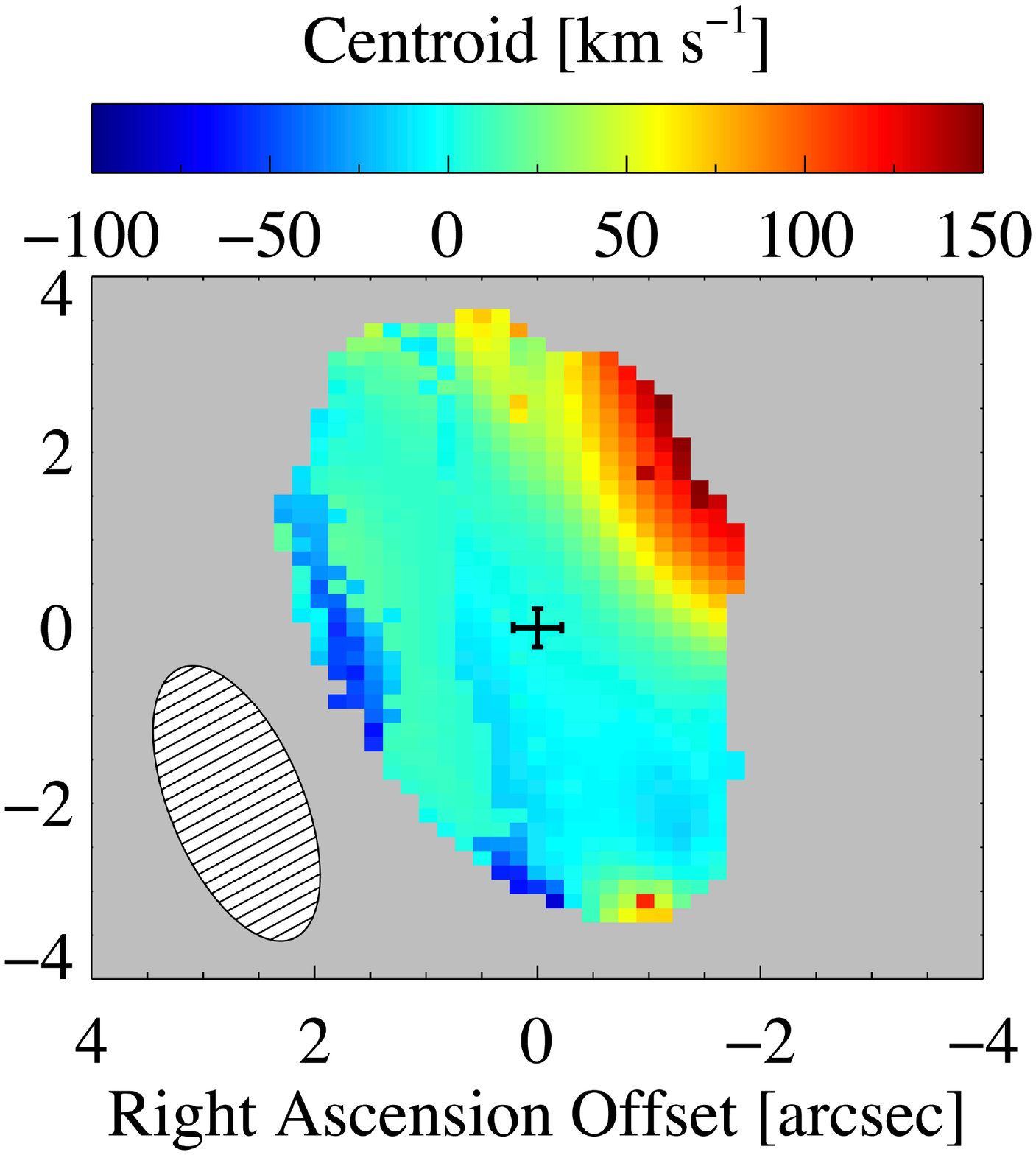}
\hspace{-0.85cm}
\begin{overpic}[width=5.1cm]{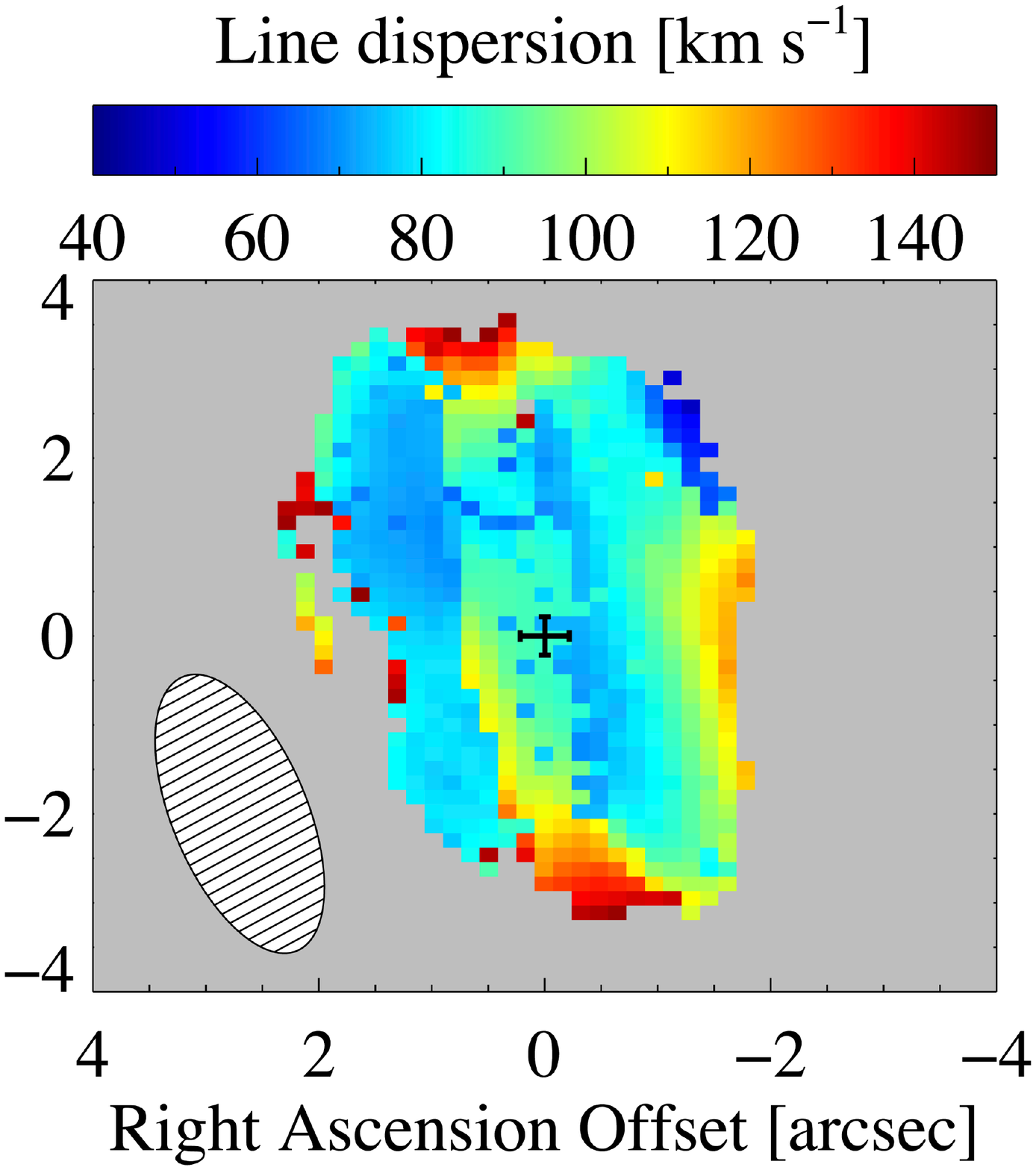}
  \put (67,69.5) {\rotatebox{-90}{\large{Z11598$-$0112}}}
\end{overpic}
}
\vspace{-1.2cm}
\centerline{
\includegraphics[width=5.1cm]{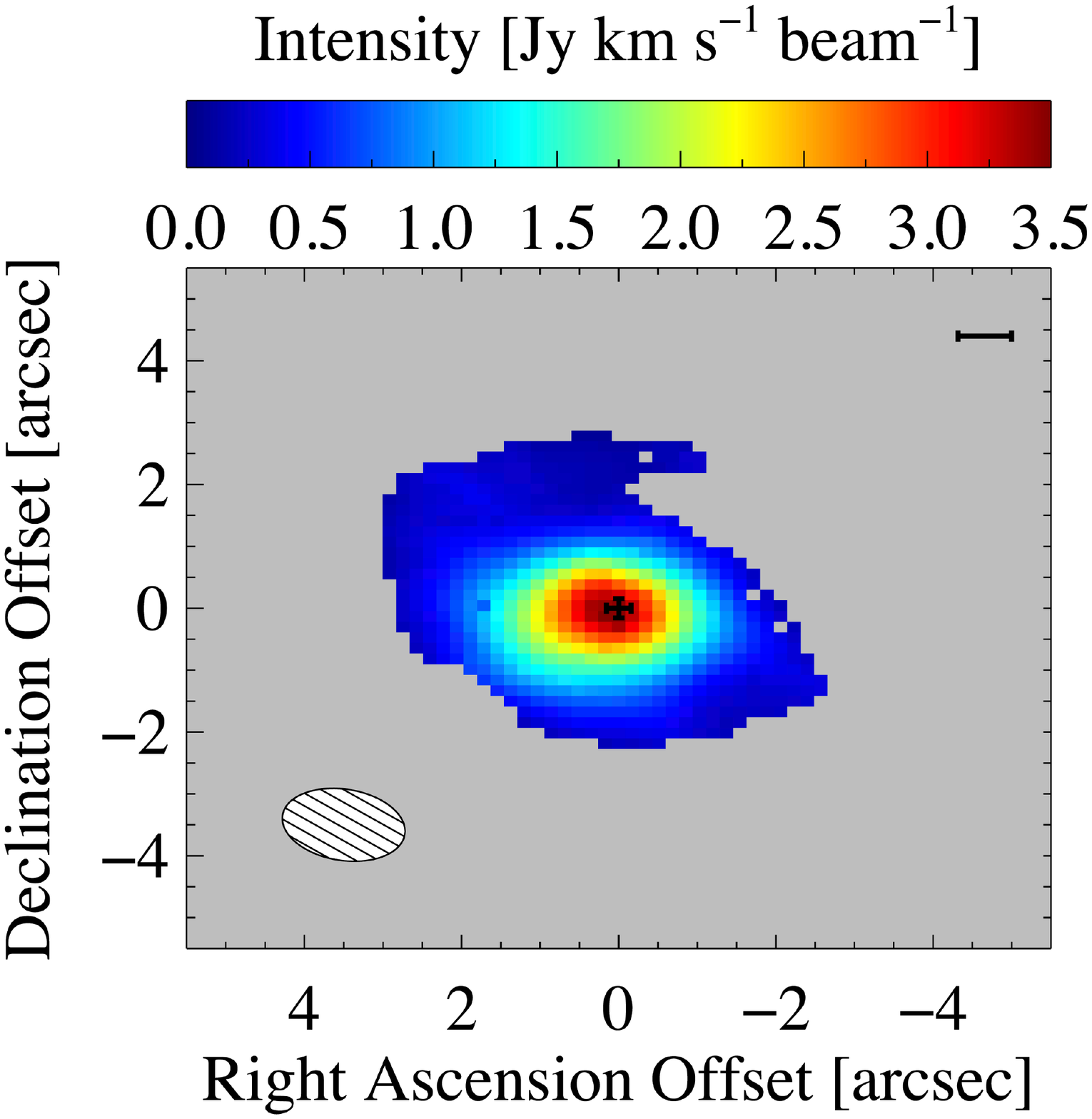}
\hspace{-0.55cm}
\includegraphics[width=5.1cm]{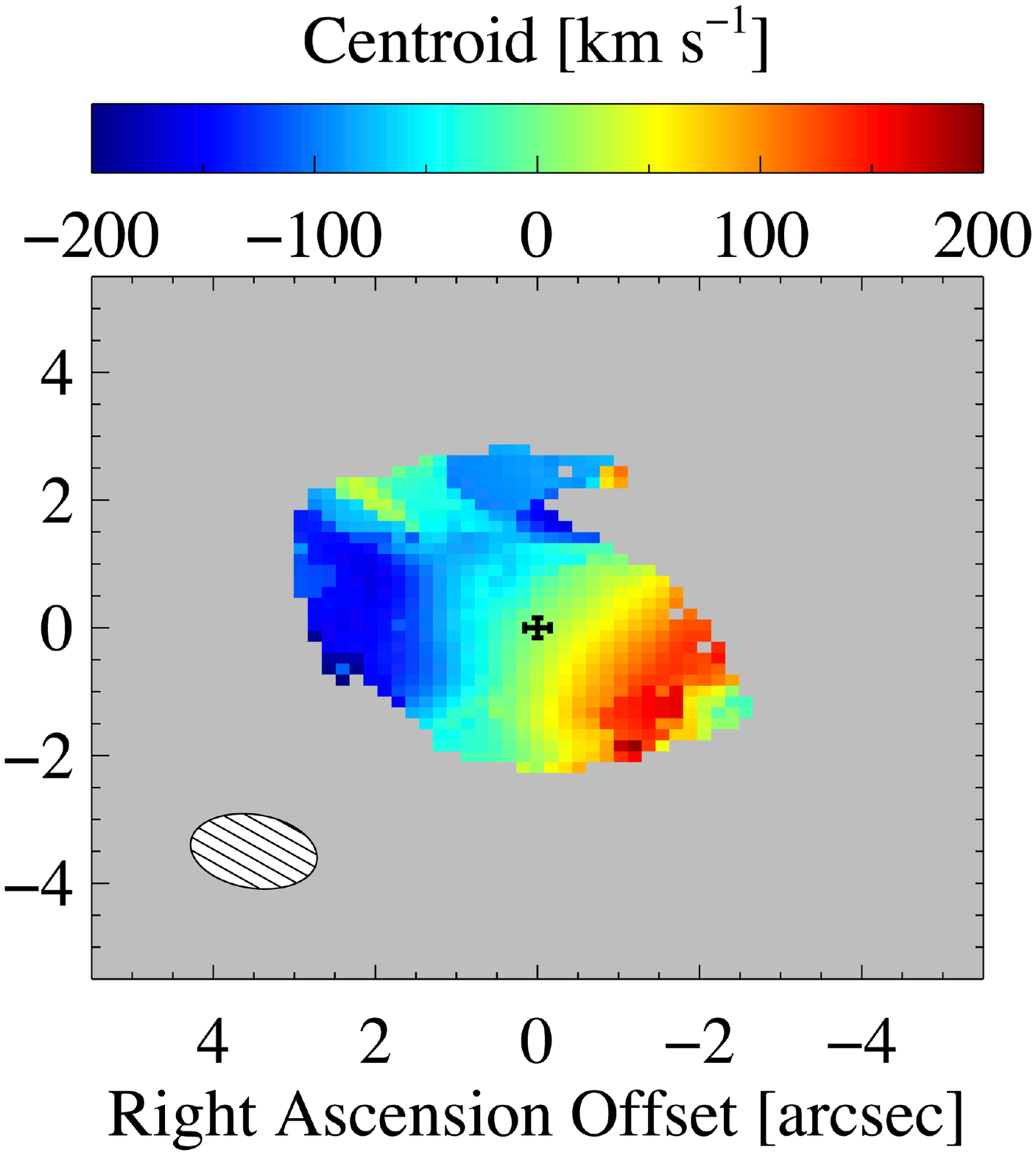}
\hspace{-0.85cm}
\begin{overpic}[width=5.1cm]{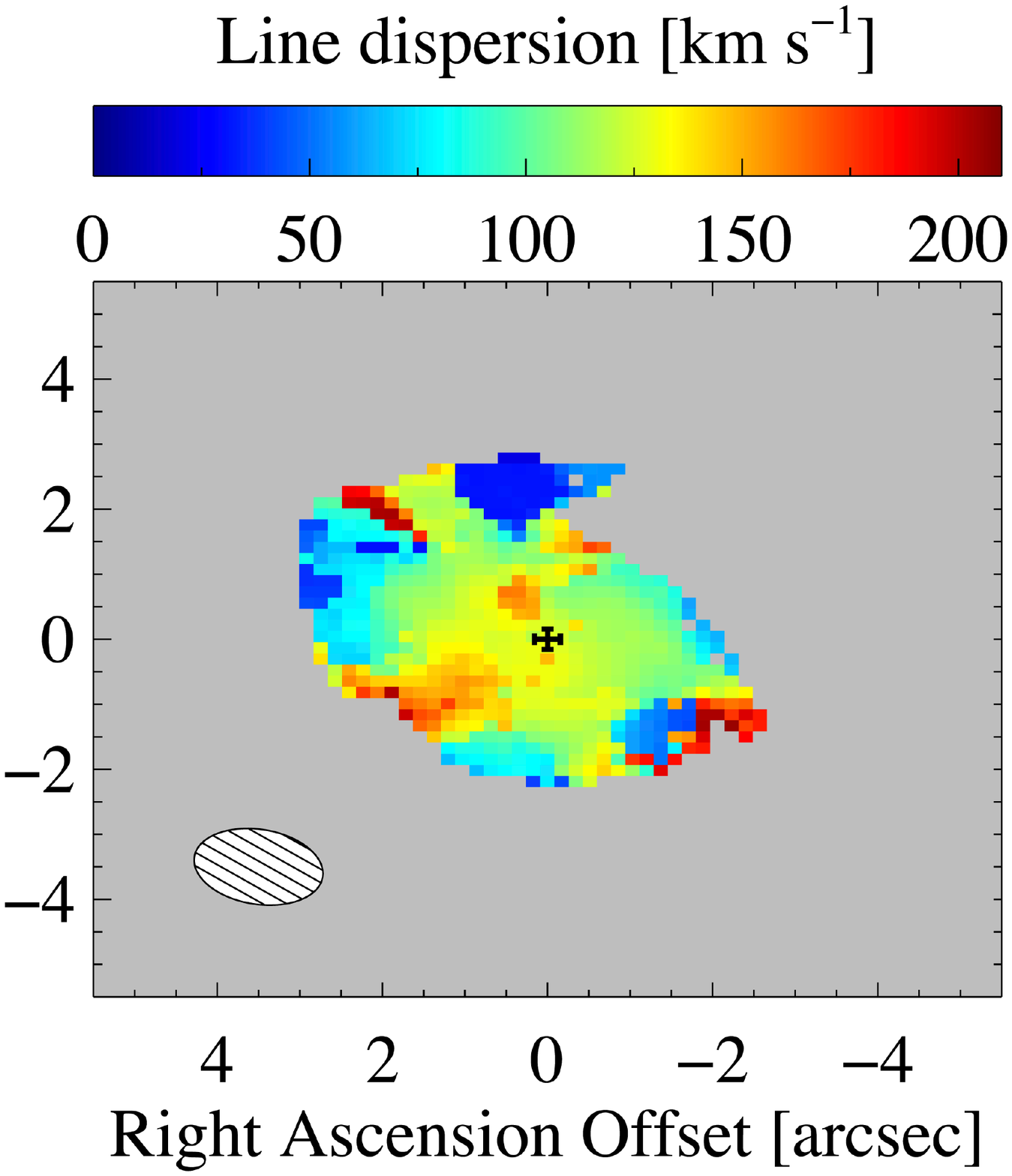}
  \put (67,69.5) {\rotatebox{-90}{\large{F13342+3932}}}
\end{overpic}
}
\vspace{-1.2cm}
\centerline{
\includegraphics[width=5.1cm]{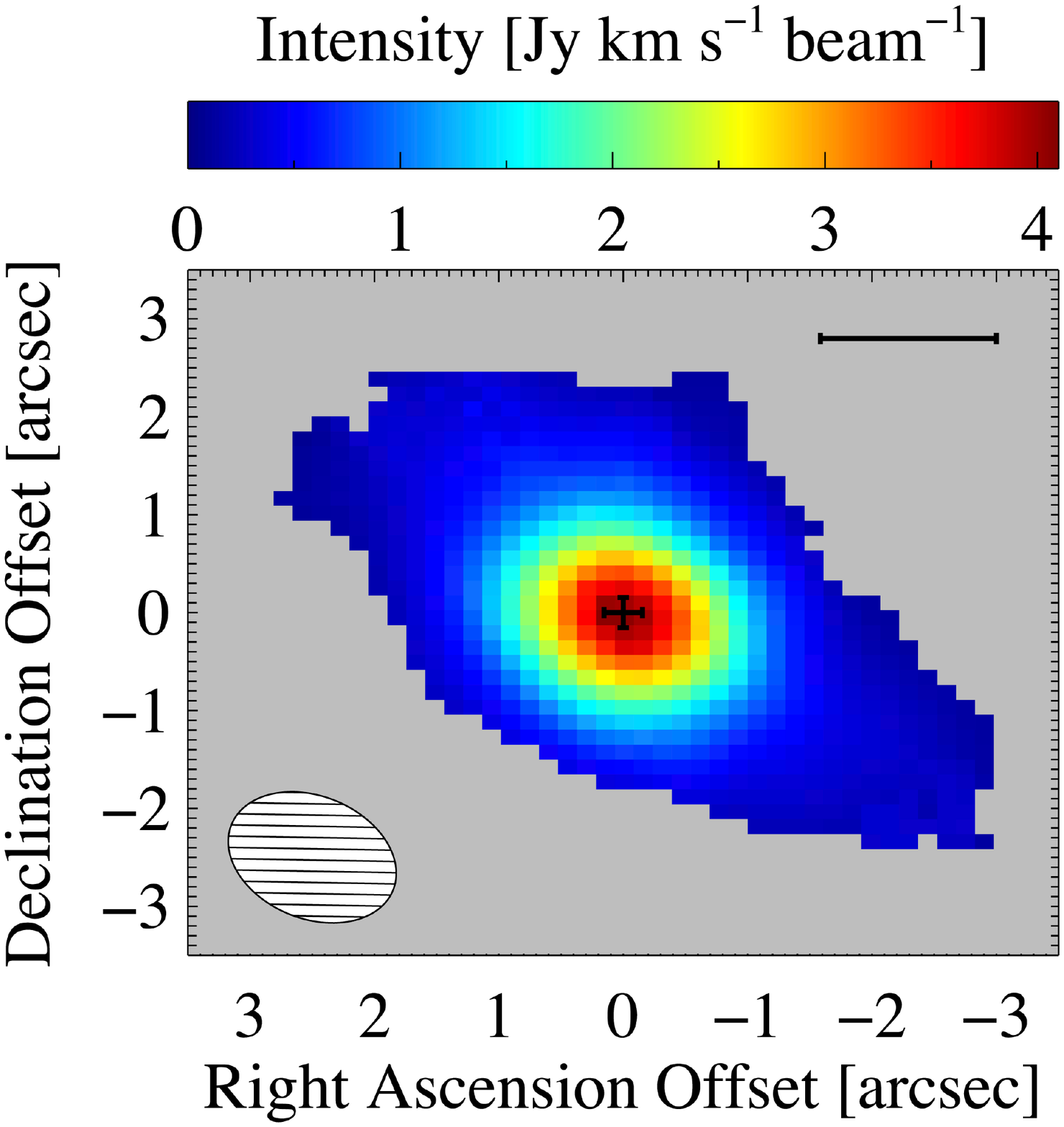}
\hspace{-0.55cm}
\includegraphics[width=5.1cm]{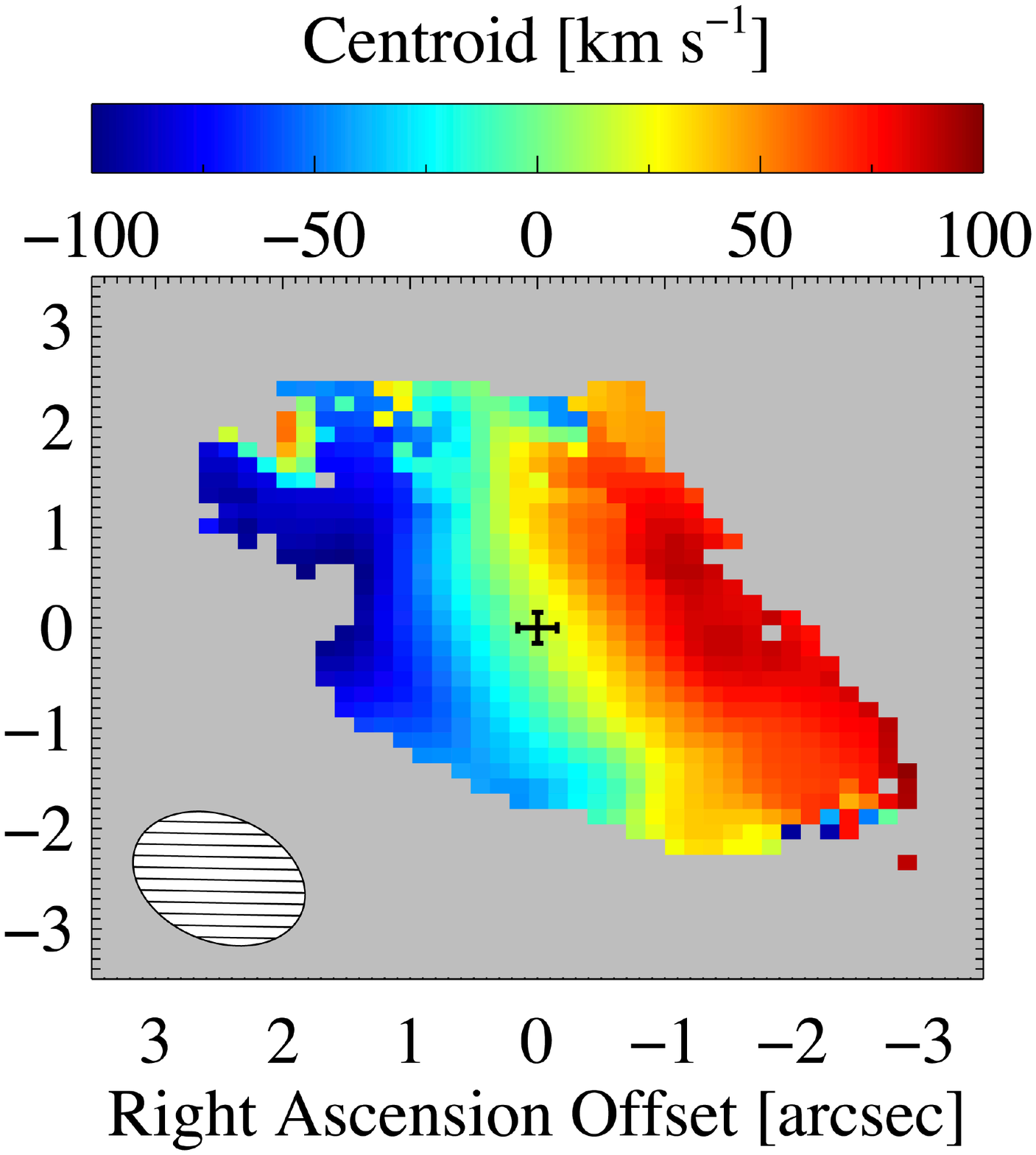}
\hspace{-0.85cm}
\begin{overpic}[width=5.1cm]{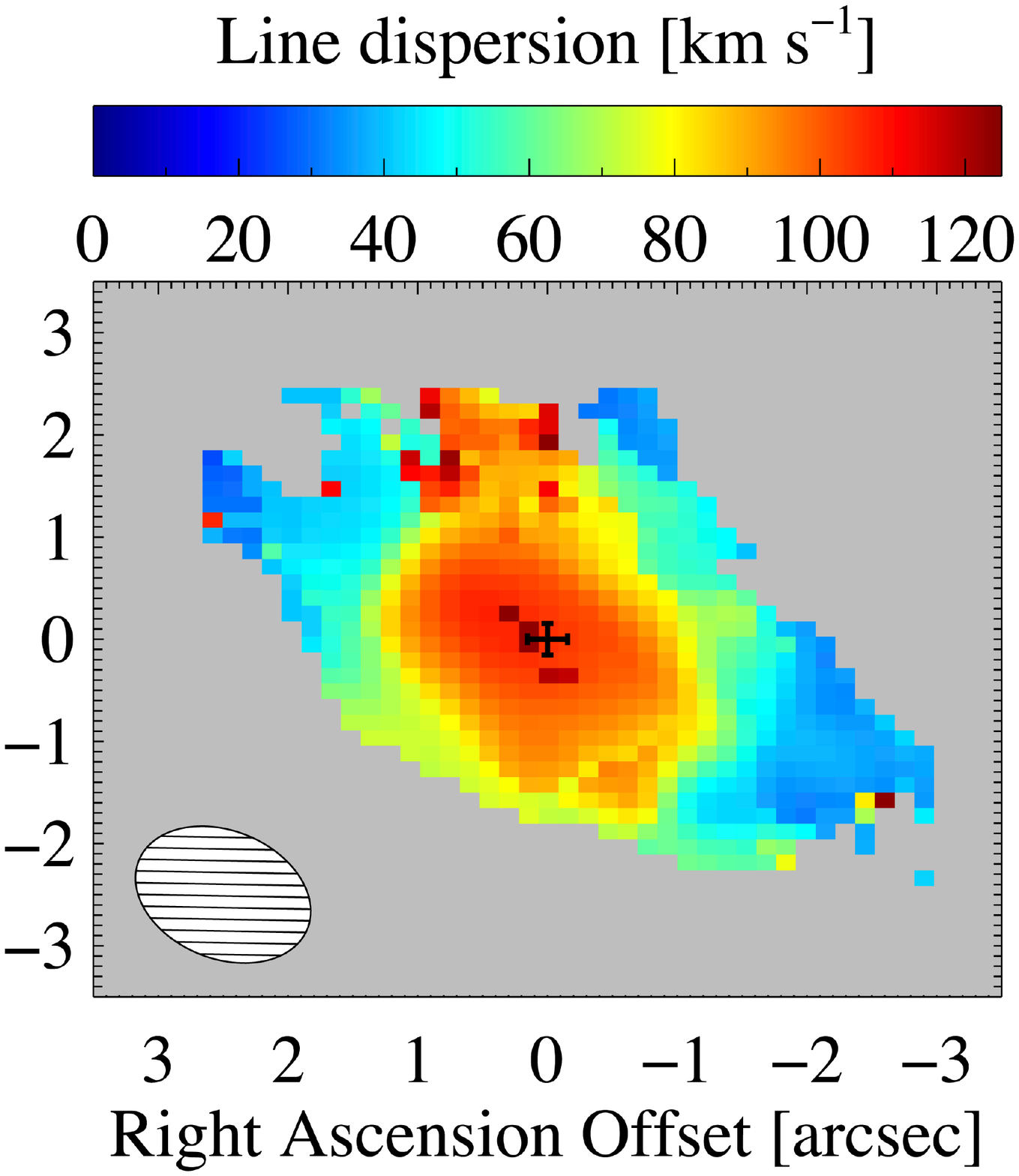}
  \put (67,69.5) {\rotatebox{-90}{\large{PG1440+356}}}
\end{overpic}
}
\vspace{-0.5cm}
\caption{Maps of the integrated flux, velocity centroid, and velocity dispersion of the \COIO\ emission line for each object.  Formal intensity uncertainties are $0.02-0.15$~Jy~beam$^{-1}$, centroid uncertainties are $2-150$~km~s$^{-1}$, and dispersion uncertainties are $1-130$~km~s$^{-1}$.  These maps include measurements from the best-fitting model for spaxels where the total line profile was detected at the 5$\sigma$ level, as described in Section~\ref{sec:analysis}.  The black cross shows the location of the coordinates in Table~\ref{tab:obslog} with NOEMA absolute positional uncertainties comparable to $10^{-1}$ of the size of the beam, which is shown as a hatched ellipse in each panel. The scale bar marks 2~kpc.\label{fig:3map}}
\end{figure*}

\section{Analysis of the NOEMA observations}
\label{sec:analysis}
In this section we present the analysis of the new \COIO\ data cubes for the four AGN listed in Table~\ref{tab:obslog}.  This includes the spectral decomposition of the data cubes and measurements made to quantify the \COIO\ luminosity and molecular gas mass.
\subsection{Spectral decomposition and physical properties}
\label{sec:specfit}
In order to measure physical properties from the \COIO\ line profile, we decomposed the continuum-subtracted spectra following \citet{runnoe18}. Briefly, each profile was modeled with up to three Gaussians, which were allowed to vary completely independently and to which no physical meaning was attributed. We imposed several constraints to prevent over-fitting of noise spikes and ensure meaningful results: we required that the fluxes of the Gaussians be positive, the widths be larger than the velocity resolution, and the position of the second and third Gaussians be within the domain of the line profile. We used the IDL package \textsc{mpfit} \citep{markwardt12} to minimize the chi-squared statistic and identify the best-fitting model.      


We measured the properties of the line profiles in spaxels where the \COIO\ line was robustly detected. Specifically, for a detection we required that the peak of the best-fitting model be at least 5 times the noise, which is taken to be the average rms of continuum channels with $1000<v<1500$~\kms. We measured the integrated flux, velocity centroid, and velocity dispersion within $\pm1000$~\kms\ of zero velocity from the best-fitting model in detected spaxels. Uncertainties on these measured properties were taken to be the standard deviation of the distribution for each measurement determined by Monte Carlo techniques for the entire decomposition 10$^3$ times. In each iteration, we decomposed a synthetic spectrum calculated by perturbing the flux in each channel according to a normal distribution with width equal to its measured noise. From this we built distributions of integrated flux, velocity centroid, and velocity dispersion whose means were comparable to our measured values in each spaxel and who standard deviations we took to be the uncertainty on that measured value. The formal decomposition in each spaxel yields intensity uncertainties that are $0.02-0.15$~Jy~beam$^{-1}$, centroid uncertainties that are $2-150$~km~s$^{-1}$, and dispersion uncertainties that are $1-130$~km~s$^{-1}$. Adding the velocity resolution in quadrature yields a minimum velocity uncertainty of $\sim7$~\kms.

In Figure~\ref{fig:3map} we show maps of the integrated flux, velocity centroid, and velocity dispersion for each object. Smooth large-scale trends in these maps are real, but we note that our methodology is still susceptible to perturbations in individual spaxels. This occurs most commonly near the edge of the detectable range and most prominently in the velocity dispersion which is the most sensitive measurement to low signal-to-noise \citep{denney09a} with the largest uncertainties.

We derived physical properties for all four sources, which we present in Table~\ref{tab:prop}.  The redshift was determined from the flux-weighted centroid (i.e. the first moment) of the best-fitting line profile in the pixel with the highest integrated \COIO\ flux. All measured properties use this \COIO\ redshift. The total \COIO\ luminosity, $L_{CO}^{\prime}$, was calculated by integrating the data within $\pm1000$~\kms\ of the line center and in a $5\times5$\arcsec\ box. Fluxes and consequently luminosities are 10--20\% lower than published values from single-dish observations \citep{xia12}, which is consistent within typical absolute flux uncertainties. The total molecular gas mass, $M$(H$_2$), follows assuming a conversion from \COIO\ of $\alpha = M(\textrm{H}2)/L_{CO}^\prime \sim 0.8 \textrm{\Msun} (\textrm{K km s}^{-1}\textrm{pc}^2)^{-1}$ \citep{cicone14,lutz20}.  The value is unconstrained in these objects, so we selected the value typical of ULIRGs consistent with other AGN \citep[e.g.,][]{lutz20}. Uncertainties are propagated from the uncertainties in the \COIO\ flux measurements. 

\input{./tables/properties_mnras.tex}

\section{Individual Objects}
\label{sec:obj}
\begin{figure*}
\begin{minipage}{\textwidth}
\vspace{0.5cm}
\begin{overpic}[width=\textwidth]{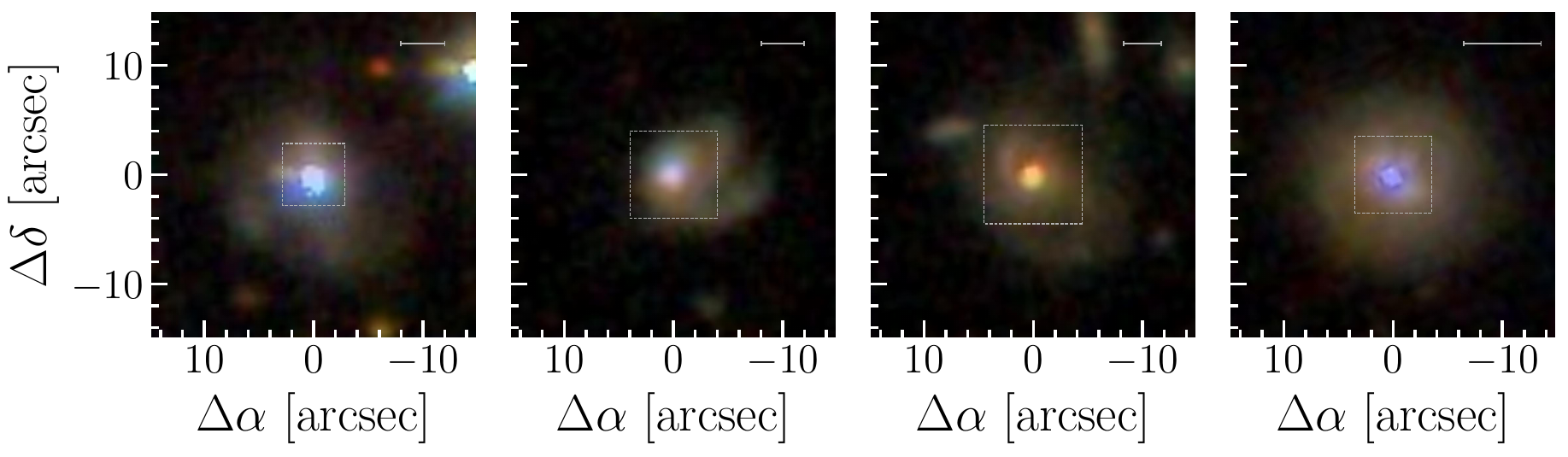}
 \put (9.8,28.4) {\large{F07599+6508}}
 \put (32.7,28.4) {\large{Z11598$-$0112}}
 \put (55.7,28.4) {\large{F13342+3932}}
 \put (78.6,28.4) {\large{PG1440+356}}
\end{overpic}
\end{minipage}
\caption{SDSS $gri$ images of the target host galaxies. In each case, the field is centered on the target using the coordinates in Table~\ref{tab:obslog}, the scale bar marks 10~kpc, and the dashed box shows the map size shown for NOEMA data in Figure~\ref{fig:3map}.  North is up and East is left.\label{fig:sdss}}
\end{figure*}
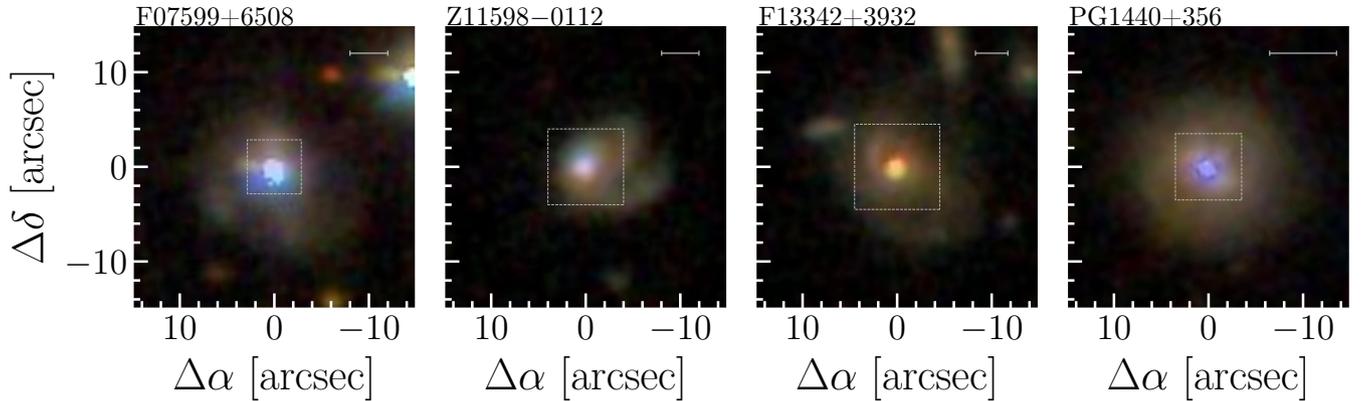

In this section, we provide a discussion of the observations of each source that, when appropriate, is given in the context of existing multi-wavelength information in the literature.  In order to put the new observations in context, we show the Sloan Digital Sky Survey data release 16 \citep[SDSS DR16,][]{dr16} host galaxy images in Figure~\ref{fig:sdss}.

\subsection{F07599+6508}
\label{sec:f07599}
F07599+6508 is a ULIRG \citep{low89} with a Type~1 low-ionization broad absorption line (loBAL) quasar \citep{lanzetta93,lipari94,hines95} nucleus and unusually strong Fe~\textsc{ii} emission \citep{lawrence88}.  In these and other respects, it is reminiscent of better-known Mrk~231 \citep{rupke05}.  The host galaxy is dominated by the bright AGN point source \citep{sanders88}, but shows some relatively symmetric extended structure in {\it Hubble Space Telescope (HST)} $H$-band imaging \citep{veilleux06}.  Outflows have already been detected in multiple gas phases.  There are two neutral outflows in Na~\textsc{i}D.  One is an unresolved, high-velocity outflow \citep{boroson92b,rupke05}, consistent with the resonance response of material entrained in the BAL outflow, and the other low-velocity outflow is uncharacteristically misaligned with the host galaxy's minor axis \citep{rupke17}.  An ionized outflow is detected in [N~\textsc{ii}]~$\lambda6583$ and other strong optical emission lines \citep{rupke17} and a warm molecular outflow via the P-Cygni profile of OH~$\lambda119$~$\mu$m \citep{veilleux13b}.  Existing single-dish \COIO\ observations from the IRAM 30~m telescope \citep[$\theta_b=24$\arcsec,][]{xia12} show a double-peaked line profile.  One narrow peak is near the systemic velocity and there is a second, broad redshifted peak.  They obtain a flux-weighted \COIO\ redshift of 0.148.

We do not detect a \COIO\ outflow in F07599+6508 on the arcsecond scales probed with our observations.  The \COIO\ emission is not well resolved spatially, detected at very high velocities ($>500$~\kms), nor do we observe clear signs of complex kinematic motions beyond a rotating disk.  There is some spatial extension in several velocity channels (e.g., $-120$~\kms, 180--240~\kms), but higher spatial resolution observations would be required to identify departures from a Keplerian motion.  

The line profile that we observe is qualitatively similar to the single-dish observation from \citet{xia12}, after integrating over a large spatial area, but the emission is less redshifted than expected from those data.  We obtain a flux-weighted redshift which is 60~\kms\ larger than the optical redshift.  This is smaller than was obtained for the \citet{xia12} single-dish observations.  They adopt an optical redshift of 0.148 and find a \COIO\ flux-weighted redshift of 0.149, a difference of $\sim\!300$~\kms.  This is in part due to the fact that we obtain the redshift from the central pixel rather than the spatially integrated profile.

\subsection{Z11598\texorpdfstring{$-$}0112}
\label{sec:z11598}
Z11598$-$0112 is a ULIRG \citep{kim98} with a radio-loud narrow-line Seyfert 1 (NLS1) nucleus \citep{condon98b,moran96}.  Optical spectra show no Na~\textsc{i}D absorption that would be indicative of a neutral outflow \citep{rupke05}, but weak absorption is detected in more sensitive data along with a miniBAL in the UV \citep{martin15}.  A warm molecular outflow is detected based on OH~$\lambda$119~$\mu$m absorption with a blueshift of $-153\pm50$~\kms\ \citep{veilleux13b}.  Single-dish CO observations with a $\theta_b=24$\arcsec beam show potential for high-velocity wings to the \COIO\ line profile \citep{xia12}. They obtain a flux-weighted \COIO\ redshift of 0.151, which is also what they tabulate for  the optical redshift.

The line profile that we observe is qualitatively similar to the single-dish observation from \citet{xia12}.  The \COIO\ redshift we obtain is also consistent, given their listed precision.  However, this is larger than the \citet{strauss92} optical redshift by $\sim150$~\kms.

\begin{figure}
\hspace{-0.6cm}
\begin{minipage}[!b]{0.5\textwidth}
\includegraphics[width=8.9cm]{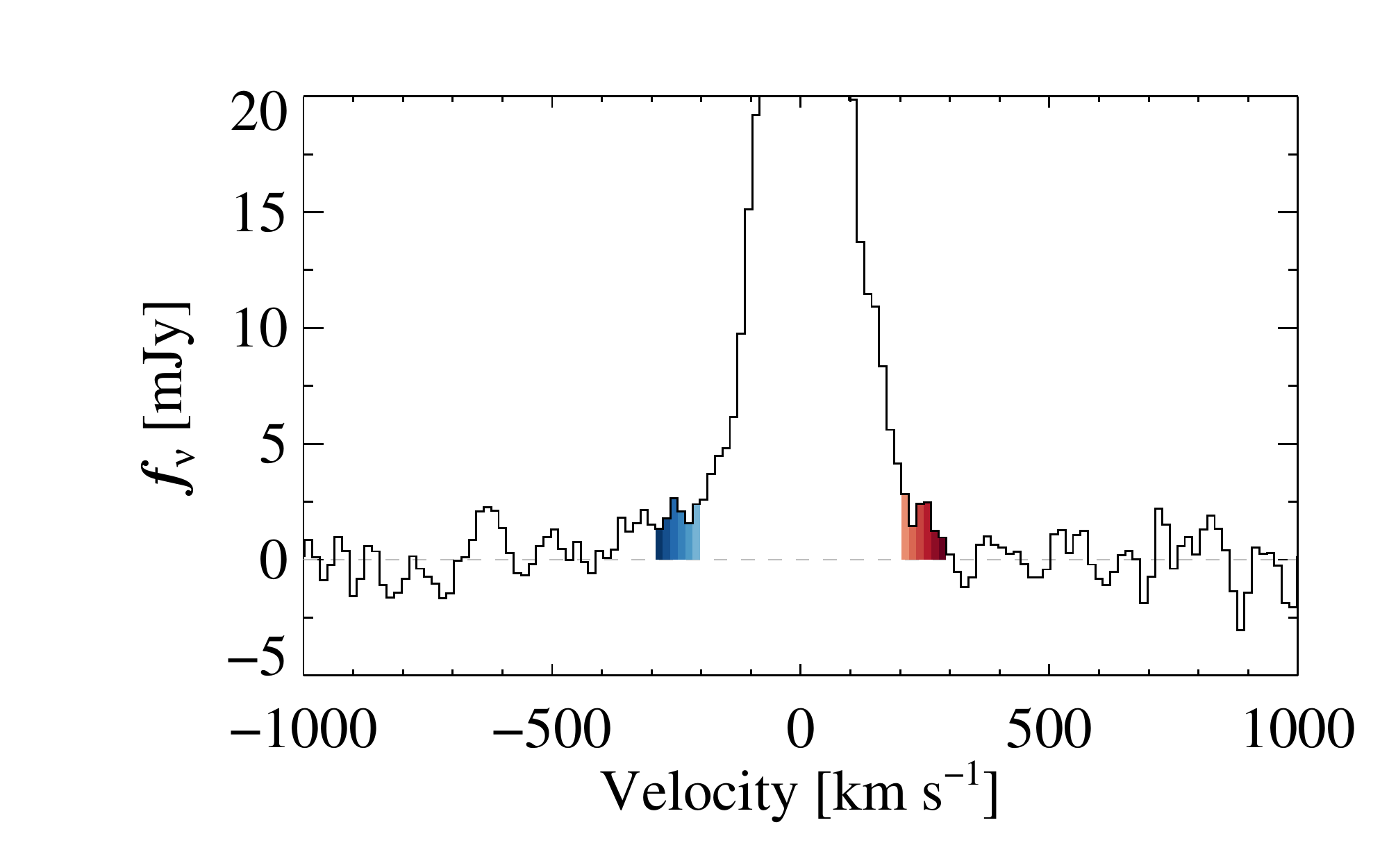}
\centering
\end{minipage}
\begin{minipage}[!b]{0.5\textwidth}
\hspace{-0.75cm}
\centering
\begin{overpic}[width=8.cm]{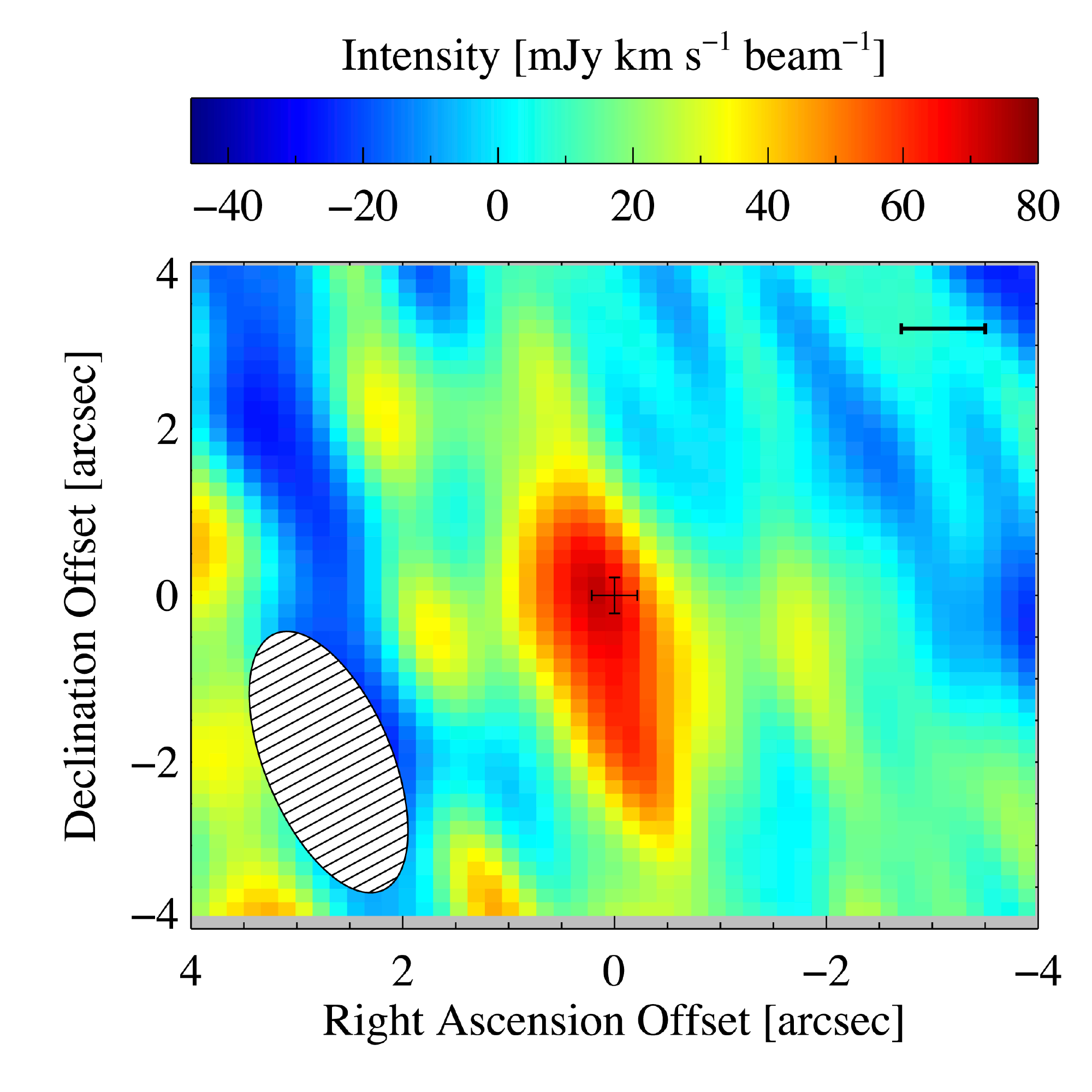}
 \put (20,70) {\large{$-300\textrm{\kms} < v < -200$\textrm{\kms}}}
\end{overpic}
\end{minipage}
\begin{minipage}[!b]{0.5\textwidth}
\vspace{-0.3cm}
\hspace{-0.75cm}
\centering
\begin{overpic}[width=8.cm]{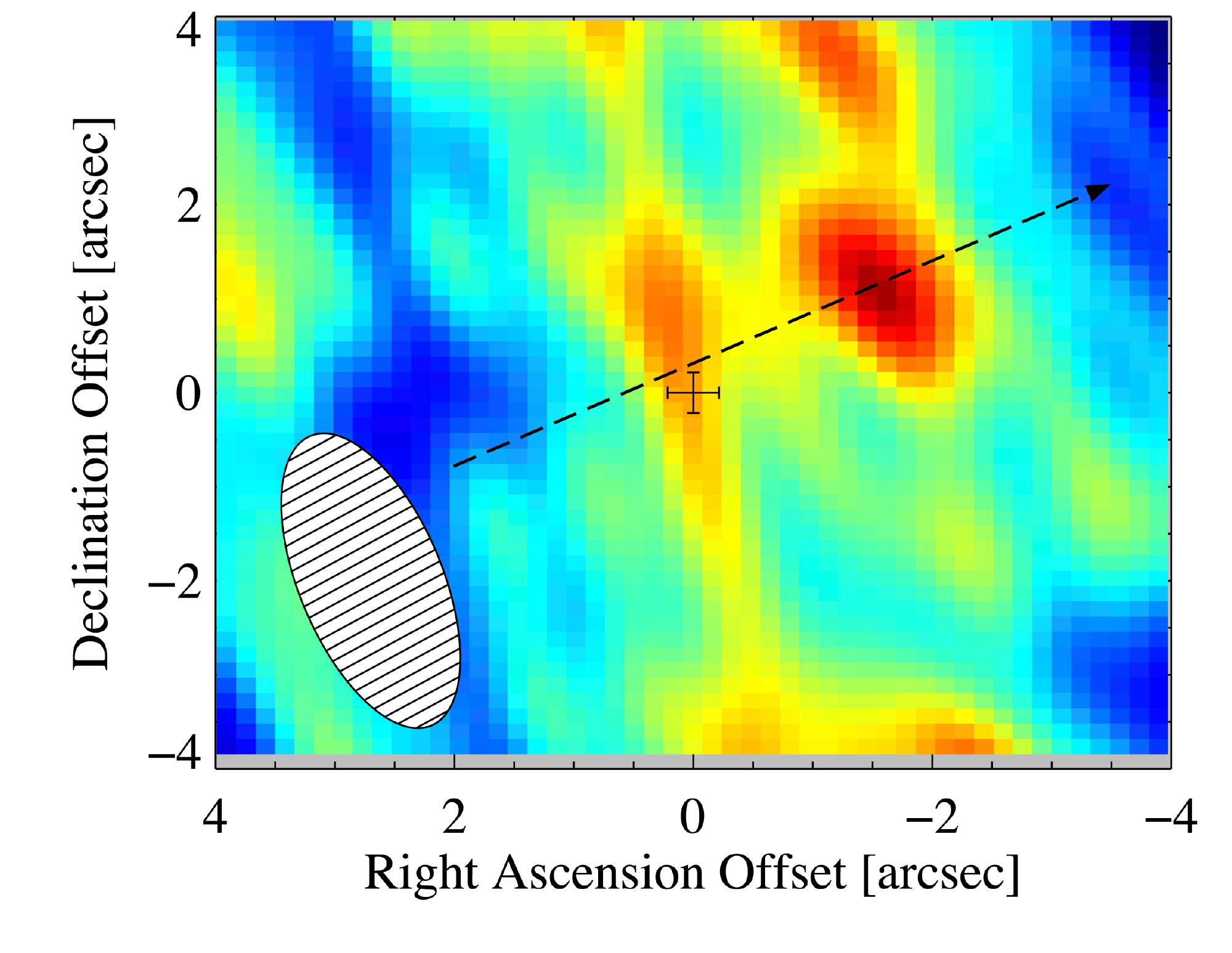}
 \put (20,70) {\large{$200\textrm{\kms} < v < 300$\textrm{\kms}}}
\end{overpic}
\end{minipage}
\caption{\COIO\ emission integrated over the wings of the line profile for Z11598$-$0112 showing the spatially off-center emission in the red wing.  The top panel shows the spatially integrated \COIO\ spectrum over a $5\arcsec\!\times5$\arcsec\ box with the integrated velocity channels highlighted in blue and red.  The middle panel shows the intensity map integrated over the blue wing, corresponding to the velocity range $-300$ to $-200$~\kms.  The scale bar marks 2~kpc. The bottom panel shows the intensity map integrated over the red wing, corresponding to the velocity range $200$ to $300$~\kms.  The position-velocity diagram in Figure~\ref{fig:pv} follows the black arrow from bottom left to top right. \label{fig:redwing}}
\end{figure}

We do not detect a high-velocity \COIO\ outflow in Z11598$-$0112 at the scales probed with our observations.  Notably, the beam is very elongated because of the low declination of this source relative to NOEMA's northern location.  However, in the channel maps of Figure~\ref{fig:cmaps}, there is \COIO\ that is near the spatial resolution perpendicular to the major axis of the beam.  In order to boost the signal of this feature, we show an image of the \COIO\ emission integrated between 200~\kms\ and 300~\kms\ in Figure~\ref{fig:redwing}, with an illustration of the spatially integrated spectrum highlighting the relevant velocity channels.  Integrating to bluer channels only serves to increase the \COIO\ flux at the position of the source.

\begin{figure}
\includegraphics[width=8.2cm]{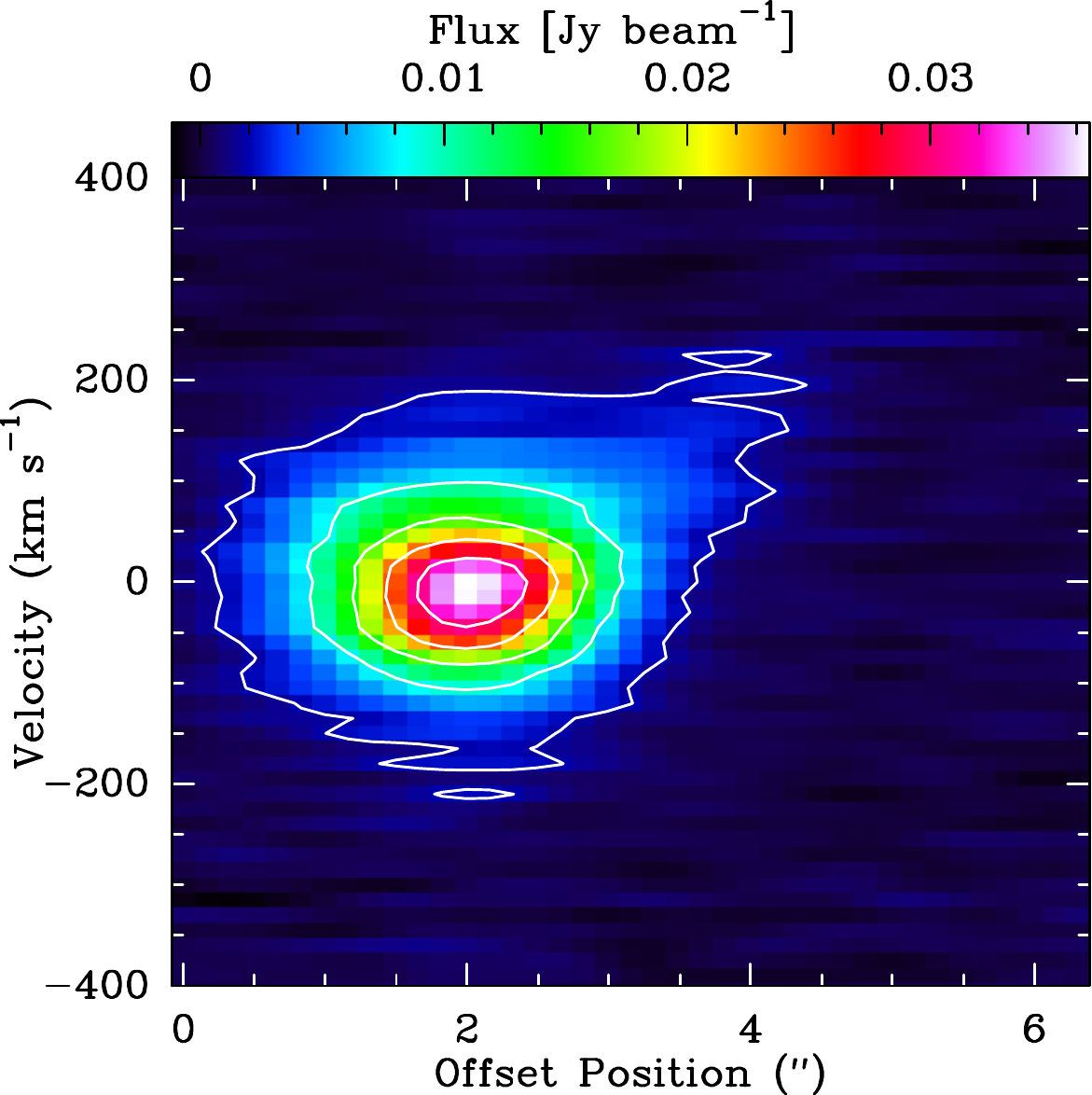}
\caption{Position-velocity diagram for Z11598$-$0112 following the arrow in Figure~\ref{fig:redwing}.  The offset is measured from the origin at the base of the arrow to its tip, so that the nucleus appears at $\sim2$\arcsec.  The \COIO\ emission at 200--300~\kms\ falls at an offset position of approximately 4\arcsec and does not appear to be in rotation. \label{fig:pv}}
\end{figure}

The single-dish \COIO\ showed the potential for high-velocity line wings.  Here, these seem to extend out to approximately 300~\kms (or slightly farther on the blue side, but no spatial extension is associated with this emission).  Binning in velocity does highlight the wings, but does not boost any signal above the noise at velocities higher than |300|~\kms.  The position-velocity diagram in Figure~\ref{fig:pv} suggests that this emission is not rotation, as it is not reflected on the blue side of the line.  This is also reflected in the channel maps.  This falls just shy of the \citet{cicone14} criteria for detecting outflows.  There is a known warm molecular outflow detected in OH~$\lambda$119~$\mu$m, but it is based on blueshifted absorption at 150~\kms\ rather than a P-Cygni profile.  It is possible that the extended redshifted CO emission represents the other side of this outflow, since the profile does cover $\sim150$~\kms.  However, this cannot be proven with the current data and this scenario does not explain why the blueshifted side of the outflow is not seen in CO.  Alternatively, the extended redshifted CO emission does align with one of the tidal stellar features in the system (see Figure~\ref{fig:sdss}), which would point to a tidal origin for the extended CO.    

\subsection{F13342+3932}
\label{sec:f13342}
F13342+3932 is a ULIRG \citep{kim98} with a radio-quiet Seyfert 1 nucleus \citep{zheng02,nagar03}.  {\it HST} WFPC2/F814W (wide $I$-band) imaging shows two prominent spiral arms in the host galaxy \citep{borne00,rupke17}.  The host galaxy environment was initially mis-classified as an interacting group \citep{borne00}, but redshift analysis shows that only the nearest companion (upper left corner in Figure~\ref{fig:sdss}) with $z=0.180$ is nearby in velocity space \citep{veilleux02}.  Gemini IFS reveals the stellar velocity field and relatively compact ionized and neutral outflows in all the strong optical emission lines (e.g., [N~\textsc{ii}]~$\lambda6583$) and Na~\textsc{i}D \citep{rupke17}.  The velocity field of the gas traced by optical emission lines is rotation dominated with a major axis position angle $\sim\!55^\circ$ measured East of North (i.e., oriented roughly NE to SW).  The outflow is roughly perpendicular to this along the minor axis, as is the wide-angle ionization cone of the AGN central engine.  The total outflow may be AGN-driven as the amount of momentum required (though not energy) is too high to attribute to starbursts.  Single-dish CO observations with a $\theta_b=24$\arcsec beam show a line profile with a narrow blue peak and redshifted broad base \citep{xia12}.  This is set to zero velocity using a redshift of 0.179 and the flux-weighted \COIO\ redshift is 0.180.

The \COIO\ observations from NOEMA show a triple-peaked line profile in the integrated spectrum of Figure~\ref{fig:intspec}.  The middle and red peaks are what blend together in the single-dish observations to create what appears to be a broad, redshifted base to the line.  The peaks in the line profile separate themselves spatially to some degree.  As seen in Figure~\ref{fig:smaps}, the bluest and reddest peaks are emitted from the NE and SW regions, whereas the central peak comes primarily from the center and SE of the source.  The red and blue peaks are coincident with the gas disk from \citet{rupke17} and the central peak falls in the vicinity of the base of one spiral arm (NE region of the NOEMA field of view in Figure~\ref{fig:sdss}).  The triple peaked \COIO\ line may result from a double-peaked line structure consistent with the stellar disk plus an additional peak related to the tidal feature or spiral arm shown to the NE in Figure~\ref{fig:sdss} potentially due to interaction with the nearby companion. We do no further modeling of this source because an interacting system is too complex given the quality of the data. 

\subsection{PG\texorpdfstring{\,}{ }1440+356}
\label{sec:pg1440}
PG\,1440+356 (Mrk~478) is a classic, bright Type-1 radio-quiet quasar \citep{schmidt83,meurs84,green86,miller93}.  The spectral energy distribution of this source is extremely well observed at radio through X-ray wavelengths \citep[e.g.,][]{shang05,shang11} and is very far-infrared bright \citep{schweitzer06}.  The host galaxy is a barred spiral with an elliptical disk \citep{veilleux09a} and potentially also faint evidence for shells \citep{surace01}.  A warm molecular outflow may be present based on OH~$\lambda$119~$\mu$m seen in emission at $\sim-150$~\kms (whereas OH absorption would have constituted an unambiguous outflow detection) \citep{veilleux13b}.  Single-dish CO observations do not resolve any spatial structure at 4\arcsec\ resolution, but the \COIO\ line profile has $z=0.078$ and is obviously complex \citep{evans01}.  

A visual inspection of the data cube shows complex velocity and spatial structure.  There is a double-peaked line profile that is indicative of an emitting disk \citep{beckwith93} with the highest velocity components distributed approximately East-West, as seen in Figure~\ref{fig:smaps}.  In addition, the \COIO\ emission is spatially extended along the NE-SW direction (Figure~\ref{fig:cmaps} at e.g., $\pm75$~\kms), with this emission corresponding to the most red and blueshifted gas in the spectral maps beyond the double-peaked line profile.  Thus, it appears that PG\,1440+356 hosts a molecular gas disk, plus additional structure.  It is difficult to determine the nature of this \COIO\ because this source does not host a known warm molecular gas outflow and the \COIO\ velocities are not high enough to unambiguously attribute it to an outflow.  An alternative is that the \COIO\ may reside in a warped disk or bar, which is particularly reasonable given that the PG\,1440+356 host galaxy is a barred spiral.  

\subsubsection{Kinematics in PG$\,$1440+356}
\label{sec:outflow}
Our goal was to detect and quantify the properties of any outflow that is present in the PG\,1440+356 \COIO\ data cube.  There is a clear pattern in the data cube for structure in the \COIO\ emitting gas that is more complex than a simple disk.  Here, we take the approach of isolating the emission not clearly attributable to a rotating disk to use it to calculate upper limits on the properties of any outflow in PG\,1440+356 \citep[e.g.,][]{veilleux17,herrera-camus19}.

To this end, we parameterized the rotating disk so that it could be subtracted to isolate the other component in the residuals. We adopted a circularly symmetric logarithmic gravitational potential which produces the flat rotation curves typically seen in CO on kpc scales similar to our observations \citep[e.g.,][]{sofue99,sofue01}. Our priority was to constrain the overall disk emission rather than estimate the disk properties.  Thus, we limited the input physics to rotation to which we applied an emissivity profile rather than performing a full treatment of the radiative transfer of a disk in local thermal equilibrium \citep[e.g.,][]{pringle81,yen14}. The model disk is constructed as follows:
\begin{equation}
\label{eqn:vrot}
V_{rot}(R) = \sqrt{\frac{v_0^2R^2}{R^2+R_{c}^2}},
\end{equation}
where 
\begin{eqnarray}
&R = r\cdot d, \\
&r = \sqrt{x^2+(y/\textrm{cos}\,i)^2}, \\
&x = - (\Delta \alpha-x_{0}) \,\textrm{cos}\,\psi+(\Delta\delta-y_{0})\,\textrm{sin}\,\psi, \\
&y = (\Delta \alpha-x_{0}) \,\textrm{sin}\,\psi+(\Delta\delta-y_{0})\,\textrm{cos}\,\psi.
\end{eqnarray}
Here $V_{rot}(R)$ is the radial profile of the rotational velocity, $v_0$ is the value of the flat rotation curve at $R>R_c$, $R$ is radius in the disk plane in physical units (e.g., kiloparsecs), $r$ is the radial positional offset in the disk plane expressed in angular units on the sky (e.g., arcseconds), $d$ is the angular diameter distance to the source, $i$ is the inclination angle of the disk to the line of sight, $x$ and $y$ are the positional offsets along the major and minor axes in the disk plane measured in angular units on the sky from the kinematic center, $x_0$ and $y_0$ are the coordinates of the kinematic disk center in angular sky units, $\Delta \alpha$ and $\Delta \delta$ are right ascension and declination offsets from the center of the flux map, and $\psi$ is the position angle of the disk major axis measured counterclockwise North of West.  The line of sight velocity is then a function of position in the disk given by,
\begin{equation}
V_{LOS} = V_{rot} \cdot \textrm{sin}\,i \cdot \frac{x}{r},
\end{equation}
where $r$ is the positional offset radius in the disk plane related to the physical radius by $d$.

The profile of the \COIO\ emission line was modeled as a Gaussian,
\begin{equation}
\label{eqn:phi}
\phi(v) \propto \textrm{exp}\left[-\frac{(v-V_{LOS})^2}{2\sigma_v^2}\right],
\end{equation}
where $\sigma_v$ is the velocity broadening parameter and a single value is adopted for the entire cube.  Due to the limitations of the data quality, the intensity was obtained by applying a power-law emissivity profile to the disk,
\begin{equation}
\label{eqn:I}
I(r) = I_{0} \left(\frac{r}{r_{0}}\right)^{-p},
\end{equation}
which is calculated between $r_{in}$ and $r_{out}$ and $r_{0} = 1$~arcsec.  The model data cube was then calculated by multiplying Equations~\ref{eqn:phi} and \ref{eqn:I}. 

We adopt the \COIO\ redshift, $z$, as an additional free parameter applied to the observed-frame data cube before comparison with the disk model. This accounts for the fact that the \COIO\ redshift does not necessarily match the one obtained from optical ionized emission lines and so is not known a priori. The \COIO\ redshift is often determined from the line peak or flux-weighted centroid (as in \S\ref{sec:specfit}), but given the complex shape of the line profile including it as a free parameter allows the more refined measurement consistent with the disk model.

Each model disk was smeared with a beam prior to comparison with the observed data cube.  We used the {\tt convolution} package in {\tt astropy} to calculate a two-dimensional Gaussian kernel from the NOEMA beam parameters for the PG\,1440+356 observation.  We then convolved this with the model data cube in every velocity channel to obtain the ``observed'' disk model cube. 

We compared the physical disk model to the data with the affine-invariant Markov Chain Monte Carlo (MCMC) ensemble sampler {\tt EMCEE} \citep{foreman13}.  This routine samples the posterior distribution of the model parameters as a consequence of the combined prior probability and likelihood functions.  It takes as inputs the data, an initial distribution of ``walkers'' (effectively an initial-guess distribution for each model parameter), a prior probability distribution for each disk model parameter, and a functional form for the likelihood.  

The walkers were initialized based on the outcome of a simple chi-squared optimization, the priors and allowed range for each disk model parameter are listed in Table~\ref{tab:priors}, and we adopted a normal likelihood function. For numerical stability, the $I_{0}$ parameter was treated in logarithmic space and its uniform prior was effectively a Jeffreys prior on the linear quantity (i.e. the prior probability distribution of $I_{0}$ was proportional to 1/$I_{0}$ to avoid unintentionally biasing the outcome towards large values). The $\sigma_{v}$ and $r_{in}$ parameters were treated with modified Jeffreys priors because values near zero were possible.  Thus, they were uniform below 10~\kms and 0.1~arcsec, respectively, and followed a Jeffreys prior above these values. The limits on each property were chosen to provide a full range in parameter space, by exploring clear limits in the data cube, and by obtaining measurements of physical properties from the literature. The range in allowed redshift encompasses various estimates from optical emission lines. The $R_c$ parameter, which controls the radius where the disk rotation curve flattens to a constant value, was fixed at 0. This was done because the data preferred sub-pixel values for $R_c$ and further exploration in this parameter detracted from other more important parameters.

\input{./tables/priors_mnras.tex}

\begin{figure*}
\hspace{-0.6cm}
\begin{minipage}[!b]{0.33\textwidth}
\centering
\includegraphics[width=6.5cm]{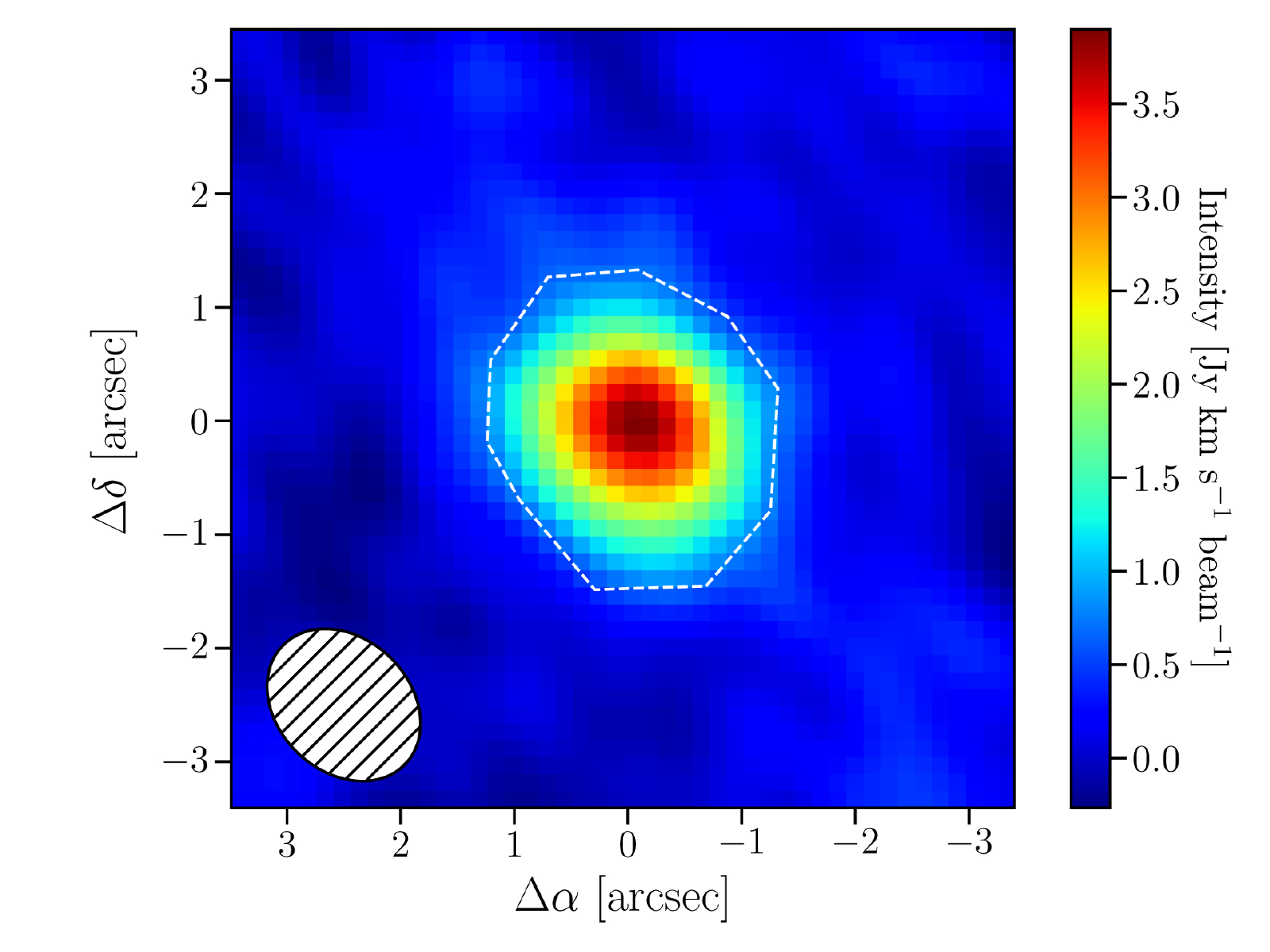}
\end{minipage}
\hspace{0.15cm}
\begin{minipage}[!b]{0.33\textwidth}
\centering
\includegraphics[width=6.5cm]{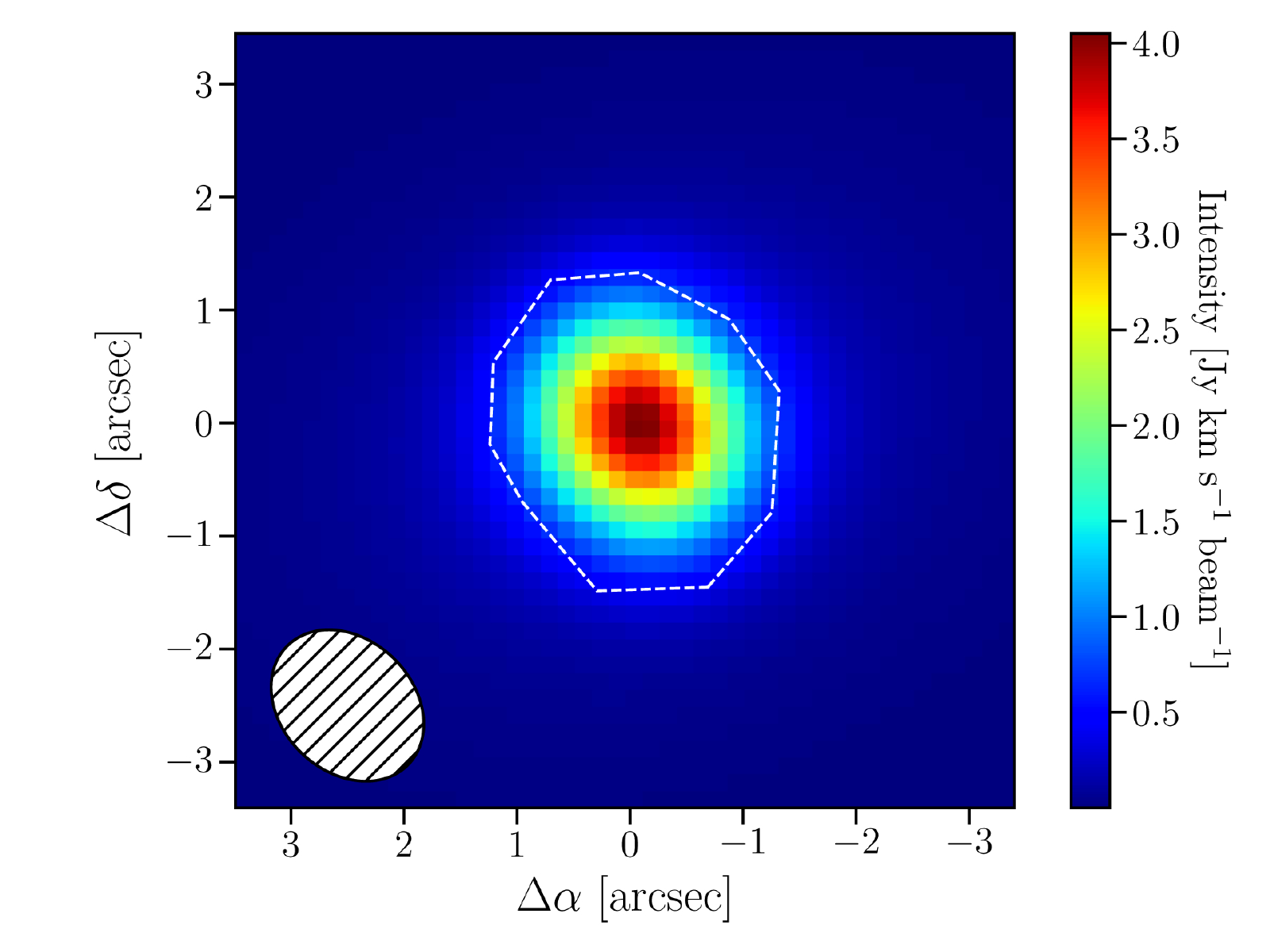}
\end{minipage}
\hspace{0.15cm}
\begin{minipage}[!b]{0.33\textwidth}
\centering
\includegraphics[width=6.5cm]{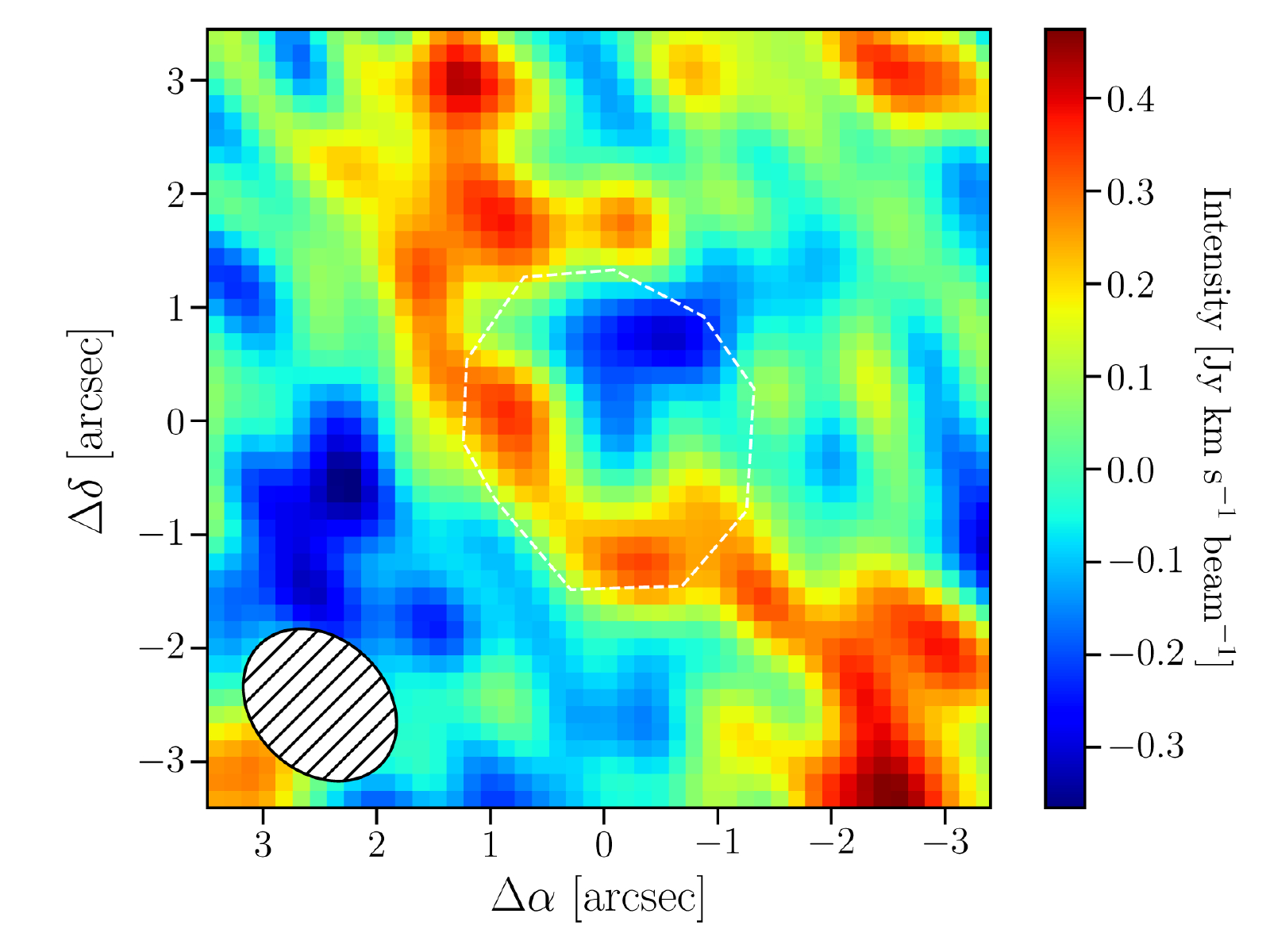}
\end{minipage}
\caption{Maps of the spatially resolved CO emission in PG\,1440+356.  The data, disk model, and residuals are shown from left to right, the beam size is indicated, and the dashed line indicates the masked region for the disk fit.  Notably, the right panel shows that the component isolated in the residuals lies along the Northeast-Southwest axis. \label{fig:comap}}
\end{figure*}

\begin{figure}
\hspace{-0.77cm}
\includegraphics[width=0.53\textwidth]{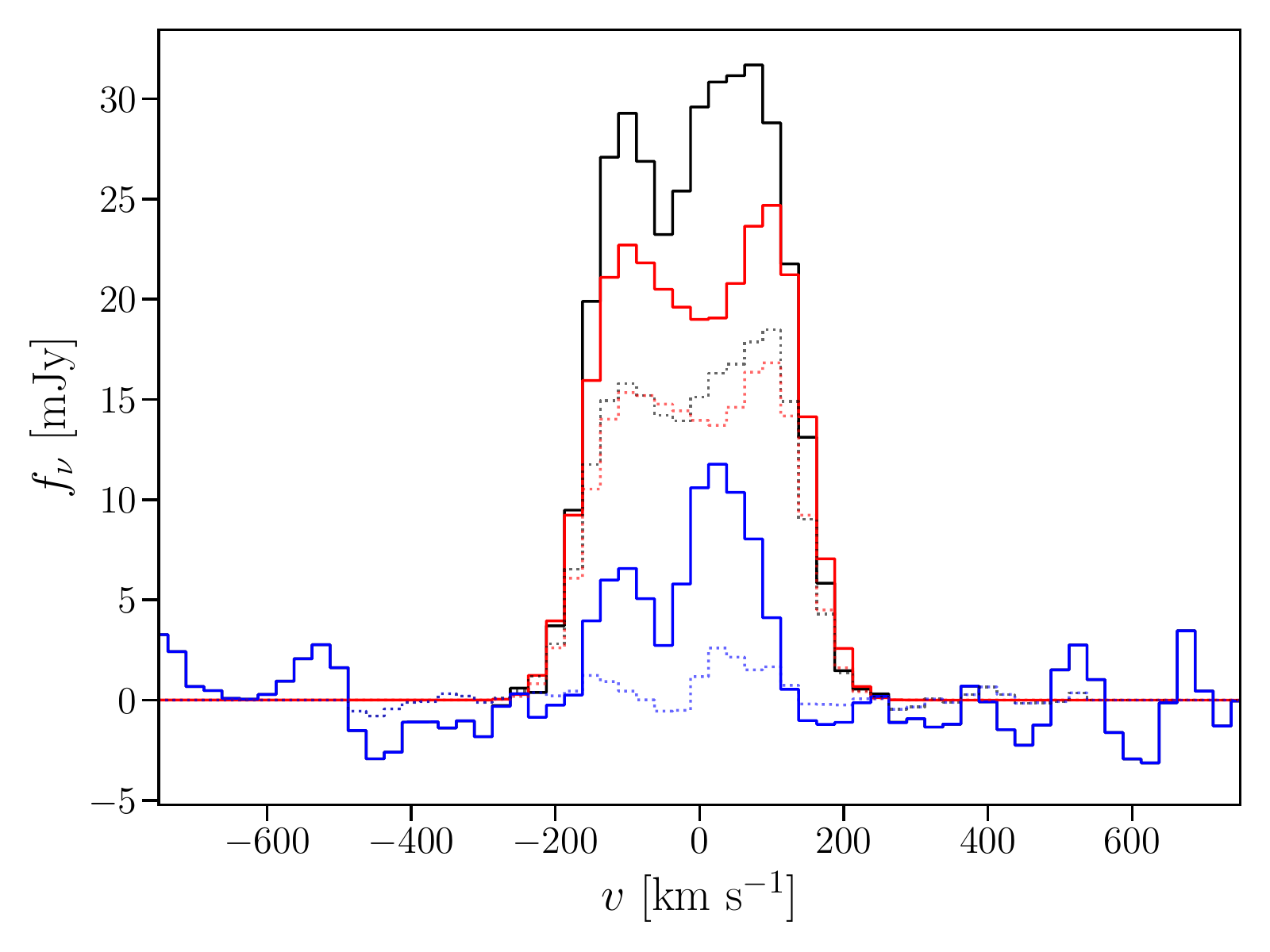}
\caption{The integrated CO spectra for PG\,1440+356.  The data, disk model, and residuals are shown in black, red, and blue, respectively.  Solid lines show spectrum from the full data cube and dotted lines show the masked region used for fitting the disk.  The residuals in the masked region (dashed-blue) are near zero, indicating that the fit was successful.  The solid blue line represents the integrated spectrum of the kinematically complex component isolated in the full data cube residuals.} \label{fig:spec}
\end{figure}

\begin{figure*}
\hspace{-2.25cm}
\centering
\includegraphics[width=1.1\textwidth]{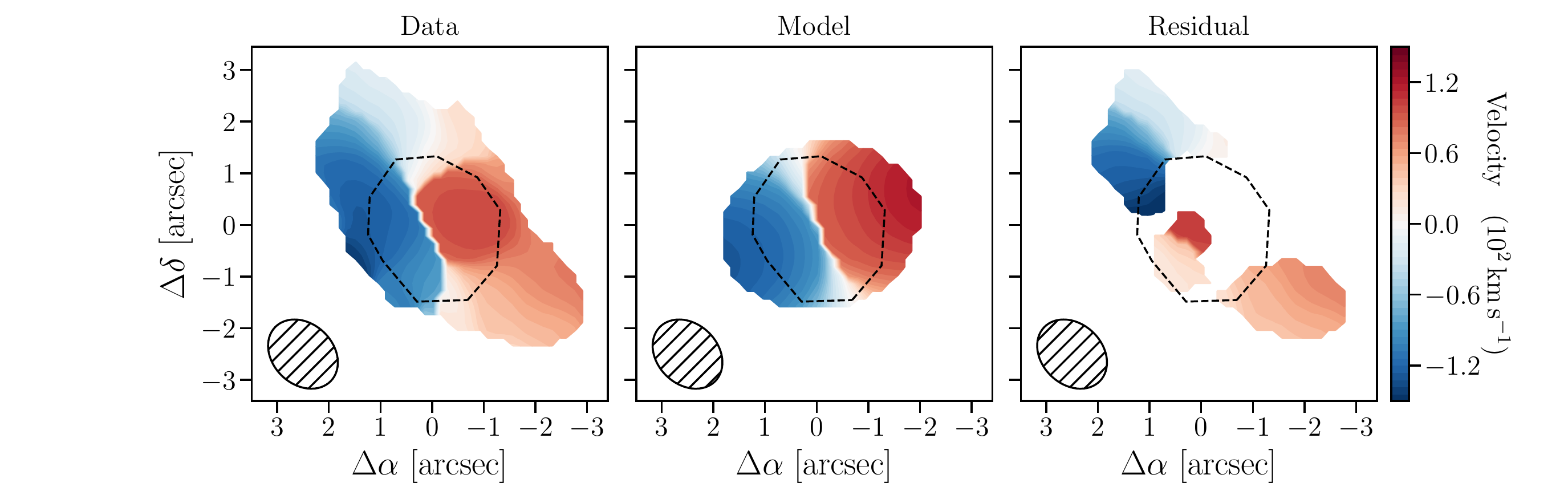}
\caption{The line-of-sight velocity map for PG\,1440~356 for the data (left), rotating disk model (middle), and residuals (right). The component isolated in the residuals has a slightly different axis than the rotating disk based on the location of the lowest velocity gas (see the end of Section~\ref{sec:outflow}), indicating it may be a warped disk or other component like an outflow. The dashed line encloses the sky area where the disk model was fit to the data, $\sigma$ is taken to be the rms in this region, and only spaxels detected at greater than the $3\sigma$ level are shown. This map is analogous to a first moment map, but is calculated using a fitting procedure that isolates the peak of the emission line in each spaxel and yields a smoother map.\label{fig:vmap}}
\end{figure*}

After a burn-in phase where the simulation transformed the initial 300-walker distributions to resemble the posterior distributions of interest, we ran the simulation for 250 iterations.  We explored running the simulation for up to 1000 iterations, but found that 250 was sufficient.  The comparison between the data and rotating disk model cubes was made within the central part of the data cube, shown in Figures~\ref{fig:comap} and \ref{fig:vmap}.  This was done to ameliorate the effects of the other \COIO\ component on the disk simulation, and the region was chosen to include the area where the rotating disk dominates emission.  Figure~\ref{fig:spec} shows the integrated spectrum of the total data cube, the total rotating disk model, and the data and disk model within the region of comparison to verify the success of this approach in providing a good comparison.  The parameters and uncertainties of the rotating disk model most likely to produce the observed data were determined from the 16, 50, and 84th percentile of the posterior distributions for each property and are listed in Table~\ref{tab:priors}.     

In Figure~\ref{fig:comap} we show the intensity maps of the data cube, the rotating disk model, and the residual between them when they are subtracted over all spatial scales (i.e. not just within the region where the simulation was run).  We emphasize that the goal here was not to obtain the best-fitting parameters to a disk model, but rather to isolate the non-disk-like \COIO\ component in the residuals.  We experimented with more complex disk models, for example by allowing the emissivity profile to be a combination of an exponential and Gaussian or allowing spatial shifts between components, but the simulation struggled to converge in this case. We also fit a Keplerian disk model to the data. Parameters shared between the two models had similar (although not identical) values and the overall outcome and conclusions of the exercise were unchanged.

The line-of-sight velocity map of the residual \COIO\ emission after rotating disk model subtraction is shown in Figure~\ref{fig:vmap}.  The line-of-sight velocity was calculated using the {\tt bettermoments} package \citep{teague18,teague18b,teague19}.  As an alternative to traditional velocity centroid maps, which represent the intensity weighted average velocity, this code fits a quadratic function to the pixel with the maximum intensity and its nearest neighbors.  This approach effectively finds the peak of the line profile and quantifies its velocity in a robust way that is less sensitive to noise and profile asymmetry than the first moment.  We also calculated the first-moment map, which was noisier but qualitatively similar.  The  map is plotted for spaxels that are significant at 3$\sigma$ or greater level.  The rms was determined from the residuals within the masked fitting region.  The left panel shows the line-of-sight velocity map for the data, the middle panel for the rotating disk model, and the right shows the residuals.  The central region where the disk simulation was performed is marked, but the resulting disk was subtracted from all spaxels.  

The \COIO\ in the right panel of Figure~\ref{fig:vmap} shows the residual emission from the rotating disk model when it is greater than the 3$\sigma$ level. Inside the masked region the residuals represent a mismatch between the data and the model that is larger than 3 times the rms determined there. That is, this shows the spatial position of the difference between the black dotted and red dotted line in Figure~\ref{fig:spec}: the biggest difference is at $0 < v < 150$~\kms, which produces a redshift in the residual velocity map. Other differences between the data and rotating disk model in the masked region are small, below the 3$\sigma$ detection threshold, and so they do not appear in the figure.

Outside of the masked region, the \COIO\ differs in spatial and velocity space from the rotating disk and is detected at the $>\!3\sigma$ level. The minor axes of the data and rotating disk model, traced by the gas with velocities nearest systemic in each component, are slightly different.  The disk has a major axis of approximately 15$^\circ$ measured North of East (see Table~\ref{tab:priors}), which places the minor axis 15$^\circ$ from North-South. On the other hand, the gas with velocities nearest systemic in the residual map is oriented North-South.

It is worthwhile to compare these results to known properties of stars in the PG\,1440+356 host.  Notably, in {\it HST} near-infrared imaging, PG\,1440+356 has an extended stellar disk with a photometric major axis position angle that is 60$^\circ$ East of North \citep{veilleux09a}. The stellar residual revealed after the subtraction of the stellar model fit shows features much like spiral arms aligned along a similar axis.  While the major axis of this stellar disk is different from that of the compact molecular disk in our model by $\sim\!30^{\circ}$, it is consistent with the residual emission in CO in terms of the direction of the residual flux. We note that the kinematic position angle of stars in the host may not be the same as the photometric position angle of the stellar component \citep[e.g.,][]{veilleux09a,rupke17}.

Ultimately, if there is a cold molecular gas outflow in PG\,1440+356, it has $L_{CO}^{\prime} < 2.2\times10^8$~K~km~s$^{-1}$~pc$^{2}$ and $M < 1.8\times10^8$~\Msun, determined from the residual CO emission and again assuming $\alpha = M(\textrm{H}2)/L_{CO}^\prime \sim 0.8 \textrm{\Msun} (\textrm{K km s}^{-1}\textrm{pc}^2)^{-1}$. For the Keplerian disk model, these values are of the same magnitude but less restrictive. Normally, the impact of an outflow on its host galaxy is determined by the velocity, size, and the value and outflow rate of the mass, momentum, and energy. However, the determination of these properties requires a clear outflow detection and knowledge of the outflow geometry so that an appropriate physical outflow model can be adopted.  These things are beyond the capabilities of these data to support.  We can only conclude that the \COIO\ in PG\,1440+356 has structure more complex than a rotating disk, which may be a moderate velocity outflow, a warped disk, or a bar. Given the comparison to the stellar emission in near-infrared imaging, the latter options appear more likely.

\section{Discussion}
\label{sec:discussion}
In this work, we present IRAM NOEMA interferometric observations of \COIO\ in four type-1 AGN known to host outflows in other gas phases.  We calculated the \COIO\ luminosity and molecular gas mass and, in the case of PG\,1440+356, we subtracted a rotating disk model to reveal more complex kinematics in the residuals. 

During the process of fitting the disk model to the PG\,1440+356 data cube, it became clear that we were working at the limitation of the data.  For example, reasonable limits on the black hole mass from other lines of evidence were important for constraining the fit. In light of this, one limitation of this work is that the best-fitting parameters of the disk model may be more uncertain than they appear in Table~\ref{tab:priors}.  There are choices (e.g., the precise region where the rotating disk model was compared to the data) that are not represented in the uncertainties.  Because our aim was to characterize a rotating disk to highlight the residuals from this model in an illustrative way, we did not pursue the impact of these steps.  But due to these considerations, we caution that the formal fit uncertainties on the best-fitting disk model parameters may be underestimated compared to the true uncertainty in these values even though the global disk is reasonable.

Identifying AGN-driven cold molecular outflows via \COIO\ is challenging.  Common methods rely on high-velocity ($>300-500$~\kms) \COIO\ emission or a known warm molecular outflow based on an observe P-Cygni profile in OH~$\lambda$119~$\mu$m \citep{cicone14} or detailed kinematic modeling \citep[e.g.,][]{ramakrishnan19}. In the case of PG\,1440+356, the OH~$\lambda119$~$\mu$m line is observed to be only in emission so while a warm molecular outflow is possible, it is not confirmed.  Since \COIO\ is not emitted at high enough velocities to require an AGN-driven wind, the picture is not clear.  The residuals from the rotating disk fit suggest that the kinematics are complex.  However, distinguishing between different physical mechanisms including tidal streams, spiral arms, or stellar or AGN driven outflows would require higher quality data.  Typical investigations that investigate fueling along spiral arms \citep{ramakrishnan19} or barred spirals \citep{combes13,combes14} are common in extremely nearby ($z<0.01$) objects with high spatial resolution as good as 24~pc.  Thus, it is not surprising that the current data cannot distinguish between flared disks, bars, spiral features, and outflows.  Of these, the flared or warped disk or outflow appear most likely due to the slight mismatch and morphology in the kinematic residuals in Figure~\ref{fig:vmap}. 

Despite these limitations and caveats, we reiterate the utility in characterizing the cold molecular gas properties of type-1 AGN outflow hosts.  To date, of order 50 local galaxies have cold molecular outflows detected and mapped \citep{veilleux20}, but only 6 these are type-1 AGN \citep{cicone14,fiore17,fluetsch19,ramakrishnan19}. Thus, characterizing the spatially resolved CO emission in four type-1 AGN known to host outflows in other phases constitutes a significant contribution to constraining their total outflow properties. Furthermore, cold molecular gas is a key ingredient in fueling both star formation and AGN activity.  The interest in constraining the molecular gas properties of AGN hosts \citep{shangguan20} reflects this.  Observations of higher order CO transitions will reveal conditions in the host galaxies of these objects. 

\section{Summary}
\label{sec:summary}

In this work, we present new IRAM NOEMA interferometric observations of \COIO\ in F07599+6508, Z11598$-$0112,  F13342+3932, and PG1440+356.  These are all type-1 AGN and ULIRGs known to host warm molecular, neutral, or ionized outflows and were observed in the spirit of characterizing the multi-phase outflow properties of nearby type-1 AGN.  We performed a spectral decomposition of the \COIO\ data cubes in order to determine the luminosity and molecular gas mass.  In the context of existing multi-wavelength observations, we discussed the morphology of \COIO\ in each source.  The new observations of PG\,1440+356 show complex kinematics, which we revealed by subtracting a best-fitting rotating disk model. 

Our results are as follows:
\begin{itemize}
    \item We detected \COIO\ in all four sources, as expected given existing single-dish observations in all cases \citep{evans01,xia12}.\\
    \item In Section~\ref{sec:analysis} we calculated \COIO\ luminosities, $L_{CO}^{\prime}$, in the range 8--11$\times 10^9$\,K~km~s$^{-1}$~pc$^{2}$ and inferred molecular gas masses, $M$(H$_2$), in the range 6.5--9~$\times 10^9$~\Msun assuming $\alpha = M(\textrm{H}2)/L_{CO}^\prime \sim 0.8 \textrm{\Msun} (\textrm{K km s}^{-1}\textrm{pc}^2)^{-1}$.  These are presented in Table~\ref{tab:prop}.\\
    \item  The improved spatial (1--3\arcsec) and velocity (15-60~\kms) resolution over single-dish observations reveals new details in the data cubes.  These are discussed in the context of existing multi-wavelength data in Section~\ref{sec:obj}.  In particular, these new observations facilitated updated \COIO\ redshifts in F07599+6508, F13342+3932, and PG\,1440+356.  F13342+3932 is an interacting system with corresponding triple-peaked \COIO\ line profile and spatial morphology.  \\
    \item Although we do not detect a high-velocity outflow in Z11598$-$0112, we do detect redshifted material at 200--300~\kms\ that does not appear to be in rotation.  It is at the level of the noise in the spectrum and falls just short of formal outflow detection criteria. \\
    \item We modeled PG\,1440+356 with a rotating disk and isolated kinematics not derived from a logarithmic potential in the \COIO\ residuals.  We placed limits on the \COIO\ luminosity and molecular gas mas that can be in an outflow in PG\,1440+356 but do not claim an outflow detection.  These residuals may represent a warped disk, bar, spiral, or outflow.  
\end{itemize}

These objects represent just a few nearby type-1 AGN hosting outflows in other gas phases.  In the spirit of measuring their global outflow properties and constraining statistical samples of cold molecular gas outflows, future work should continue in this vein.  

\section*{Acknowledgements}
The authors thank the anonymous referee for their valuable suggestions which improved the manuscript. JCR acknowledges the support of Vincent Pi{\'e}tu in reducing one of the NOEMA datasets, as well as the hospitality of IRAM where some of the data reduction was carried out.  DSNR acknowledges financial support from the J. Lester Crain chair at Rhodes College.  This work was based on observations carried out under project numbers W15CS, W16BQ, W17CS, and W18CV with the IRAM NOEMA interferometer. IRAM is supported by INSU/CNRS (France), MPG (Germany) and IGN (Spain).

\section*{Data Availability}
The data underlying this article will be shared on reasonable request to the corresponding author.

\appendix
\section{Supplementary data visualizations}
\label{app:maps}
This appendix includes supplemental visualizations of the \COIO\ data cubes.  They were obtained directly from the data with minimal spectral or spatial modeling or decomposition.  To obtain these data, the continuum was subtracted to isolate the \COIO\ line emission and the data cube was cleaned.  All visualizations are centered on the coordinates in Table~\ref{tab:obslog} and the \COIO\ redshift in Table~\ref{tab:prop}. We discuss these in the context of further multi-wavelength data for each object in Section~\ref{sec:obj}. 

\begin{figure*}
\begin{overpic}[width=0.9\textwidth]{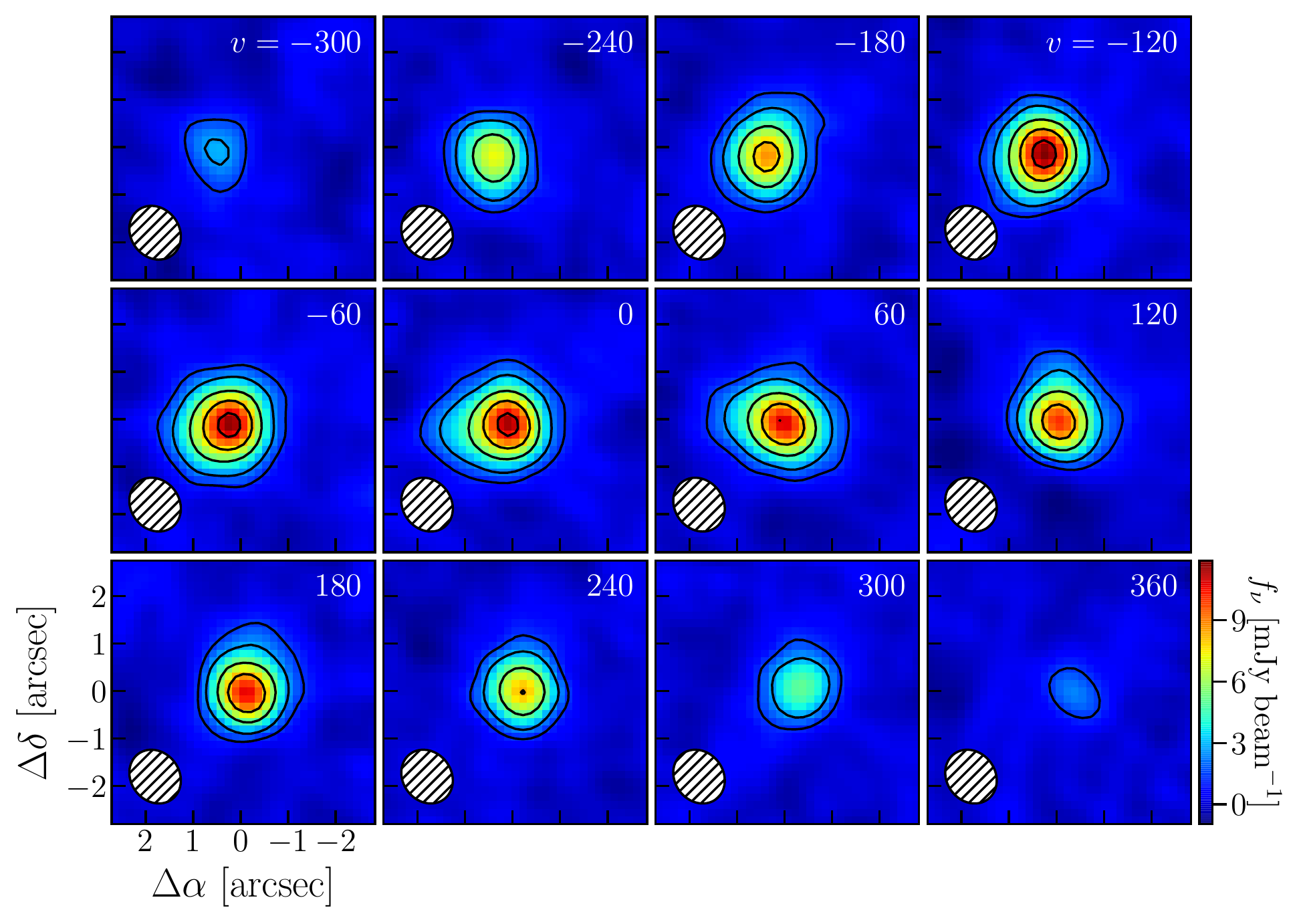}
 \put (8.3,70.25) {\large{F07599+6508}}
\end{overpic}
\caption{Channel maps, which show the image of the object in the velocity bin indicated by the \kms\ label located in the upper right corner of each panel.  Contours are at 5, 10, 20, 30, 40, and 50$\sigma$, where $\sigma$ is taken to be the measured noise in each channel.  The beam is shown in the lower left corner and beam properties are listed in Table~\ref{tab:obslog}. \label{fig:cmaps}}
\end{figure*}

\begin{figure*}
\begin{overpic}[width=1.0\textwidth]{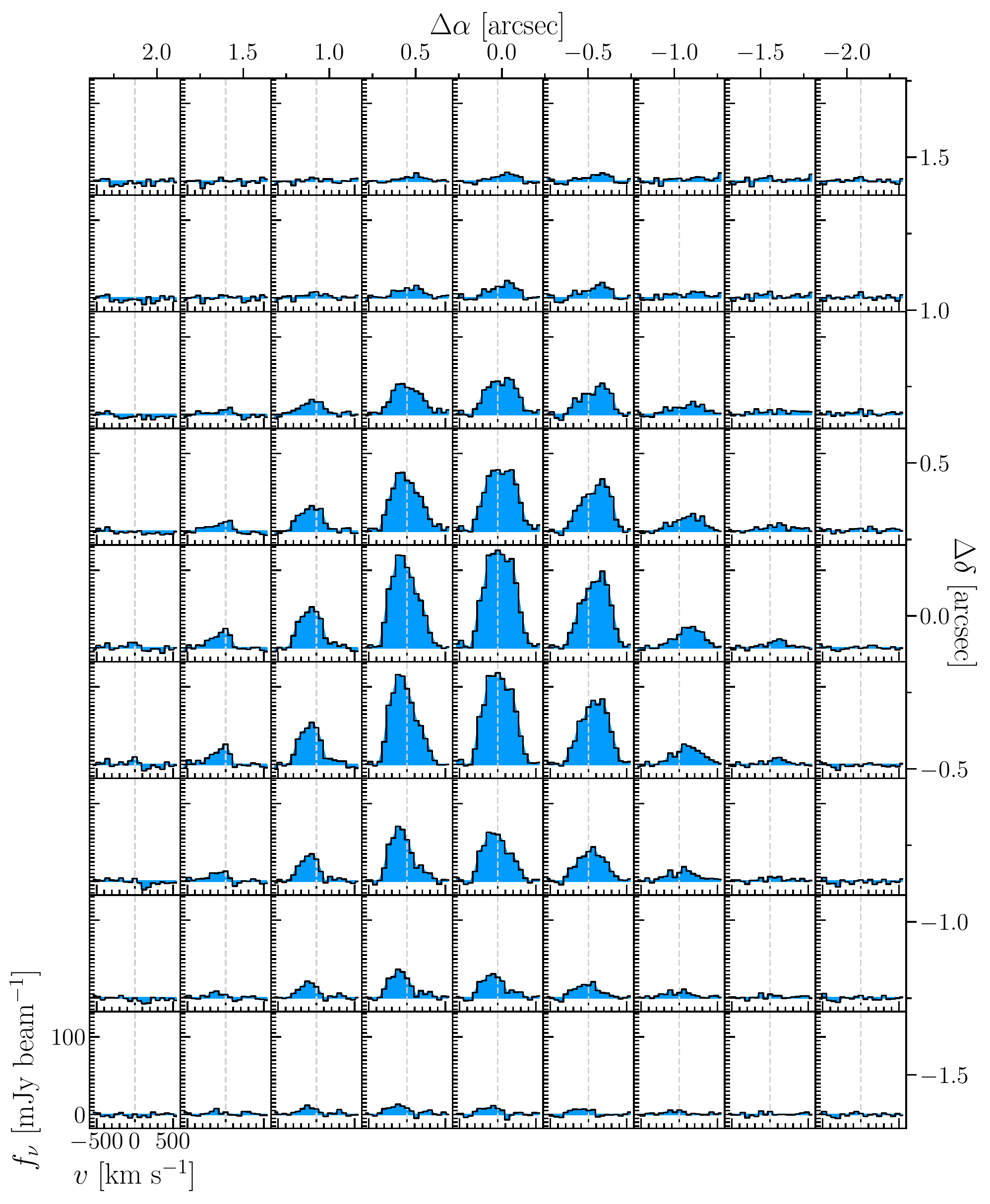}
 \put (63.25,4.5) {\large{F07599+6508}}
\end{overpic}
\caption{Spectral maps for each object.  This figure shows the shape of the \COIO\ emission-line profile (as indicated by the left and bottom axes) at different spatial positions (as indicated by the top and right axes) and in each case, the spectrum is integrated over the extent of the spatial panel.  The dashed gray line shows the systemic velocity set by the redshift in Table~\ref{tab:prop}. \label{fig:smaps}}
\end{figure*}

\begin{figure*}
\centerline{
\hspace{-2.5cm}
\begin{minipage}{0.5\textwidth}
\begin{overpic}[width=10cm]{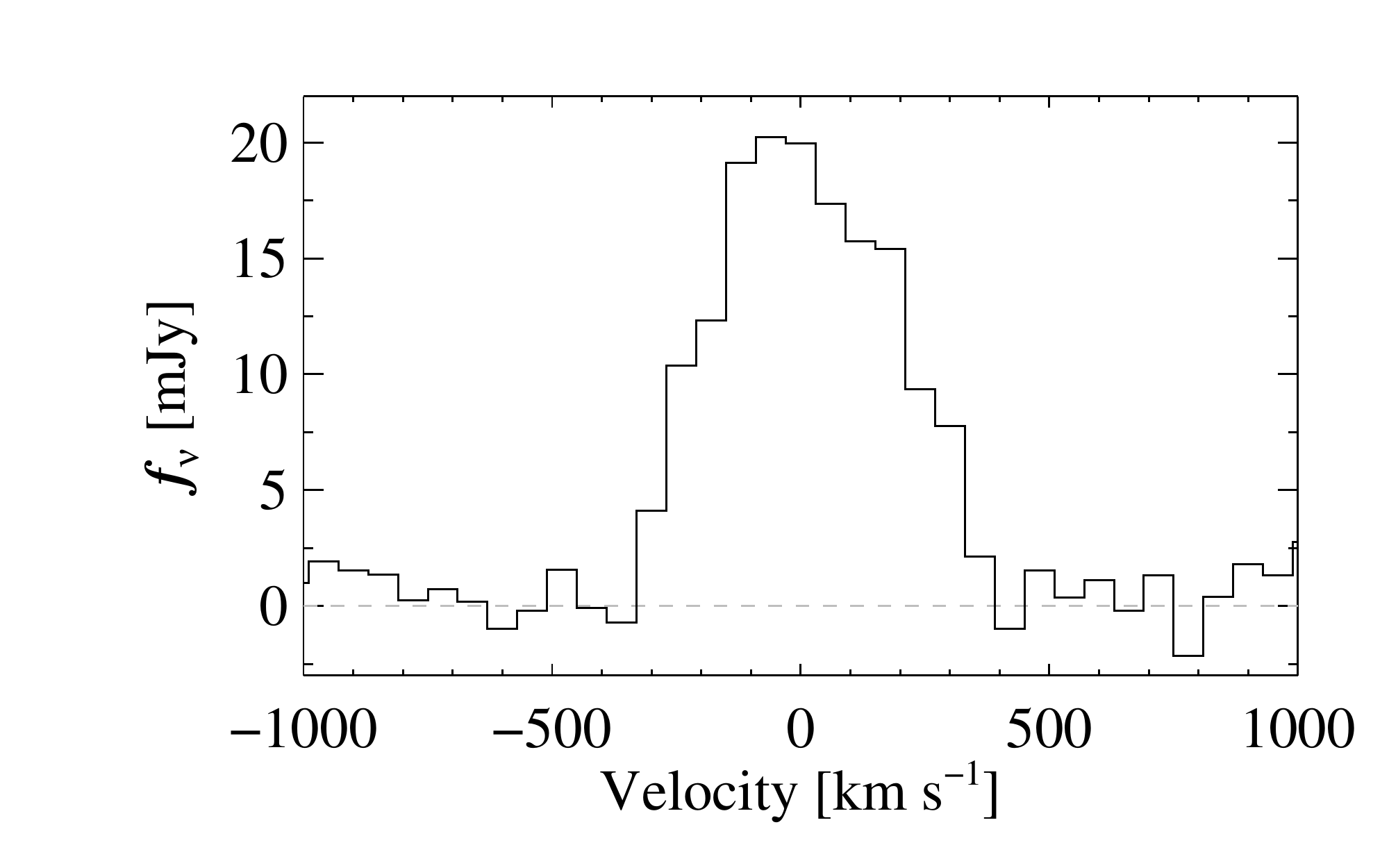}
 \put (22,56.2) {\large{F07599+6508}}
\end{overpic}
\end{minipage}
\begin{minipage}{0.5\textwidth}
\begin{overpic}[width=10cm]{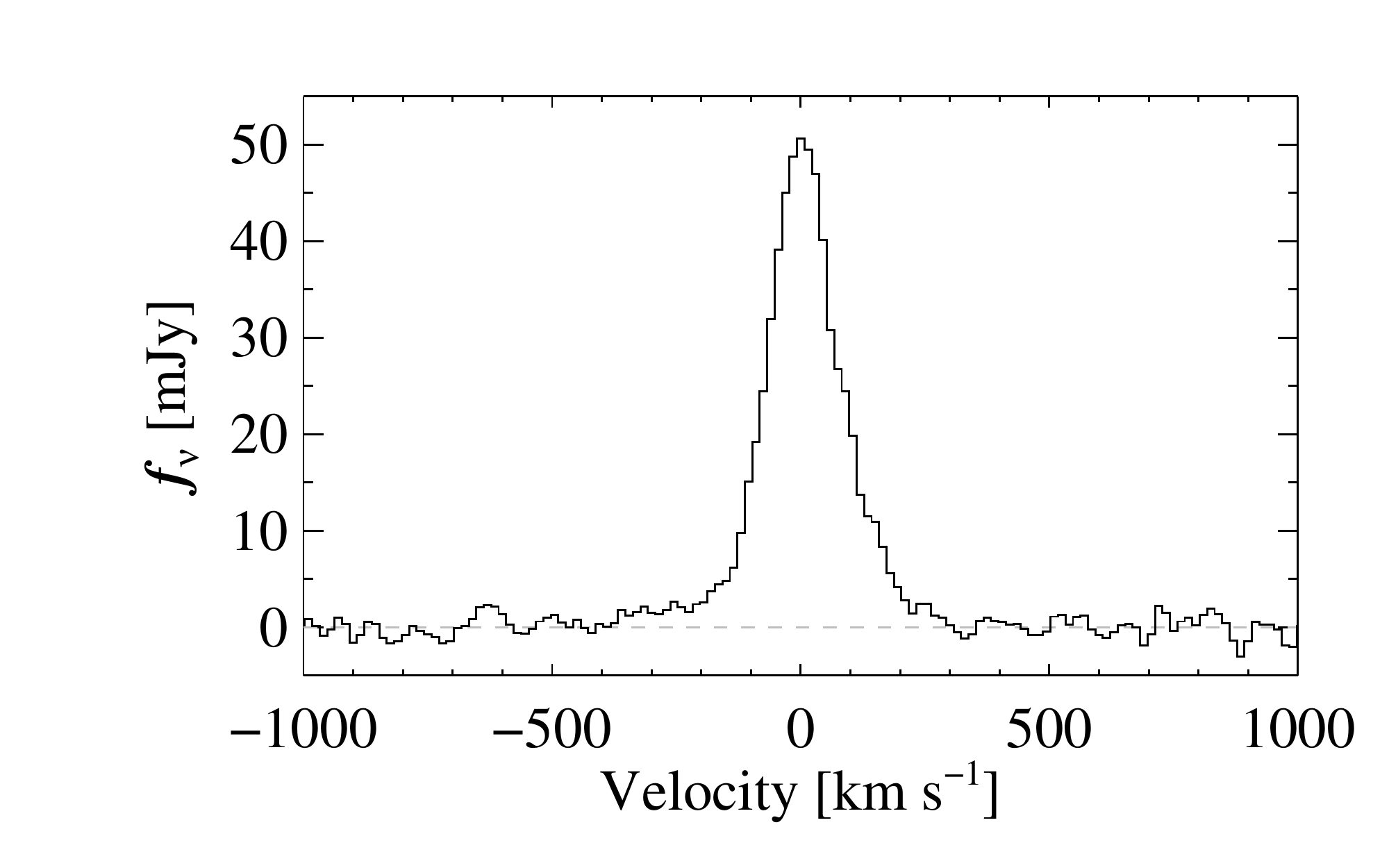}
 \put (22,56.2) {\large{Z11598$-$0112}}
\end{overpic}
\end{minipage}
}
\centerline{
\hspace{-2.5cm}
\begin{minipage}{0.5\textwidth}
\begin{overpic}[width=10cm]{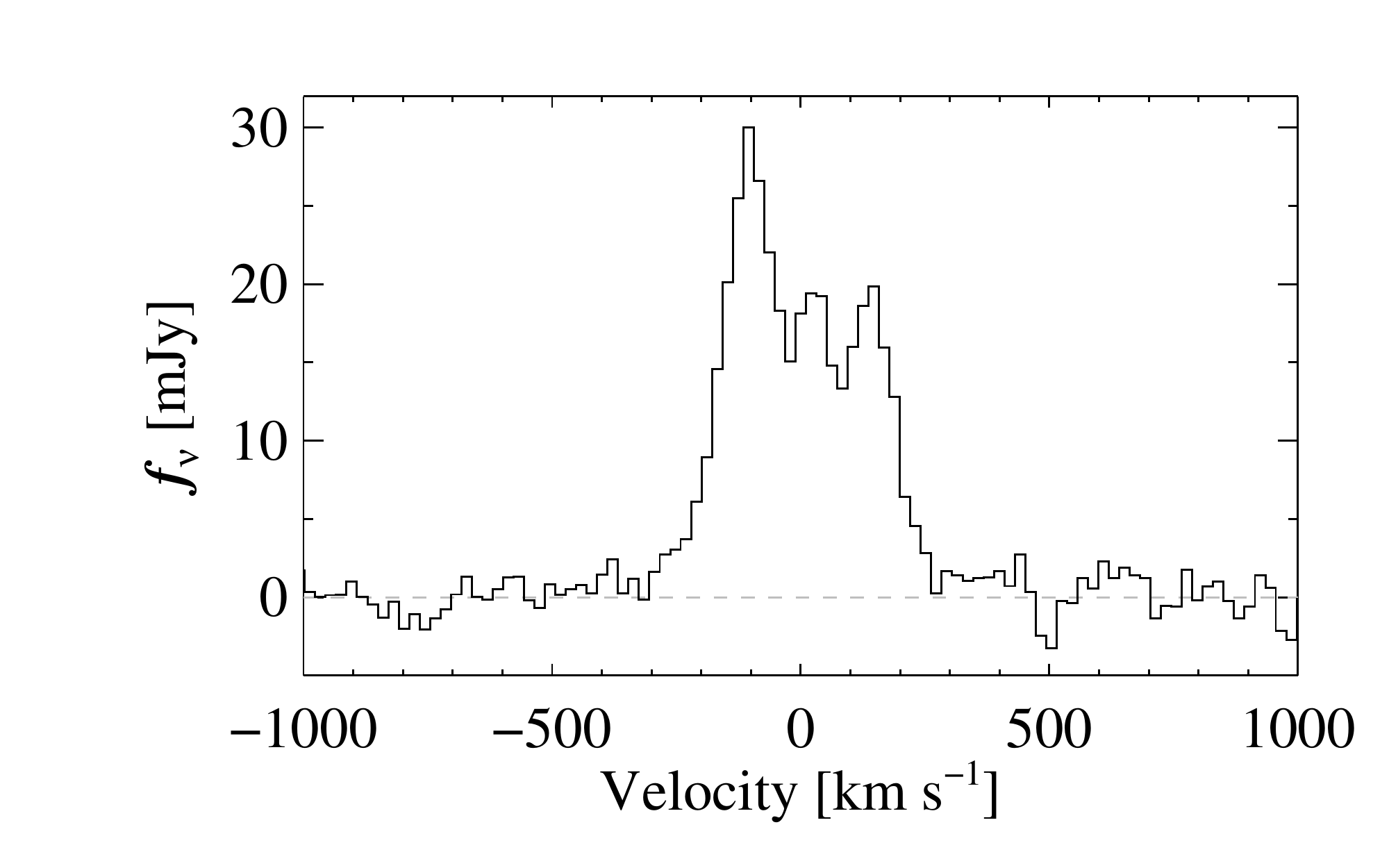}
 \put (22,56.2) {\large{F13342+3932}}
\end{overpic}
\end{minipage}
\begin{minipage}{0.5\textwidth}
\begin{overpic}[width=10cm]{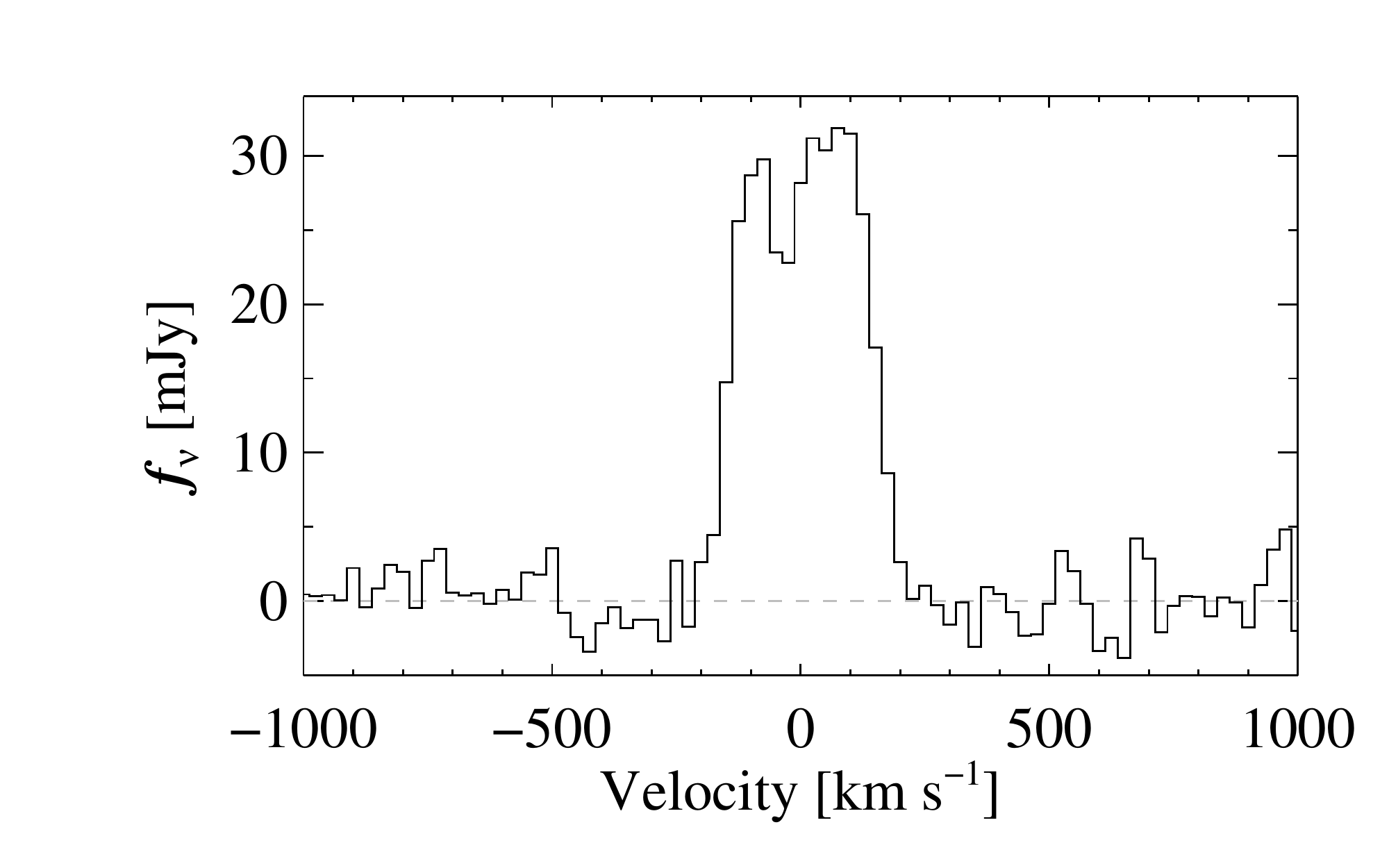}
 \put (22,56.2) {\large{PG1440+356}}
\end{overpic}
\end{minipage}
}
\caption{Spatially integrated spectra.  Spatial integration was done over a $5\arcsec\!\times5$\arcsec\ box at the phase center.  The spectral resolution is 60, 15, 21, and 25~\kms\ for F07599+6508, Z11598$-$0112,  F13342+3932, and PG1440+356, respectively.  The velocity scale was set with the \COIO\ redshifts in Table~\ref{tab:prop} to obtain a systemic value near the center of the line profile, despite multiple peaks in some objects. \label{fig:intspec}}
\end{figure*}

In Figure~\ref{fig:cmaps}, we show channel maps for each object, which display the image of the object in each velocity bin indicated in white text over the detectable range.  In Figure~\ref{fig:smaps} we show the integrated spectrum as a function of spatial position in the data cube.  Here, a cell is made up of several spaxels from the original data cube for visual clarity.  The spectrum shown in one cell is obtained from the spatially integrated spectrum of all the spaxels that contribute to it. Finally, in Figure~\ref{fig:intspec}, we show the \COIO\ spectrum integrated spatially over the entire data cube.  

\begin{figure*}
\begin{overpic}[width=0.89\textwidth]{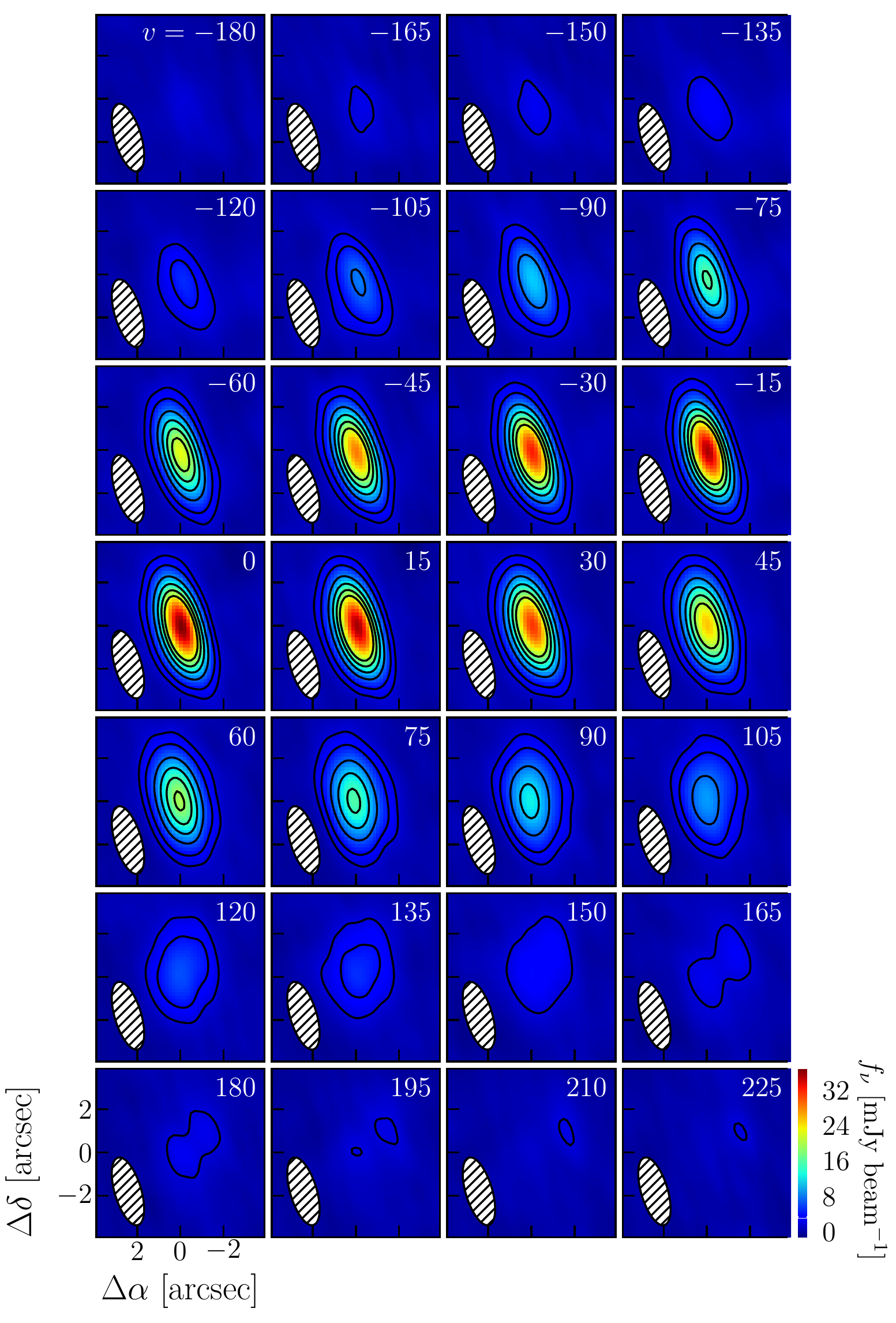}
 \put (7.5,99.25) {\large{Z11598$-$0112}}
\end{overpic}
\centerline{Figure~\ref{fig:cmaps}. -- Continued.}
\end{figure*}

\begin{figure*}
\begin{overpic}[width=1.0\textwidth]{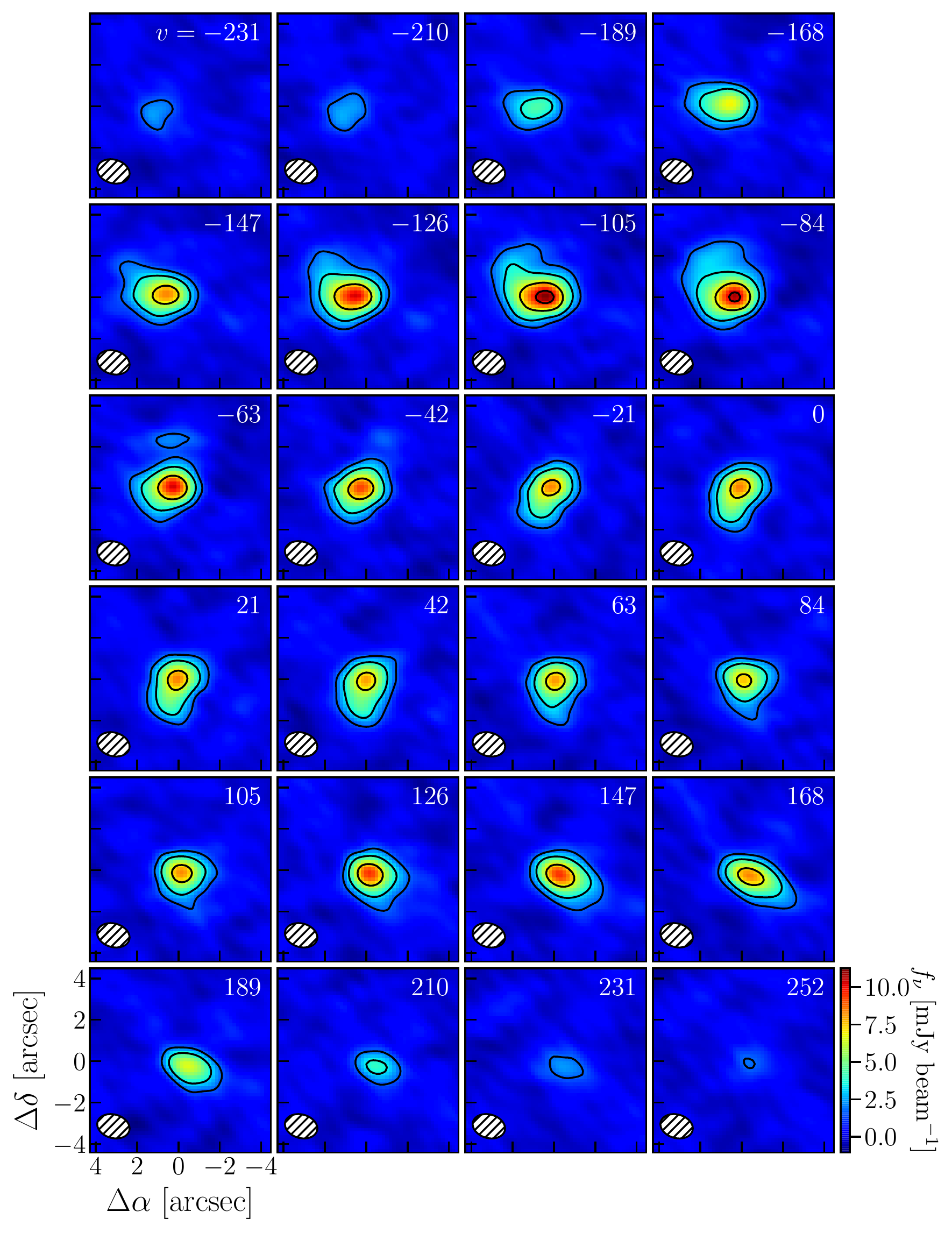}
 \put (7.5,99.25) {\large{F13342+3932}}
\end{overpic}
\centerline{Figure~\ref{fig:cmaps}. -- Continued.}
\end{figure*}

\begin{figure*}
\begin{overpic}[width=1.0\textwidth]{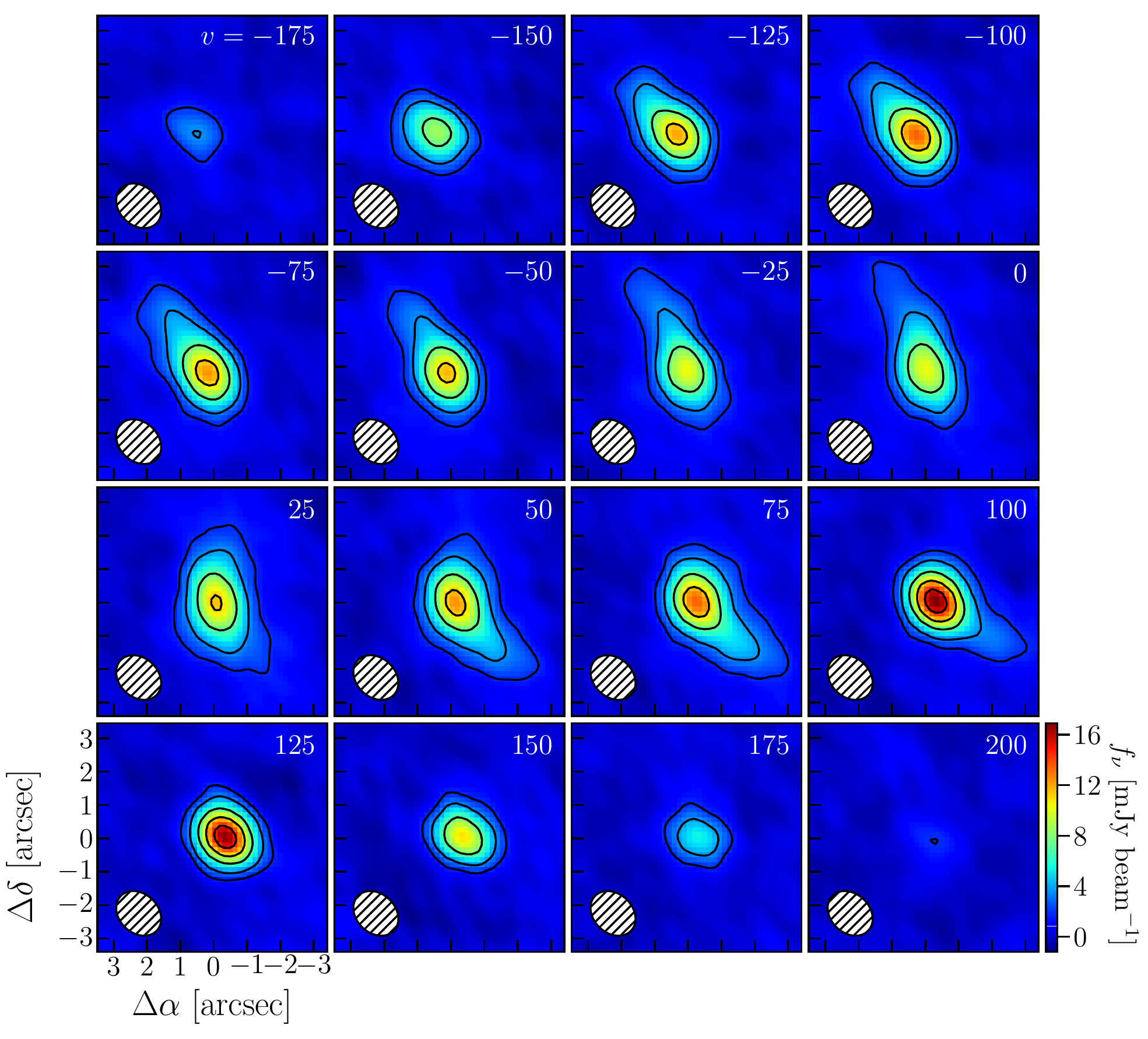}
 \put (8.5,90.2) {\large{PG1440+356}}
\end{overpic}
\centerline{Figure~\ref{fig:cmaps}. -- Continued.}
\end{figure*}

\begin{figure*}
\begin{overpic}[width=1.0\textwidth]{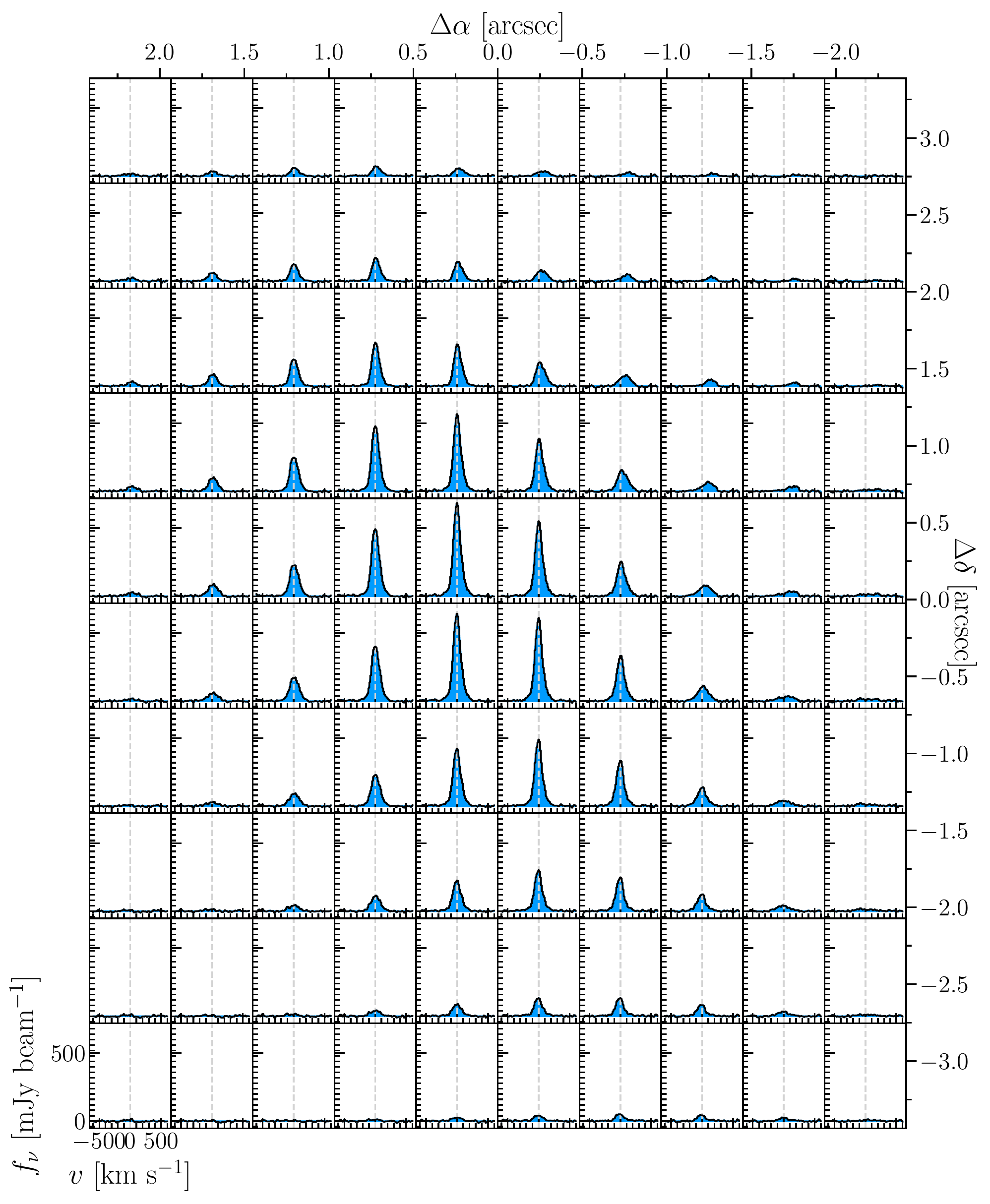}
 \put (64.25,4.5) {\large{Z11598$-$0112}}
\end{overpic}
\centerline{Figure~\ref{fig:smaps}. -- Continued.}
\end{figure*}

\begin{figure*}
\begin{overpic}[width=1.0\textwidth]{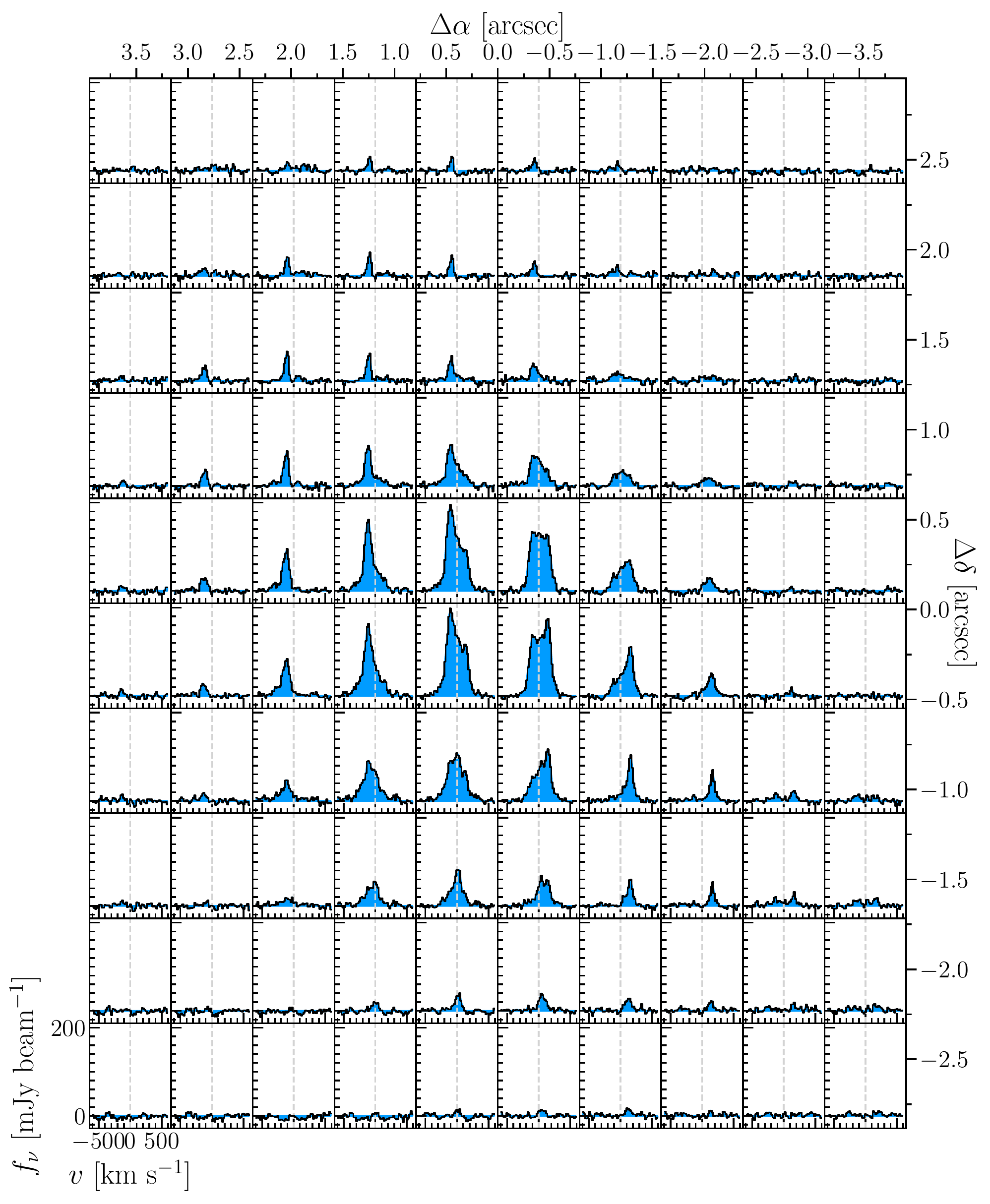}
 \put (64.25,4.5) {\large{F13342+3932}}
\end{overpic}
\centerline{Figure~\ref{fig:smaps}. -- Continued.}
\end{figure*}

\begin{figure*}
\begin{overpic}[width=1.0\textwidth]{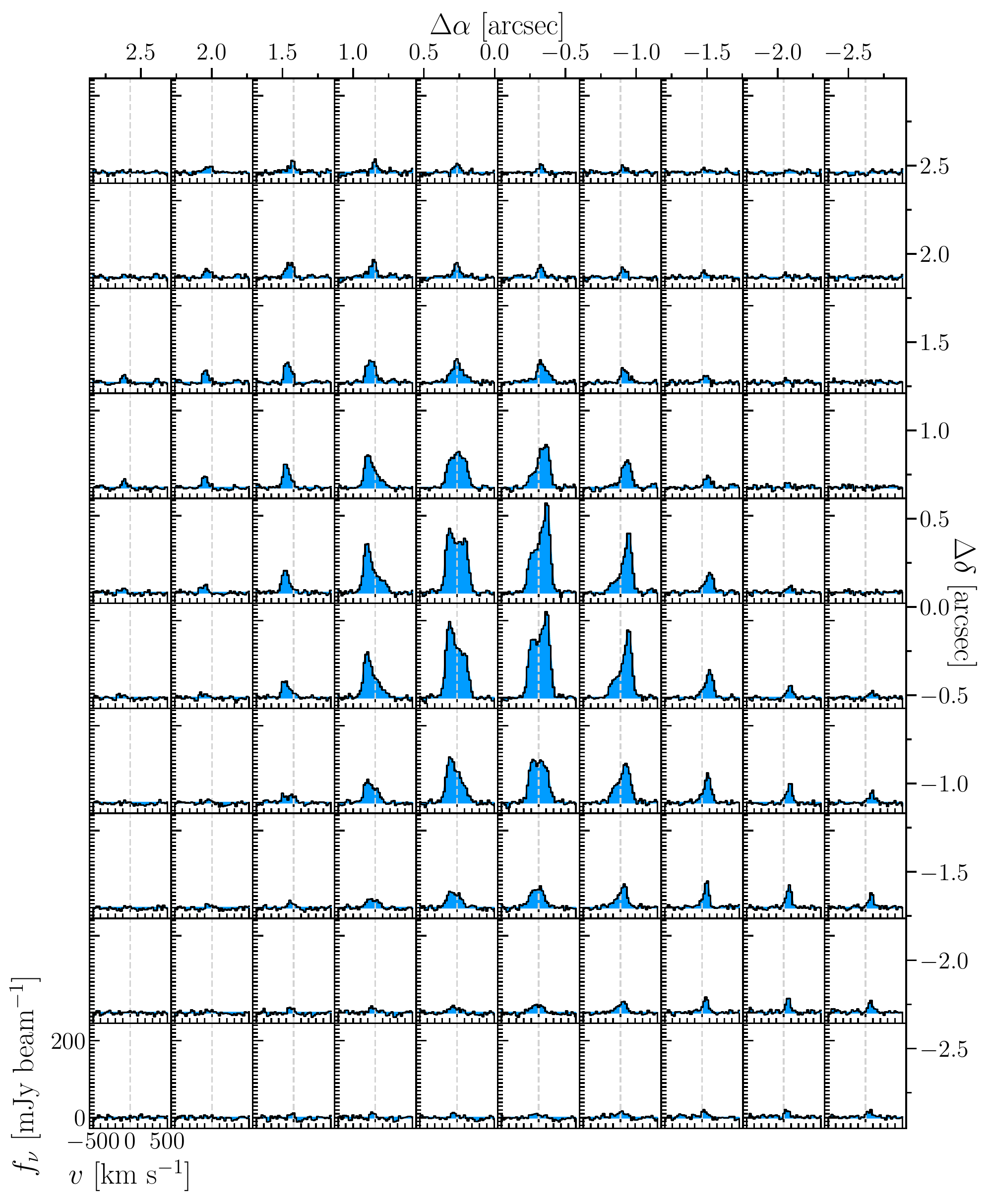}
 \put (63.5,4.5) {\large{PG1440+356}}
\end{overpic}
\centerline{Figure~\ref{fig:smaps}. -- Continued.}
\end{figure*}

\bibliographystyle{mnras}
\bibliography{all.102416}
\label{lastpage}
\end{document}

%% file: commands.tex
\newcounter{species}
\def\ion#1#2{\setcounter{species}{#2}#1$\;${\sc\roman{species}}\relax}

\newcommand\edit[1]{#1}

\newcommand{\kms}{km~s$^{-1}$}

\def\arcsec{\hbox{$^{\prime\prime}$}}
\newcommand{\Msun}{M$_{\odot}$}

\newcommand{\COIO}{CO$\,(1{-}0)$}

\def\lsim{\lower0.3em\hbox{$\,\buildrel <\over\sim\,$}}
\def\gsim{\lower0.3em\hbox{$\,\buildrel >\over\sim\,$}}

\def\kms{\,km~s$^{-1}$}      

\def\lesssim{\mathrel{\hbox{\rlap{\hbox{%
 \lower4pt\hbox{$\sim$}}}\hbox{$<$}}}}
\def\gtrsim{\mathrel{\hbox{\rlap{\hbox{%
 \lower4pt\hbox{$\sim$}}}\hbox{$>$}}}}

\def\arcsec{\hbox{$^{\prime\prime}$}}

\def\farcs{\hbox{$.\!\!^{\prime\prime}$}}

%% file: tables/obslog_mnras.tex
\begin{table*}
\centering
\begin{minipage}[]{\textwidth}
\caption{Description of observations\label{tab:obslog}}
\setlength{\tabcolsep}{2pt}
\renewcommand{\thefootnote}{\alph{footnote}}
\hspace{-0.5cm}
\scalebox{0.9}{
{\footnotesize
\begin{tabular}{ccccrccccccccc}
 & RA & Dec. & & & & & & $\nu_{obs}$ & $t_{OS}$  & Min, max & Prim. beam & Synt. beam & Beam \\
Object & [J2000] &  [J2000] & Redshift & Obs. date\phantom{0} & Config. & $N_{ant}$ & Corr. & [GHz] & [hr] & baseline [m] & [arcsec] & [arcsec] & PA [deg] \\
(1) &  (2) & (3) & (4) & (5)\phantom{0001} & (6) & (7) & (8) & (9) & (10) & (11) & (12) & (13) &  (14) \\
\hline
F07599+6508     & 08:04:30.46 & $+$64:59:52.88 & 0.1483   & 01--03/16    & A+B & 7  & W & 100.384           & 15.8 & 39.8--756.1 & 50.2 & 1.19$\times$1.02 & 36.91 \\
Z11598$-$0112   & 12:02:26.7575 & $-$01:29:15.418 & 0.1512045  & 12/18--01/19 & A+C & 10 & P & 100.131           & 8.5  & 15.9--760.0 & 50.3 & 3.26$\times$1.23 & 16.54 \\
F13342+3932     & 13:36:24.0589 & $+$39:17:31.131 & 0.1797   & 02--04/18    & A   & 9  & P & \phantom{1}97.712 & 7.5  & 18.2--737.5 & 51.6 & 1.60$\times$1.14 & 74.13 \\
PG\,1440+356    & 14:42:07.4732 & $+$35:26:22.953 & 0.077756 & 12/16        & A   & 8  & W & 106.780           & 9.0  & 23.6--758.5 & 47.1 & 1.50$\times$1.17 & 45.62 \\
\hline
\vspace{-.6cm}
\footnotetext{\textsc{NOTE --} (1) Object name; (2)--(3) Phase center coordinates determined from the NASA Extragalactic Database; (4) Redshift used to set \COIO\ velocity scale, see Section~\ref{sec:data} for references; (5) Date of observations in mm--mm/yy format; (6) Array configuration, C is the most compact; (7) Number of antennas; (8) Correlator, either WideX (W) or PolyFiX (P); (9) Observed frequency; (10) Time on source; (11) Minimum and maximum baselines; (12) The half power of the circular primary beam; (13) Synthesized beam major and minor axes; (14) Beam position angle measured East of North.}
\end{tabular}
}}
\end{minipage}
\end{table*}

%% file: tables/properties_mnras.tex
\begin{table}
\begin{minipage}[2cm]{\columnwidth}
\caption{Physical properties\label{tab:prop}}
\setlength{\tabcolsep}{2pt}
\renewcommand{\thefootnote}{\alph{footnote}}
\hspace{-0.5cm}
\scalebox{0.85}{
{\footnotesize
\begin{tabular}{ccccc}
 & \COIO & $S_{CO}\Delta\nu$ & $L_{CO}^{\prime}$ & $M$(H$_2$) \\ 
Object & Redshift & [Jy~km~s$^{-1}$] & [10$^9\,$K~km~s$^{-1}$~pc$^{2}$] & [10$^9\,$\Msun]\\ 
(1) &  (2) & (3) & (4) & (5) \\
\hline
F07599+6508                  & 0.1485$\pm$0.0002 & $10.0\pm0.1$  & \phantom{1}$9.6\pm0.1$   & $7.7\pm0.1$ \\
Z11598$-$0112                & 0.15118$\pm$0.00005 & \phantom{1}$9.46\pm0.02$  &\phantom{1}$9.45\pm0.02$   & $7.56\pm0.02$ \\
F13342+3932                  & 0.17965$\pm$0.00007 &  \phantom{1}$8.2\pm0.1$ &$11.6\pm0.1$  & $9.3\pm0.1$ \\
PG\,1440+356                  & 0.07777$\pm$0.00008 & \phantom{1}$9.2\pm0.1$  &\phantom{1}$2.35\pm0.02$   & $1.88\pm0.01$ \\
\hline
\vspace{-.6cm}
\footnotetext{\textsc{NOTE --} (1) Object name; (2) Flux-weighted CO redshift from the pixel with the maximum integrated flux, uncertainties are formal fit uncertainties added in quadrature with the velocity resolution; (3) Integrated flux from the \edit{data (\S\ref{sec:specfit}) within $\pm1000$~km~s$^{-1}$ in a $5\times5$\arcsec\ box}; (4) Total \COIO\ luminosity assuming redshift from column 2; (5) Total gas mass assuming $\alpha = M(\textrm{H}2)/L_{CO}^\prime \sim 0.8 \textrm{\Msun} (\textrm{K km s}^{-1}\textrm{pc}^2)^{-1}$.}
\end{tabular}
}}
\end{minipage}
\end{table}

%% file: tables/priors_mnras.tex
\begin{table*}
\begin{minipage}[2cm]{\textwidth}
\caption{Parameters of PG~1440+356 disk
\label{tab:priors}}
\setlength{\tabcolsep}{2pt}
\renewcommand{\thefootnote}{\alph{footnote}}
\hspace{-0.5cm}
\begin{tabular}{ccccl}
 & \multicolumn{3}{c}{Prior Distribution} & Adopted\\ 
 Parameter & Distribution & Lower bound & Upper bound & Value \\
\hline
$i$ (deg).........................    & Uniform           &  0     & 90     & 47$^{+7}_{-1}$ \\ 
$\psi$ (deg)........................  & Uniform           &  0     & 180    & 19$^{+7}_{-1}$ \\
$v_0$ (\kms).................         & Jeffreys          & 1.0    & 1000.  & 184$^{+3}_{-29}$ \\   
$R_c$ (kpc)......................     & Fixed             & ...    & ...    & 0.0 \\   
$x_0$ (arcsec)...................     & Uniform           & $-$1.0 & 1.0    & 0.092$^{+0.002}_{-0.020}$ \\   
$y_0$ (arcsec)...................     & Uniform           & $-$0.5 & 0.5    & 0.027$^{+0.001}_{-0.023}$ \\
$\sigma_{v}$ (\kms)................   & Mod. Jeffreys     & 0.001  & 400.   & 40.7$^{+4}_{-0.2}$ \\
log$(I_0)$ (Jy beam$^{-1}$)...        & Uniform           & $-$3.0 & $-$1.0 & $-$2.06$^{+0.007}_{-0.7}$ \\
$p$.................................  & Uniform           & 0.0    & 6.0    & 2.34$^{+1.9}_{-0.01}$\\
$r_{in}$ (arcsec).................    & Mod. Jeffreys     & 0.0    & 5.0    & 0.13$^{+0.11}_{-0.01}$\\
$r_{out}$ (arcsec)...............     & Jeffreys          & 1.0    & 20.0   & 4.7$^{+0.5}_{-0.4}$ \\
$z$.................................  & Uniform           & 0.075  & 0.080  & 0.0777142$^{+9\times10^{-7}}_{-5\times10^{-7}}$\\
\hline
\vspace{-.6cm}
\footnotetext{\textsc{NOTE --} See \S\ref{sec:outflow} of the text for a detailed discussion.}
\end{tabular}
\end{minipage}
\end{table*}

%% file: noutflows-comp.mnras.bbl
\begin{thebibliography}{}
\makeatletter
\relax
\def\mn@urlcharsother{\let\do\@makeother \do\$\do\&\do\#\do\^\do\_\do\%\do\~}
\def\mn@doi{\begingroup\mn@urlcharsother \@ifnextchar [ {\mn@doi@}
  {\mn@doi@[]}}
\def\mn@doi@[#1]#2{\def\@tempa{#1}\ifx\@tempa\@empty \href
  {http://dx.doi.org/#2} {doi:#2}\else \href {http://dx.doi.org/#2} {#1}\fi
  \endgroup}
\def\mn@eprint#1#2{\mn@eprint@#1:#2::\@nil}
\def\mn@eprint@arXiv#1{\href {http://arxiv.org/abs/#1} {{\tt arXiv:#1}}}
\def\mn@eprint@dblp#1{\href {http://dblp.uni-trier.de/rec/bibtex/#1.xml}
  {dblp:#1}}
\def\mn@eprint@#1:#2:#3:#4\@nil{\def\@tempa {#1}\def\@tempb {#2}\def\@tempc
  {#3}\ifx \@tempc \@empty \let \@tempc \@tempb \let \@tempb \@tempa \fi \ifx
  \@tempb \@empty \def\@tempb {arXiv}\fi \@ifundefined
  {mn@eprint@\@tempb}{\@tempb:\@tempc}{\expandafter \expandafter \csname
  mn@eprint@\@tempb\endcsname \expandafter{\@tempc}}}

\bibitem[\protect\citeauthoryear{{Ahumada} et~al.,}{{Ahumada}
  et~al.}{2020}]{dr16}
{Ahumada} R.,  et~al., 2020, \mn@doi [\apjs] {10.3847/1538-4365/ab929e}, \href
  {https://ui.adsabs.harvard.edu/abs/2020ApJS..249....3A} {249, 3}

\bibitem[\protect\citeauthoryear{{Alatalo} et~al.,}{{Alatalo}
  et~al.}{2011}]{alatalo11}
{Alatalo} K.,  et~al., 2011, \mn@doi [\apj] {10.1088/0004-637X/735/2/88}, \href
  {https://ui.adsabs.harvard.edu/abs/2011ApJ...735...88A} {735, 88}

\bibitem[\protect\citeauthoryear{{Alexander} \& {Hickox}}{{Alexander} \&
  {Hickox}}{2012}]{alexander12}
{Alexander} D.~M.,  {Hickox} R.~C.,  2012, \mn@doi [\nar]
  {10.1016/j.newar.2011.11.003}, \href
  {https://ui.adsabs.harvard.edu/abs/2012NewAR..56...93A} {56, 93}

\bibitem[\protect\citeauthoryear{{Alonso-Herrero} et~al.,}{{Alonso-Herrero}
  et~al.}{2019}]{alonso-herrero19}
{Alonso-Herrero} A.,  et~al., 2019, \mn@doi [\aap]
  {10.1051/0004-6361/201935431}, \href
  {https://ui.adsabs.harvard.edu/abs/2019A&A...628A..65A} {628, A65}

\bibitem[\protect\citeauthoryear{{Arribas}, {Colina}, {Bellocchi}, {Maiolino}
  \& {Villar-Mart{\'\i}n}}{{Arribas} et~al.}{2014}]{arribas14}
{Arribas} S.,  {Colina} L.,  {Bellocchi} E.,  {Maiolino} R.,
  {Villar-Mart{\'\i}n} M.,  2014, \mn@doi [\aap] {10.1051/0004-6361/201323324},
  \href {https://ui.adsabs.harvard.edu/abs/2014A&A...568A..14A} {568, A14}

\bibitem[\protect\citeauthoryear{{Beckwith} \& {Sargent}}{{Beckwith} \&
  {Sargent}}{1993}]{beckwith93}
{Beckwith} S. V.~W.,  {Sargent} A.~I.,  1993, \mn@doi [\apj] {10.1086/172131},
  \href {https://ui.adsabs.harvard.edu/abs/1993ApJ...402..280B} {402, 280}

\bibitem[\protect\citeauthoryear{{Borne}, {Bushouse}, {Lucas}  \&
  {Colina}}{{Borne} et~al.}{2000}]{borne00}
{Borne} K.~D.,  {Bushouse} H.,  {Lucas} R.~A.,   {Colina} L.,  2000, \mn@doi
  [\apjl] {10.1086/312461}, \href
  {https://ui.adsabs.harvard.edu/abs/2000ApJ...529L..77B} {529, L77}

\bibitem[\protect\citeauthoryear{{Boroson} \& {Meyers}}{{Boroson} \&
  {Meyers}}{1992}]{boroson92b}
{Boroson} T.~A.,  {Meyers} K.~A.,  1992, \mn@doi [\apj] {10.1086/171800}, \href
  {https://ui.adsabs.harvard.edu/abs/1992ApJ...397..442B} {397, 442}

\bibitem[\protect\citeauthoryear{{Carniani} et~al.,}{{Carniani}
  et~al.}{2015}]{carniani15}
{Carniani} S.,  et~al., 2015, \mn@doi [\aap] {10.1051/0004-6361/201526557},
  \href {https://ui.adsabs.harvard.edu/abs/2015A&A...580A.102C} {580, A102}

\bibitem[\protect\citeauthoryear{{Castro-Carrizo} \& {Neri}}{{Castro-Carrizo}
  \& {Neri}}{2010}]{castro-carrizo10}
{Castro-Carrizo} A.,  {Neri} R.,  2010, IRAM NOEMA Data Reduction CookBook

\bibitem[\protect\citeauthoryear{{Cazzoli}, {Arribas}, {Maiolino}  \&
  {Colina}}{{Cazzoli} et~al.}{2016}]{cazzoli16}
{Cazzoli} S.,  {Arribas} S.,  {Maiolino} R.,   {Colina} L.,  2016, \mn@doi
  [\aap] {10.1051/0004-6361/201526788}, \href
  {https://ui.adsabs.harvard.edu/abs/2016A&A...590A.125C} {590, A125}

\bibitem[\protect\citeauthoryear{{Cicone} et~al.,}{{Cicone}
  et~al.}{2014}]{cicone14}
{Cicone} C.,  et~al., 2014, \mn@doi [\aap] {10.1051/0004-6361/201322464}, \href
  {https://ui.adsabs.harvard.edu/abs/2014A&A...562A..21C} {562, A21}

\bibitem[\protect\citeauthoryear{{Cicone}, {Brusa}, {Ramos Almeida}, {Cresci},
  {Husemann}  \& {Mainieri}}{{Cicone} et~al.}{2018}]{cicone18}
{Cicone} C.,  {Brusa} M.,  {Ramos Almeida} C.,  {Cresci} G.,  {Husemann} B.,
  {Mainieri} V.,  2018, \mn@doi [Nature Astronomy] {10.1038/s41550-018-0406-3},
  \href {https://ui.adsabs.harvard.edu/abs/2018NatAs...2..176C} {2, 176}

\bibitem[\protect\citeauthoryear{{Combes} et~al.,}{{Combes}
  et~al.}{2013}]{combes13}
{Combes} F.,  et~al., 2013, \mn@doi [\aap] {10.1051/0004-6361/201322288}, \href
  {https://ui.adsabs.harvard.edu/abs/2013A&A...558A.124C} {558, A124}

\bibitem[\protect\citeauthoryear{{Combes} et~al.,}{{Combes}
  et~al.}{2014}]{combes14}
{Combes} F.,  et~al., 2014, \mn@doi [\aap] {10.1051/0004-6361/201423433}, \href
  {https://ui.adsabs.harvard.edu/abs/2014A&A...565A..97C} {565, A97}

\bibitem[\protect\citeauthoryear{{Combes} et~al.,}{{Combes}
  et~al.}{2019}]{combes19}
{Combes} F.,  et~al., 2019, \mn@doi [\aap] {10.1051/0004-6361/201834560}, \href
  {https://ui.adsabs.harvard.edu/abs/2019A&A...623A..79C} {623, A79}

\bibitem[\protect\citeauthoryear{{Condon}, {Yin}, {Thuan}  \&
  {Boller}}{{Condon} et~al.}{1998}]{condon98b}
{Condon} J.~J.,  {Yin} Q.~F.,  {Thuan} T.~X.,   {Boller} T.,  1998, \mn@doi
  [\aj] {10.1086/300624}, \href
  {https://ui.adsabs.harvard.edu/abs/1998AJ....116.2682C} {116, 2682}

\bibitem[\protect\citeauthoryear{{Dasyra} \& {Combes}}{{Dasyra} \&
  {Combes}}{2012}]{dasyra12}
{Dasyra} K.~M.,  {Combes} F.,  2012, \mn@doi [\aap]
  {10.1051/0004-6361/201219229}, \href
  {https://ui.adsabs.harvard.edu/abs/2012A&A...541L...7D} {541, L7}

\bibitem[\protect\citeauthoryear{{Denney}, {Peterson}, {Dietrich},
  {Vestergaard}  \& {Bentz}}{{Denney} et~al.}{2009}]{denney09a}
{Denney} K.~D.,  {Peterson} B.~M.,  {Dietrich} M.,  {Vestergaard} M.,   {Bentz}
  M.~C.,  2009, \mn@doi [\apj] {10.1088/0004-637X/692/1/246}, \href
  {http://adsabs.harvard.edu/abs/2009ApJ...692..246D} {692, 246}

\bibitem[\protect\citeauthoryear{{Di Matteo}, {Springel}  \& {Hernquist}}{{Di
  Matteo} et~al.}{2005}]{dimatteo05}
{Di Matteo} T.,  {Springel} V.,   {Hernquist} L.,  2005, \mn@doi [\nat]
  {10.1038/nature03335}, \href
  {http://adsabs.harvard.edu/abs/2005Natur.433..604D} {433, 604}

\bibitem[\protect\citeauthoryear{{DiPompeo}, {Myers}, {Hickox}, {Geach}  \&
  {Hainline}}{{DiPompeo} et~al.}{2014}]{dipompeo14b}
{DiPompeo} M.~A.,  {Myers} A.~D.,  {Hickox} R.~C.,  {Geach} J.~E.,   {Hainline}
  K.~N.,  2014, \mn@doi [\mnras] {10.1093/mnras/stu1115}, \href
  {http://adsabs.harvard.edu/abs/2014MNRAS.442.3443D} {442, 3443}

\bibitem[\protect\citeauthoryear{{DiPompeo}, {Myers}, {Hickox}, {Geach},
  {Holder}, {Hainline}  \& {Hall}}{{DiPompeo} et~al.}{2015}]{dipompeo15a}
{DiPompeo} M.~A.,  {Myers} A.~D.,  {Hickox} R.~C.,  {Geach} J.~E.,  {Holder}
  G.,  {Hainline} K.~N.,   {Hall} S.~W.,  2015, \mn@doi [\mnras]
  {10.1093/mnras/stu2341}, \href
  {http://adsabs.harvard.edu/abs/2015MNRAS.446.3492D} {446, 3492}

\bibitem[\protect\citeauthoryear{{DiPompeo}, {Hickox}, {Carroll}, {Runnoe},
  {Mullaney}  \& {Fischer}}{{DiPompeo} et~al.}{2018}]{dipompeo18}
{DiPompeo} M.~A.,  {Hickox} R.~C.,  {Carroll} C.~M.,  {Runnoe} J.~C.,
  {Mullaney} J.~R.,   {Fischer} T.~C.,  2018, \mn@doi [\apj]
  {10.3847/1538-4357/aab365}, \href
  {https://ui.adsabs.harvard.edu/abs/2018ApJ...856...76D} {856, 76}

\bibitem[\protect\citeauthoryear{{Evans}, {Frayer}, {Surace}  \&
  {Sanders}}{{Evans} et~al.}{2001}]{evans01}
{Evans} A.~S.,  {Frayer} D.~T.,  {Surace} J.~A.,   {Sanders} D.~B.,  2001,
  \mn@doi [\aj] {10.1086/319972}, \href
  {http://adsabs.harvard.edu/abs/2001AJ....121.1893E} {121, 1893}

\bibitem[\protect\citeauthoryear{{Evans} et~al.,}{{Evans}
  et~al.}{2009}]{evans09}
{Evans} A.~S.,  et~al., 2009, \mn@doi [\aj] {10.1088/0004-6256/138/1/262},
  \href {http://adsabs.harvard.edu/abs/2009AJ....138..262E} {138, 262}

\bibitem[\protect\citeauthoryear{{Feruglio}, {Maiolino}, {Piconcelli}, {Menci},
  {Aussel}, {Lamastra}  \& {Fiore}}{{Feruglio} et~al.}{2010}]{feruglio10}
{Feruglio} C.,  {Maiolino} R.,  {Piconcelli} E.,  {Menci} N.,  {Aussel} H.,
  {Lamastra} A.,   {Fiore} F.,  2010, \mn@doi [\aap]
  {10.1051/0004-6361/201015164}, \href
  {https://ui.adsabs.harvard.edu/abs/2010A&A...518L.155F} {518, L155}

\bibitem[\protect\citeauthoryear{{Fiore} et~al.,}{{Fiore}
  et~al.}{2017}]{fiore17}
{Fiore} F.,  et~al., 2017, \mn@doi [\aap] {10.1051/0004-6361/201629478}, \href
  {https://ui.adsabs.harvard.edu/abs/2017A&A...601A.143F} {601, A143}

\bibitem[\protect\citeauthoryear{{Fischer} et~al.,}{{Fischer}
  et~al.}{2010}]{fischer10}
{Fischer} J.,  et~al., 2010, \mn@doi [\aap] {10.1051/0004-6361/201014676},
  \href {https://ui.adsabs.harvard.edu/abs/2010A&A...518L..41F} {518, L41}

\bibitem[\protect\citeauthoryear{{Fischer} et~al.,}{{Fischer}
  et~al.}{2017}]{fischer17}
{Fischer} T.~C.,  et~al., 2017, \mn@doi [\apj] {10.3847/1538-4357/834/1/30},
  \href {https://ui.adsabs.harvard.edu/abs/2017ApJ...834...30F} {834, 30}

\bibitem[\protect\citeauthoryear{{Fischer} et~al.,}{{Fischer}
  et~al.}{2018}]{fischer18}
{Fischer} T.~C.,  et~al., 2018, \mn@doi [\apj] {10.3847/1538-4357/aab03e},
  \href {https://ui.adsabs.harvard.edu/abs/2018ApJ...856..102F} {856, 102}

\bibitem[\protect\citeauthoryear{{Fluetsch} et~al.,}{{Fluetsch}
  et~al.}{2019}]{fluetsch19}
{Fluetsch} A.,  et~al., 2019, \mn@doi [\mnras] {10.1093/mnras/sty3449}, \href
  {https://ui.adsabs.harvard.edu/abs/2019MNRAS.483.4586F} {483, 4586}

\bibitem[\protect\citeauthoryear{{Fluetsch} et~al.,}{{Fluetsch}
  et~al.}{2020}]{fluetsch20}
{Fluetsch} A.,  et~al., 2020, arXiv e-prints, \href
  {https://ui.adsabs.harvard.edu/abs/2020arXiv200613232F} {p. arXiv:2006.13232}

\bibitem[\protect\citeauthoryear{{Foreman-Mackey}, {Hogg}, {Lang}  \&
  {Goodman}}{{Foreman-Mackey} et~al.}{2013}]{foreman13}
{Foreman-Mackey} D.,  {Hogg} D.~W.,  {Lang} D.,   {Goodman} J.,  2013, \mn@doi
  [\pasp] {10.1086/670067}, \href
  {http://adsabs.harvard.edu/abs/2013PASP..125..306F} {125, 306}

\bibitem[\protect\citeauthoryear{{Garc{\'\i}a-Burillo}
  et~al.,}{{Garc{\'\i}a-Burillo} et~al.}{2015}]{garcia-burillo15}
{Garc{\'\i}a-Burillo} S.,  et~al., 2015, \mn@doi [\aap]
  {10.1051/0004-6361/201526133}, \href
  {https://ui.adsabs.harvard.edu/abs/2015A&A...580A..35G} {580, A35}

\bibitem[\protect\citeauthoryear{{Gonz{\'a}lez-Alfonso}
  et~al.,}{{Gonz{\'a}lez-Alfonso} et~al.}{2017}]{gonzalez-alfonso17}
{Gonz{\'a}lez-Alfonso} E.,  et~al., 2017, \mn@doi [\apj]
  {10.3847/1538-4357/836/1/11}, \href
  {https://ui.adsabs.harvard.edu/abs/2017ApJ...836...11G} {836, 11}

\bibitem[\protect\citeauthoryear{{Green}, {Schmidt}  \& {Liebert}}{{Green}
  et~al.}{1986}]{green86}
{Green} R.~F.,  {Schmidt} M.,   {Liebert} J.,  1986, \mn@doi [\apjs]
  {10.1086/191115}, \href
  {https://ui.adsabs.harvard.edu/abs/1986ApJS...61..305G} {61, 305}

\bibitem[\protect\citeauthoryear{{Harrison}, {Alexander}, {Mullaney}  \&
  {Swinbank}}{{Harrison} et~al.}{2014}]{harrison14}
{Harrison} C.~M.,  {Alexander} D.~M.,  {Mullaney} J.~R.,   {Swinbank} A.~M.,
  2014, \mn@doi [\mnras] {10.1093/mnras/stu515}, \href
  {https://ui.adsabs.harvard.edu/abs/2014MNRAS.441.3306H} {441, 3306}

\bibitem[\protect\citeauthoryear{{Harrison}, {Costa}, {Tadhunter},
  {Fl{\"u}tsch}, {Kakkad}, {Perna}  \& {Vietri}}{{Harrison}
  et~al.}{2018}]{harrison18}
{Harrison} C.~M.,  {Costa} T.,  {Tadhunter} C.~N.,  {Fl{\"u}tsch} A.,  {Kakkad}
  D.,  {Perna} M.,   {Vietri} G.,  2018, \mn@doi [Nature Astronomy]
  {10.1038/s41550-018-0403-6}, \href
  {https://ui.adsabs.harvard.edu/abs/2018NatAs...2..198H} {2, 198}

\bibitem[\protect\citeauthoryear{{Herrera-Camus} et~al.,}{{Herrera-Camus}
  et~al.}{2019}]{herrera-camus19}
{Herrera-Camus} R.,  et~al., 2019, \mn@doi [\apj] {10.3847/1538-4357/aaf6a7},
  \href {https://ui.adsabs.harvard.edu/abs/2019ApJ...871...37H} {871, 37}

\bibitem[\protect\citeauthoryear{{Hines} \& {Wills}}{{Hines} \&
  {Wills}}{1995}]{hines95}
{Hines} D.~C.,  {Wills} B.~J.,  1995, \mn@doi [\apjl] {10.1086/309611}, \href
  {https://ui.adsabs.harvard.edu/abs/1995ApJ...448L..69H} {448, L69}

\bibitem[\protect\citeauthoryear{{Kang} \& {Woo}}{{Kang} \&
  {Woo}}{2018}]{kang18}
{Kang} D.,  {Woo} J.-H.,  2018, \mn@doi [\apj] {10.3847/1538-4357/aad561},
  \href {https://ui.adsabs.harvard.edu/abs/2018ApJ...864..124K} {864, 124}

\bibitem[\protect\citeauthoryear{{Karouzos}, {Woo}  \& {Bae}}{{Karouzos}
  et~al.}{2016}]{karouzos16}
{Karouzos} M.,  {Woo} J.-H.,   {Bae} H.-J.,  2016, \mn@doi [\apj]
  {10.3847/0004-637X/819/2/148}, \href
  {https://ui.adsabs.harvard.edu/abs/2016ApJ...819..148K} {819, 148}

\bibitem[\protect\citeauthoryear{{Kim} \& {Sanders}}{{Kim} \&
  {Sanders}}{1998}]{kim98}
{Kim} D.~C.,  {Sanders} D.~B.,  1998, \mn@doi [\apjs] {10.1086/313148}, \href
  {https://ui.adsabs.harvard.edu/abs/1998ApJS..119...41K} {119, 41}

\bibitem[\protect\citeauthoryear{{King}}{{King}}{2003}]{king03}
{King} A.,  2003, \mn@doi [\apjl] {10.1086/379143}, \href
  {http://adsabs.harvard.edu/abs/2003ApJ...596L..27K} {596, L27}

\bibitem[\protect\citeauthoryear{{Lanzetta}, {Turnshek}  \&
  {Sandoval}}{{Lanzetta} et~al.}{1993}]{lanzetta93}
{Lanzetta} K.~M.,  {Turnshek} D.~A.,   {Sandoval} J.,  1993, \mn@doi [\apjs]
  {10.1086/191749}, \href
  {https://ui.adsabs.harvard.edu/abs/1993ApJS...84..109L} {84, 109}

\bibitem[\protect\citeauthoryear{{Lawrence}, {Saunders}, {Rowan-Robinson},
  {Crawford}, {Ellis}, {Frenk}, {Efstathiou}  \& {Kaiser}}{{Lawrence}
  et~al.}{1988}]{lawrence88}
{Lawrence} A.,  {Saunders} W.,  {Rowan-Robinson} M.,  {Crawford} J.,  {Ellis}
  R.~S.,  {Frenk} C.~S.,  {Efstathiou} G.,   {Kaiser} N.,  1988, \mn@doi
  [\mnras] {10.1093/mnras/235.1.261}, \href
  {https://ui.adsabs.harvard.edu/abs/1988MNRAS.235..261L} {235, 261}

\bibitem[\protect\citeauthoryear{{Leung} et~al.,}{{Leung}
  et~al.}{2019}]{leung19}
{Leung} G. C.~K.,  et~al., 2019, \mn@doi [\apj] {10.3847/1538-4357/ab4a7c},
  \href {https://ui.adsabs.harvard.edu/abs/2019ApJ...886...11L} {886, 11}

\bibitem[\protect\citeauthoryear{{Lipari}}{{Lipari}}{1994}]{lipari94}
{Lipari} S.,  1994, \mn@doi [\apj] {10.1086/174884}, \href
  {https://ui.adsabs.harvard.edu/abs/1994ApJ...436..102L} {436, 102}

\bibitem[\protect\citeauthoryear{{Low}, {Cutri}, {Kleinmann}  \&
  {Huchra}}{{Low} et~al.}{1989}]{low89}
{Low} F.~J.,  {Cutri} R.~M.,  {Kleinmann} S.~G.,   {Huchra} J.~P.,  1989,
  \mn@doi [\apjl] {10.1086/185424}, \href
  {https://ui.adsabs.harvard.edu/abs/1989ApJ...340L...1L} {340, L1}

\bibitem[\protect\citeauthoryear{{Lutz} et~al.,}{{Lutz} et~al.}{2020}]{lutz20}
{Lutz} D.,  et~al., 2020, \mn@doi [\aap] {10.1051/0004-6361/201936803}, \href
  {https://ui.adsabs.harvard.edu/abs/2020A&A...633A.134L} {633, A134}

\bibitem[\protect\citeauthoryear{{Markwardt}}{{Markwardt}}{2012}]{markwardt12}
{Markwardt} C.,  2012, {MPFIT: Robust non-linear least squares curve fitting},
  Astrophysics Source Code Library (\mn@eprint {ascl} {1208.019})

\bibitem[\protect\citeauthoryear{{Martin}, {Dijkstra}, {Henry}, {Soto},
  {Danforth}  \& {Wong}}{{Martin} et~al.}{2015}]{martin15}
{Martin} C.~L.,  {Dijkstra} M.,  {Henry} A.,  {Soto} K.~T.,  {Danforth} C.~W.,
   {Wong} J.,  2015, \mn@doi [\apj] {10.1088/0004-637X/803/1/6}, \href
  {https://ui.adsabs.harvard.edu/abs/2015ApJ...803....6M} {803, 6}

\bibitem[\protect\citeauthoryear{{McElroy}, {Croom}, {Pracy}, {Sharp}, {Ho}  \&
  {Medling}}{{McElroy} et~al.}{2015}]{mcelroy15}
{McElroy} R.,  {Croom} S.~M.,  {Pracy} M.,  {Sharp} R.,  {Ho} I.~T.,
  {Medling} A.~M.,  2015, \mn@doi [\mnras] {10.1093/mnras/stu2224}, \href
  {https://ui.adsabs.harvard.edu/abs/2015MNRAS.446.2186M} {446, 2186}

\bibitem[\protect\citeauthoryear{{Meurs} \& {Wilson}}{{Meurs} \&
  {Wilson}}{1984}]{meurs84}
{Meurs} E.~J.~A.,  {Wilson} A.~S.,  1984, \aap, \href
  {https://ui.adsabs.harvard.edu/abs/1984A&A...136..206M} {136, 206}

\bibitem[\protect\citeauthoryear{{Miller}, {Rawlings}  \& {Saunders}}{{Miller}
  et~al.}{1993}]{miller93}
{Miller} P.,  {Rawlings} S.,   {Saunders} R.,  1993, \mn@doi [\mnras]
  {10.1093/mnras/263.2.425}, \href
  {https://ui.adsabs.harvard.edu/abs/1993MNRAS.263..425M} {263, 425}

\bibitem[\protect\citeauthoryear{{Moran}, {Halpern}  \& {Helfand}}{{Moran}
  et~al.}{1996}]{moran96}
{Moran} E.~C.,  {Halpern} J.~P.,   {Helfand} D.~J.,  1996, \mn@doi [\apjs]
  {10.1086/192341}, \href
  {https://ui.adsabs.harvard.edu/abs/1996ApJS..106..341M} {106, 341}

\bibitem[\protect\citeauthoryear{{Morganti}, {Tadhunter}  \&
  {Oosterloo}}{{Morganti} et~al.}{2005}]{morganti05}
{Morganti} R.,  {Tadhunter} C.~N.,   {Oosterloo} T.~A.,  2005, \mn@doi [\aap]
  {10.1051/0004-6361:200500197}, \href
  {http://adsabs.harvard.edu/abs/2005A%26A...444L...9M} {444, L9}

\bibitem[\protect\citeauthoryear{{Morganti}, {Veilleux}, {Oosterloo}, {Teng}
  \& {Rupke}}{{Morganti} et~al.}{2016}]{morganti16}
{Morganti} R.,  {Veilleux} S.,  {Oosterloo} T.,  {Teng} S.~H.,   {Rupke} D.,
  2016, \mn@doi [\aap] {10.1051/0004-6361/201628978}, \href
  {https://ui.adsabs.harvard.edu/abs/2016A&A...593A..30M} {593, A30}

\bibitem[\protect\citeauthoryear{{Nagar}, {Wilson}, {Falcke}, {Veilleux}  \&
  {Maiolino}}{{Nagar} et~al.}{2003}]{nagar03}
{Nagar} N.~M.,  {Wilson} A.~S.,  {Falcke} H.,  {Veilleux} S.,   {Maiolino} R.,
  2003, \mn@doi [\aap] {10.1051/0004-6361:20031069}, \href
  {https://ui.adsabs.harvard.edu/abs/2003A&A...409..115N} {409, 115}

\bibitem[\protect\citeauthoryear{{Nesvadba} et~al.,}{{Nesvadba}
  et~al.}{2010}]{nesvadba10}
{Nesvadba} N.~P.~H.,  et~al., 2010, \mn@doi [\aap]
  {10.1051/0004-6361/200913333}, \href
  {https://ui.adsabs.harvard.edu/abs/2010A&A...521A..65N} {521, A65}

\bibitem[\protect\citeauthoryear{{Netzer} et~al.,}{{Netzer}
  et~al.}{2007}]{netzer07}
{Netzer} H.,  et~al., 2007, \mn@doi [\apj] {10.1086/520716}, \href
  {http://adsabs.harvard.edu/abs/2007ApJ...666..806N} {666, 806}

\bibitem[\protect\citeauthoryear{{Pereira-Santaella}
  et~al.,}{{Pereira-Santaella} et~al.}{2016}]{pereira-santaella16}
{Pereira-Santaella} M.,  et~al., 2016, \mn@doi [\aap]
  {10.1051/0004-6361/201628875}, \href
  {https://ui.adsabs.harvard.edu/abs/2016A&A...594A..81P} {594, A81}

\bibitem[\protect\citeauthoryear{{Pereira-Santaella}
  et~al.,}{{Pereira-Santaella} et~al.}{2018}]{pereira-santaella18}
{Pereira-Santaella} M.,  et~al., 2018, \mn@doi [\aap]
  {10.1051/0004-6361/201833089}, \href
  {https://ui.adsabs.harvard.edu/abs/2018A&A...616A.171P} {616, A171}

\bibitem[\protect\citeauthoryear{{Pringle}}{{Pringle}}{1981}]{pringle81}
{Pringle} J.~E.,  1981, \mn@doi [\araa] {10.1146/annurev.aa.19.090181.001033},
  \href {https://ui.adsabs.harvard.edu/abs/1981ARA&A..19..137P} {19, 137}

\bibitem[\protect\citeauthoryear{{Ramakrishnan} et~al.,}{{Ramakrishnan}
  et~al.}{2019}]{ramakrishnan19}
{Ramakrishnan} V.,  et~al., 2019, \mn@doi [\mnras] {10.1093/mnras/stz1244},
  \href {https://ui.adsabs.harvard.edu/abs/2019MNRAS.487..444R} {487, 444}

\bibitem[\protect\citeauthoryear{{Runnoe}, {G{\"u}ltekin}  \& {Rupke}}{{Runnoe}
  et~al.}{2018}]{runnoe18}
{Runnoe} J.~C.,  {G{\"u}ltekin} K.,   {Rupke} D. S.~N.,  2018, \mn@doi [\apj]
  {10.3847/1538-4357/aa9934}, \href
  {https://ui.adsabs.harvard.edu/abs/2018ApJ...852....8R} {852, 8}

\bibitem[\protect\citeauthoryear{{Rupke} \& {Veilleux}}{{Rupke} \&
  {Veilleux}}{2013a}]{rupke13a}
{Rupke} D.~S.~N.,  {Veilleux} S.,  2013a, \mn@doi [\apj]
  {10.1088/0004-637X/768/1/75}, \href
  {http://adsabs.harvard.edu/abs/2013ApJ...768...75R} {768, 75}

\bibitem[\protect\citeauthoryear{{Rupke} \& {Veilleux}}{{Rupke} \&
  {Veilleux}}{2013b}]{rupke13b}
{Rupke} D.~S.~N.,  {Veilleux} S.,  2013b, \mn@doi [\apjl]
  {10.1088/2041-8205/775/1/L15}, \href
  {http://adsabs.harvard.edu/abs/2013ApJ...775L..15R} {775, L15}

\bibitem[\protect\citeauthoryear{{Rupke}, {Veilleux}  \& {Sanders}}{{Rupke}
  et~al.}{2005}]{rupke05}
{Rupke} D.~S.,  {Veilleux} S.,   {Sanders} D.~B.,  2005, \mn@doi [\apj]
  {10.1086/444451}, \href
  {https://ui.adsabs.harvard.edu/abs/2005ApJ...632..751R} {632, 751}

\bibitem[\protect\citeauthoryear{{Rupke}, {G{\"u}ltekin}  \&
  {Veilleux}}{{Rupke} et~al.}{2017}]{rupke17}
{Rupke} D. S.~N.,  {G{\"u}ltekin} K.,   {Veilleux} S.,  2017, \mn@doi [\apj]
  {10.3847/1538-4357/aa94d1}, \href
  {https://ui.adsabs.harvard.edu/abs/2017ApJ...850...40R} {850, 40}

\bibitem[\protect\citeauthoryear{{Salak}, {Nakai}, {Hatakeyama}  \&
  {Miyamoto}}{{Salak} et~al.}{2016}]{salak16}
{Salak} D.,  {Nakai} N.,  {Hatakeyama} T.,   {Miyamoto} Y.,  2016, \mn@doi
  [\apj] {10.3847/0004-637X/823/1/68}, \href
  {https://ui.adsabs.harvard.edu/abs/2016ApJ...823...68S} {823, 68}

\bibitem[\protect\citeauthoryear{{Sanders}, {Soifer}, {Elias}, {Neugebauer}  \&
  {Matthews}}{{Sanders} et~al.}{1988}]{sanders88}
{Sanders} D.~B.,  {Soifer} B.~T.,  {Elias} J.~H.,  {Neugebauer} G.,
  {Matthews} K.,  1988, \mn@doi [\apjl] {10.1086/185155}, \href
  {https://ui.adsabs.harvard.edu/abs/1988ApJ...328L..35S} {328, L35}

\bibitem[\protect\citeauthoryear{{Schmidt} \& {Green}}{{Schmidt} \&
  {Green}}{1983}]{schmidt83}
{Schmidt} M.,  {Green} R.~F.,  1983, \mn@doi [\apj] {10.1086/161048}, \href
  {http://adsabs.harvard.edu/abs/1983ApJ...269..352S} {269, 352}

\bibitem[\protect\citeauthoryear{{Scholtz} et~al.,}{{Scholtz}
  et~al.}{2020}]{scholtz20}
{Scholtz} J.,  et~al., 2020, \mn@doi [\mnras] {10.1093/mnras/staa030}, \href
  {https://ui.adsabs.harvard.edu/abs/2020MNRAS.492.3194S} {492, 3194}

\bibitem[\protect\citeauthoryear{{Schweitzer} et~al.,}{{Schweitzer}
  et~al.}{2006}]{schweitzer06}
{Schweitzer} M.,  et~al., 2006, \mn@doi [\apj] {10.1086/506510}, \href
  {https://ui.adsabs.harvard.edu/abs/2006ApJ...649...79S} {649, 79}

\bibitem[\protect\citeauthoryear{{Shang} et~al.,}{{Shang}
  et~al.}{2005}]{shang05}
{Shang} Z.,  et~al., 2005, \mn@doi [\apj] {10.1086/426134}, \href
  {http://adsabs.harvard.edu/abs/2005ApJ...619...41S} {619, 41}

\bibitem[\protect\citeauthoryear{{Shang} et~al.,}{{Shang}
  et~al.}{2011}]{shang11}
{Shang} Z.,  et~al., 2011, \mn@doi [\apjs] {10.1088/0067-0049/196/1/2}, \href
  {http://adsabs.harvard.edu/abs/2011ApJS..196....2S} {196, 2}

\bibitem[\protect\citeauthoryear{{Shangguan}, {Ho}, {Bauer}, {Wang}  \&
  {Treister}}{{Shangguan} et~al.}{2020}]{shangguan20}
{Shangguan} J.,  {Ho} L.~C.,  {Bauer} F.~E.,  {Wang} R.,   {Treister} E.,
  2020, \mn@doi [\apjs] {10.3847/1538-4365/ab5db2}, \href
  {https://ui.adsabs.harvard.edu/abs/2020ApJS..247...15S} {247, 15}

\bibitem[\protect\citeauthoryear{{Silk} \& {Rees}}{{Silk} \&
  {Rees}}{1998}]{silk98}
{Silk} J.,  {Rees} M.~J.,  1998, \aap, \href
  {http://adsabs.harvard.edu/abs/1998A%26A...331L...1S} {331, L1}

\bibitem[\protect\citeauthoryear{{Slater} et~al.,}{{Slater}
  et~al.}{2019}]{slater19}
{Slater} R.,  et~al., 2019, \mn@doi [\aap] {10.1051/0004-6361/201730634}, \href
  {https://ui.adsabs.harvard.edu/abs/2019A&A...621A..83S} {621, A83}

\bibitem[\protect\citeauthoryear{{Sofue} \& {Rubin}}{{Sofue} \&
  {Rubin}}{2001}]{sofue01}
{Sofue} Y.,  {Rubin} V.,  2001, \mn@doi [\araa]
  {10.1146/annurev.astro.39.1.137}, \href
  {https://ui.adsabs.harvard.edu/abs/2001ARA&A..39..137S} {39, 137}

\bibitem[\protect\citeauthoryear{{Sofue}, {Tutui}, {Honma}, {Tomita},
  {Takamiya}, {Koda}  \& {Takeda}}{{Sofue} et~al.}{1999}]{sofue99}
{Sofue} Y.,  {Tutui} Y.,  {Honma} M.,  {Tomita} A.,  {Takamiya} T.,  {Koda} J.,
    {Takeda} Y.,  1999, \mn@doi [\apj] {10.1086/307731}, \href
  {https://ui.adsabs.harvard.edu/abs/1999ApJ...523..136S} {523, 136}

\bibitem[\protect\citeauthoryear{{Somerville}, {Hopkins}, {Cox}, {Robertson}
  \& {Hernquist}}{{Somerville} et~al.}{2008}]{somerville08}
{Somerville} R.~S.,  {Hopkins} P.~F.,  {Cox} T.~J.,  {Robertson} B.~E.,
  {Hernquist} L.,  2008, \mn@doi [\mnras] {10.1111/j.1365-2966.2008.13805.x},
  \href {http://adsabs.harvard.edu/abs/2008MNRAS.391..481S} {391, 481}

\bibitem[\protect\citeauthoryear{{Spoon} et~al.,}{{Spoon}
  et~al.}{2013}]{spoon13}
{Spoon} H.~W.~W.,  et~al., 2013, \mn@doi [\apj] {10.1088/0004-637X/775/2/127},
  \href {https://ui.adsabs.harvard.edu/abs/2013ApJ...775..127S} {775, 127}

\bibitem[\protect\citeauthoryear{{Strauss}, {Huchra}, {Davis}, {Yahil},
  {Fisher}  \& {Tonry}}{{Strauss} et~al.}{1992}]{strauss92}
{Strauss} M.~A.,  {Huchra} J.~P.,  {Davis} M.,  {Yahil} A.,  {Fisher} K.~B.,
  {Tonry} J.,  1992, \mn@doi [\apjs] {10.1086/191730}, \href
  {https://ui.adsabs.harvard.edu/abs/1992ApJS...83...29S} {83, 29}

\bibitem[\protect\citeauthoryear{{Sturm} et~al.,}{{Sturm}
  et~al.}{2011}]{sturm11}
{Sturm} E.,  et~al., 2011, \mn@doi [\apjl] {10.1088/2041-8205/733/1/L16}, \href
  {https://ui.adsabs.harvard.edu/abs/2011ApJ...733L..16S} {733, L16}

\bibitem[\protect\citeauthoryear{{Surace}, {Sanders}  \& {Evans}}{{Surace}
  et~al.}{2001}]{surace01}
{Surace} J.~A.,  {Sanders} D.~B.,   {Evans} A.~S.,  2001, \mn@doi [\aj]
  {10.1086/324462}, \href
  {https://ui.adsabs.harvard.edu/abs/2001AJ....122.2791S} {122, 2791}

\bibitem[\protect\citeauthoryear{{Teague} \& {Foreman-Mackey}}{{Teague} \&
  {Foreman-Mackey}}{2018a}]{teague18b}
{Teague} R.,  {Foreman-Mackey} D.,  2018a, {Bettermoments: A Robust Method To
  Measure Line Centroids}, \mn@doi{10.5281/zenodo.1419754}

\bibitem[\protect\citeauthoryear{{Teague} \& {Foreman-Mackey}}{{Teague} \&
  {Foreman-Mackey}}{2018b}]{teague18}
{Teague} R.,  {Foreman-Mackey} D.,  2018b, \mn@doi [Research Notes of the
  American Astronomical Society] {10.3847/2515-5172/aae265}, \href
  {https://ui.adsabs.harvard.edu/abs/2018RNAAS...2..173T} {2, 173}

\bibitem[\protect\citeauthoryear{{Teague} \& {Foreman-Mackey}}{{Teague} \&
  {Foreman-Mackey}}{2019}]{teague19}
{Teague} R.,  {Foreman-Mackey} D.,  2019, {bettermoments: Line-of-sight
  velocity calculation} (\mn@eprint {ascl} {1901.009})

\bibitem[\protect\citeauthoryear{{Veilleux}, {Kim}  \& {Sanders}}{{Veilleux}
  et~al.}{2002}]{veilleux02}
{Veilleux} S.,  {Kim} D.~C.,   {Sanders} D.~B.,  2002, \mn@doi [\apjs]
  {10.1086/343844}, \href
  {https://ui.adsabs.harvard.edu/abs/2002ApJS..143..315V} {143, 315}

\bibitem[\protect\citeauthoryear{{Veilleux}, {Cecil}  \&
  {Bland-Hawthorn}}{{Veilleux} et~al.}{2005}]{veilleux05}
{Veilleux} S.,  {Cecil} G.,   {Bland-Hawthorn} J.,  2005, \mn@doi [\araa]
  {10.1146/annurev.astro.43.072103.150610}, \href
  {https://ui.adsabs.harvard.edu/abs/2005ARA&A..43..769V} {43, 769}

\bibitem[\protect\citeauthoryear{{Veilleux} et~al.,}{{Veilleux}
  et~al.}{2006}]{veilleux06}
{Veilleux} S.,  et~al., 2006, \mn@doi [\apj] {10.1086/503188}, \href
  {https://ui.adsabs.harvard.edu/abs/2006ApJ...643..707V} {643, 707}

\bibitem[\protect\citeauthoryear{{Veilleux} et~al.,}{{Veilleux}
  et~al.}{2009a}]{veilleux09b}
{Veilleux} S.,  et~al., 2009a, \mn@doi [\apjs] {10.1088/0067-0049/182/2/628},
  \href {https://ui.adsabs.harvard.edu/abs/2009ApJS..182..628V} {182, 628}

\bibitem[\protect\citeauthoryear{{Veilleux} et~al.,}{{Veilleux}
  et~al.}{2009b}]{veilleux09a}
{Veilleux} S.,  et~al., 2009b, \mn@doi [\apj] {10.1088/0004-637X/701/1/587},
  \href {https://ui.adsabs.harvard.edu/abs/2009ApJ...701..587V} {701, 587}

\bibitem[\protect\citeauthoryear{{Veilleux} et~al.,}{{Veilleux}
  et~al.}{2013}]{veilleux13b}
{Veilleux} S.,  et~al., 2013, \mn@doi [\apj] {10.1088/0004-637X/776/1/27},
  \href {http://adsabs.harvard.edu/abs/2013ApJ...776...27V} {776, 27}

\bibitem[\protect\citeauthoryear{{Veilleux}, {Bolatto}, {Tombesi},
  {Mel{\'e}ndez}, {Sturm}, {Gonz{\'a}lez-Alfonso}, {Fischer}  \&
  {Rupke}}{{Veilleux} et~al.}{2017}]{veilleux17}
{Veilleux} S.,  {Bolatto} A.,  {Tombesi} F.,  {Mel{\'e}ndez} M.,  {Sturm} E.,
  {Gonz{\'a}lez-Alfonso} E.,  {Fischer} J.,   {Rupke} D.~S.~N.,  2017, \mn@doi
  [\apj] {10.3847/1538-4357/aa767d}, \href
  {https://ui.adsabs.harvard.edu/abs/2017ApJ...843...18V} {843, 18}

\bibitem[\protect\citeauthoryear{{Veilleux}, {Maiolino}, {Bolatto}  \&
  {Aalto}}{{Veilleux} et~al.}{2020}]{veilleux20}
{Veilleux} S.,  {Maiolino} R.,  {Bolatto} A.~D.,   {Aalto} S.,  2020, \mn@doi
  [\aapr] {10.1007/s00159-019-0121-9}, \href
  {https://ui.adsabs.harvard.edu/abs/2020A&ARv..28....2V} {28, 2}

\bibitem[\protect\citeauthoryear{{Villar-Mart{\'\i}n}, {Arribas}, {Emonts},
  {Humphrey}, {Tadhunter}, {Bessiere}, {Cabrera Lavers}  \& {Ramos
  Almeida}}{{Villar-Mart{\'\i}n} et~al.}{2016}]{villar-martin16}
{Villar-Mart{\'\i}n} M.,  {Arribas} S.,  {Emonts} B.,  {Humphrey} A.,
  {Tadhunter} C.,  {Bessiere} P.,  {Cabrera Lavers} A.,   {Ramos Almeida} C.,
  2016, \mn@doi [\mnras] {10.1093/mnras/stw901}, \href
  {https://ui.adsabs.harvard.edu/abs/2016MNRAS.460..130V} {460, 130}

\bibitem[\protect\citeauthoryear{{Westmoquette}, {Clements}, {Bendo}  \&
  {Khan}}{{Westmoquette} et~al.}{2012}]{westmoquette12}
{Westmoquette} M.~S.,  {Clements} D.~L.,  {Bendo} G.~J.,   {Khan} S.~A.,  2012,
  \mn@doi [\mnras] {10.1111/j.1365-2966.2012.21214.x}, \href
  {https://ui.adsabs.harvard.edu/abs/2012MNRAS.424..416W} {424, 416}

\bibitem[\protect\citeauthoryear{{Wylezalek} \& {Morganti}}{{Wylezalek} \&
  {Morganti}}{2018}]{wylezalek18}
{Wylezalek} D.,  {Morganti} R.,  2018, \mn@doi [Nature Astronomy]
  {10.1038/s41550-018-0409-0}, \href
  {https://ui.adsabs.harvard.edu/abs/2018NatAs...2..181W} {2, 181}

\bibitem[\protect\citeauthoryear{{Xia} et~al.,}{{Xia} et~al.}{2012}]{xia12}
{Xia} X.~Y.,  et~al., 2012, \mn@doi [\apj] {10.1088/0004-637X/750/2/92}, \href
  {https://ui.adsabs.harvard.edu/abs/2012ApJ...750...92X} {750, 92}

\bibitem[\protect\citeauthoryear{{Yen} et~al.,}{{Yen} et~al.}{2014}]{yen14}
{Yen} H.-W.,  et~al., 2014, \mn@doi [\apj] {10.1088/0004-637X/793/1/1}, \href
  {https://ui.adsabs.harvard.edu/abs/2014ApJ...793....1Y} {793, 1}

\bibitem[\protect\citeauthoryear{{Zheng}, {Xia}, {Mao}, {Wu}  \&
  {Deng}}{{Zheng} et~al.}{2002}]{zheng02}
{Zheng} X.~Z.,  {Xia} X.~Y.,  {Mao} S.,  {Wu} H.,   {Deng} Z.~G.,  2002,
  \mn@doi [\aj] {10.1086/340964}, \href
  {https://ui.adsabs.harvard.edu/abs/2002AJ....124...18Z} {124, 18}

\bibitem[\protect\citeauthoryear{{Zubovas} \& {King}}{{Zubovas} \&
  {King}}{2012}]{zubovas12}
{Zubovas} K.,  {King} A.,  2012, \mn@doi [\apjl] {10.1088/2041-8205/745/2/L34},
  \href {https://ui.adsabs.harvard.edu/abs/2012ApJ...745L..34Z} {745, L34}

\makeatother
\end{thebibliography}
